\documentclass[11pt, a4paper]{article}
\usepackage{jheparxiv}
\usepackage{wasysym}
\usepackage[utf8]{inputenc}
\usepackage{amsmath}
\usepackage{amsfonts}
\usepackage{amssymb}
\usepackage{latexsym}
\usepackage{mathrsfs}
\usepackage{mathtools}
\usepackage{braket}		
\usepackage{graphicx}
\usepackage{color}
\usepackage{xcolor}
\usepackage{slashed}
\usepackage{twistor}
\usepackage[all]{xy}
\usepackage{tikz-cd}
\usepackage{bbm}

\def\ndelta{\delta\hspace{-0.50em}\slash\hspace{-0.05em} }
\renewcommand{\by}{{\bar y}}

\newcommand{\veps}{\varepsilon}

\newcommand{\til}[1]{\widetilde{#1}}

\newcommand{\q}{{\mathbf{Q}}}

\renewcommand{\d}{\mathrm{d}}

\newcommand{\ip}{\,\lrcorner\,}
\newcommand{\eps}{\epsilon}
\newcommand{\Dbar}{\bar{D}}
\newcommand{\g}{\mathfrak{g}}

\newcommand{\mfk}[1]{\mathfrak{#1}}

\newcommand{\bigma}{\boldsymbol{\sigma}}

\newcommand{\vphi}{\varphi}
 
\renewcommand{\bz}{{\bar z}}
\renewcommand{\dal}{{\dot\alpha}}

\newcommand{\Res}[1]{\underset{#1}{\text{Res}}}

\newcommand{\br}[1]{\overline{#1}}

\newcommand{\rD}{{\mathrm{D}}}
\newcommand{\bi}{{\bar\imath}}
\newcommand{\bj}{{\bar\jmath}}

\newcommand{\Tr}{{\mathrm{Tr}}}

\newcommand{\bM}{{\mathbb{M}}}

\newcommand{\bu}{{\bar u}}
\newcommand{\bq}{{\bar q}}
\newcommand{\sJ}{{\mathsf{J}}}
\newcommand{\cK}{{\mathcal{K}}}
\newcommand{\sR}{{\mathscr{R}}}

\title{S-algebra in Gauge Theory:\\ 
Twistor, Spacetime and Holographic Perspectives}

\author[a]{Adam Kmec,}
\author[a]{Lionel Mason,}
\author[a]{Romain Ruzziconi,}
\author[b]{Atul Sharma}

\affiliation[a]{The Mathematical Institute, University of Oxford,\\ Woodstock Road, Oxford OX2 6GG, United
Kingdom\vspace{0.1cm}}
\emailAdd{adam.kmec@maths.ox.ac.uk}
\emailAdd{lmason@maths.ox.ac.uk}
\emailAdd{Romain.Ruzziconi@maths.ox.ac.uk}

\affiliation[b]{Center for the Fundamental Laws of Nature \& Black Hole Initiative,\\Harvard University, Cambridge, MA, 02138, USA \vspace{0.1cm}}
\emailAdd{atulsharma@fas.harvard.edu}

\abstract{The celestial $S$-algebra arose from a reinterpretation of collinear limits of the Yang-Mills S-matrix as  OPEs in celestial holography.   It was subsequently represented via asymptotic charge aspects defined in the Yang-Mills radiative phase space defined at null infinity on the one hand, and via a twisted holography vertex algebra construction in twistor space on the other. Here we first identify it with the traditional symmetry algebra of self-dual Yang-Mills theory as an integrable system via its hierarchies of conserved quantities and associated flows; the self-dual phase space can be canonically identified with that of full Yang-Mills at null infinity $\scri$.   We derive the associated canonical generators from the twistor space action, identifying two infinite towers of charges corresponding to the two gluon helicities. These expressions are translated into spacetime data at null infinity using twistor integral formulae. Examining the charge algebra at spacelike infinity reveals the vertex algebras studied in the context of twisted holography. Our discussion extends directly to the celestial  $\text{LHam}(\C^2)$ symmetries of self-dual gravity. This analysis provides a unified framework for celestial symmetries, connecting twistor, spacetime, and holographic approaches and culminating in a nonlinear extrapolate dictionary for self-dual gauge theory.}

\begin{document}

\maketitle

\section{Introduction}
\label{sec:intro}

Asymptotic symmetries have played a vital role in our understanding of flat space holography. Conventionally, they arise as gauge symmetries that are non-vanishing at infinity so that they produce non-vanishing Noether charges, taken modulo those gauge symmetries that do not. In contrast, in recent years, a myriad of new asymptotic and holographic symmetries have been discovered by analyzing the infrared properties of scattering amplitudes. These new symmetries are yearning for a Hamiltonian formulation in terms of Noether charges and conservation or flux-balance laws.

Celestial symmetries such as the $\text{LHam}(\C^2)$ algebra for gravity\footnote{Also known as $\text{L}w_{1+\infty}$ in celestial holography literature, this is the loop algebra of the algebra of canonical transformations of $\C^2$ equipped with its standard holomorphic Poisson bracket.} and the $S$-algebra for Yang-Mills arise from universal properties such as the soft and collinear limits of scattering amplitudes. They underpin the definition of soft graviton and gluon operators in the holographically dual celestial conformal field theories (CCFTs) \cite{Strominger:2021mtt,Guevara:2021abz}.
In \cite{Adamo:2021lrv}, the appearance of  such
$\text{LHam}(\C^2)$ symmetries in asymptotically flat spacetimes was traced back to Penrose’s nonlinear graviton construction of self-dual spacetimes from twistor space data \cite{Penrose:1976jq,Penrose:1976js}.  There, the structure preserving diffeomorphisms of twistor space reduce to the loop algebra of area-preserving diffeomorphisms, giving a geometric realization of $\text{LHam}(\C^2)$. The construction uses such diffeomorphisms to deform the complex structure on twistor space and establishes a correspondence with self-dual 4-manifolds.

The corresponding analysis for Yang-Mills theory yields the celestial OPE algebra of soft gluon operators, known as the $S$\emph{-algebra},
\begin{equation}
    [\xi_{k,l,r}^a , \xi_{m,n,s}^b] = f^{ab}{}_c\,\xi_{k+m,l+n,r+s}^c 
\end{equation}
with $f^{ab}{}_c$ the structure constants of the Lie algebra $\g$ of the gauge group, $a,b,c$ the color indices, and labels $k, l,m,n\in\Z_{\geq0}$, $r,s\in\Z$. This is the loop algebra of the Lie algebra of holomorphic maps $\C^2\to\g$ and, as remarked in \cite{Adamo:2021lrv} and made explicit in \cite{Costello:2022wso}, can be represented as the space of local maps from twistor space to the gauge group as in \eqref{twistor-S} underpinning Ward's twistor description \cite{Ward:1977ta} of self-dual Yang-Mills fields.

Both celestial symmetry algebras, $\text{LHam}(\C^2)$ for gravity and $S$-algebra for Yang-Mills, admit three different interpretations that can be discussed in parallel: 
\begin{enumerate} 

\item {\bf In twistor space:}  The action of  $\text{LHam}(\C^2)$ as deformations of twistor space was introduced in \cite{Penrose:1976jq,Penrose:1976js},  reformulated in the context of integrability in \cite{Boyer:1985aj,Mason:1990,Dunajski:2000iq} and in celestial holography in
\cite{Adamo:2021lrv,Mason:2022hly,Bu:2022iak,Kmec:2024nmu}. In this context, the celestial symmetries have a clear geometric interpretation: they correspond to holomorphic diffeomorphisms of  twistor space that preserve a fibration and a Poisson bracket and can be used to define deformations. 

Ward's twistor construction for self-dual Yang-Mills \cite{Ward:1977ta} provides the  analogous vehicle for the interpretation of the $S$-algebra: they  are local holomorphic gauge transformations of holomorphic bundles on twistor space that define deformations  corresponding to self-dual gluons. These ideas were revisited in the context of integrability in \cite{Chau:1982mj,Takasaki:1984tp, Mason:1988xd, Mason:1991rf}, and in ambitwistor-string and twistor-string  approaches to soft theorems and celestial holography in \cite{Geyer:2014lca, Costello:2022wso,Adamo:2021zpw,Adamo:2022wjo,Bittleston:2022jeq}.  

\item {\bf At null infinity, $\scri$:} A spacetime interpretation of $\text{LHam}(\C^2)$ was provided in \cite{Freidel:2021ytz,Geiller:2024bgf,Cresto:2024fhd,Cresto:2024mne,Miller:2025wpq,Ruzziconi:2025fct}. In this case, the symmetries are realized as a charge algebra in the radiative phase space at null infinity, using the Ashtekar-Streubel symplectic structure \cite{Ashtekar:1981bq}. The analogous discussion for the $S$-algebra was given in \cite{Freidel:2023gue,Cresto:2025bfo,Miller:2024zmc}, together with an interpretation in terms of overleading gauge transformations provided in  \cite{Nagy:2024dme,Nagy:2024jua}. 

\item {\bf Celestial OPEs and twisted holography:} As already mentioned, the bottom-up holographic construction of celestial symmetries was presented in \cite{Strominger:2021mtt,Guevara:2021abz}, followed up in \cite{Ball:2021tmb,Himwich:2021dau,Mago:2021wje,Hu:2023geb,Mason:2023mti,Ruzziconi:2024kzo}. A top-down perspective on celestial symmetries via twisted holography in twistor space was developed in \cite{Costello:2018zrm,Costello:2022wso,Costello:2022upu,Costello:2023hmi} within the framework of twisted holography and Koszul duality.
\end{enumerate}

Although these three approaches, in principle, deal with the same symmetries in asymptotically flat spacetimes, their interrelationships have been far from clear, reflecting different starting points, twistor space, versus spacetime versus (twisted) holography. A key step toward relating the twistor and spacetime approaches was achieved in \cite{Kmec:2024nmu}: the spacetime expressions for the $\text{LHam}(\C^2)$ charges at null infinity were obtained from first principles, starting with a twistor space action for self-dual gravity.\footnote{Here by self-dual gravity, we mean self-dual gravity together with an anti-self-dual linear field propagating on the self-dual background; its phase space can be identified with that of full gravity at $\scri$. We will use a similar description for self-dual Yang-Mills.} However, the connection with the top-down holographic description of these symmetries remained unclear.

In this work, we push this unification further, focusing on Yang-Mills theory: we offer a common framework to deal with the $S$-algebra symmetries from twistor, spacetime, and holographic perspectives. In particular, we connect to integrability structures arising from the identification of the Yang-Mills phase space at $\scri$ with that with only  self-dual interactions.  This theory is completely integrable, and   exhibits a hierarchy of conserved quantities and their corresponding flows.  They preserve the flows with only self-dual interactions, but their conservation in the full theory may be obstructed.

The rest of the paper is organized as follows. In Section \ref{sec:twistor}, we review the uplift of self-dual Yang-Mills (sdYM) to a theory living on twistor space. We provide a brief discussion of the associated phase space and action of the S-algebra.  We then introduce the frames that bridge twistor space to spacetime, and in particular at $\scri$; in particular, at $\scri$, the frames can be understood as gauge transformations that map twistor data for solutions to the self-duality equations to radiative gauge data at $\scri$.

This is followed by an introduction to the Freidel-Pranzetti-Raclariu (FPR) charge aspects. These charge aspects are a family of asymptotic Penrose integrals of the kind studied in \cite{Bramson:1977edc}. They act as building blocks of Noether charges of the $S$-algebra \cite{Freidel:2023gue,Cresto:2025bfo}. They are functions of Bondi coordinates $(u,z,\bz)$ on $\scri$ and encode asymptotic data of spacetime fields at null infinity. We construct both positive and negative helicity charge aspects, denoted $R_s$ and $\til R_s$ respectively, and labeled by a `spin' $s\in\Z_{\geq-1}$. Using the twistor uplift of sdYM, we show that for $s\geq0$, the charge aspects obey the recursion relations,
\be
\begin{split}
    \p_uR_s + \p_\bz R_{s-1} + [\bar\cA,R_{s-1}] &= 0\,,\\
    \p_u\til R_s + \p_\bz\til R_{s-1} + [\bar\cA,\til R_{s-1}] &= 0\,.
\end{split}
\ee
Here, $\bar\cA=\lim_{r\to\infty}A_\bz$ is the radiative data of the spacetime gauge field $A$. The seed $R_{-1}$ for the positive helicity recursion is the asymptotic data of the $B_{uz}$ component of the anti-self-dual field strength $B$. Similarly, the seed for the negative helicity recursion is $\til R_{-1}=\bar\cA$. 

In Section \ref{sec:space}, we show how radiative gauge arises as a  light-cone gauge of `$\cK$-matrix-type'  \cite{Mason:1991rf} adapted to Bondi coordinates. Our set-up enlarges the previous analysis of positive helicity charge aspects in \cite{Kmec:2024nmu} to incorporate both gluon helicities. On spacetime, we show that the negative helicity aspects arise as ``conserved quantities'' in a Lax formulation of sdYM in the $\cK$-matrix gauge. We also find that for both helicities, the charge aspects can be interpreted as the higher conserved quantities that generate  the hierarchies of sdYM viewed as an integrable system. In this language, the FPR recursions emerge naturally from the recursion operator of sdYM as described in chapter 8 of  \cite{Mason:1991rf} but expressed at $\scri$.

In Section \ref{sec:holo}, we come to the construction of the conserved charges of the $S$-algebra. We begin by discussing boundary conditions for the fields on twistor space and find that soft gluons provide a basis of overleading modes violating these boundary conditions. It follows that they can be interpreted as generators of large gauge symmetries constituting the $S$-algebra. 

Part of this analysis was already carried out in \cite{Costello:2022wso,Costello:2023hmi} in linear theory. In the nonlinear theory, we describe a scheme to construct soft gluons in terms of a sequence of functions $(\xi_s,\phi_s)$, $s\in\Z_{\geq0}$, that also live on null infinity. We show that in radiative gauge, every solution of the `dual' recursion relations,
\be
\begin{split}
    \p_u\xi_s + \p_\bz\xi_{s+1} + [\bar\cA,\xi_{s+1}] &= 0\,,\\
    \p_u\phi_s + \p_\bz\phi_{s+1} + [\bar\cA,\phi_{s+1}] &= 0\,,
\end{split}
\ee
gives rise to soft gluon wavefunctions on twistor space. The $\xi_s$ build up soft gluons of positive helicity, and the $\phi_s$ of negative helicity. Usually the collection of $\xi_s$ or $\phi_s$ associated to a soft gluon will truncate at some highest order $s$. This identifies it as a soft gluon that occurs at order $\omega^{s-1}$ in a soft expansion of an amplitude involving a gluon of energy $\omega$.

Given a solution set $\{\xi_s\}$ or $\{\phi_s\}$ representing a specific soft gluon, we derive the associated Noether charge using the symplectic structure of the twistorial theory. There are two kinds of Noether charges to consider, depending on whether one is interested in a Carrollian or a celestial dual to flat space physics. In fact, we can state the final expressions for these charges without reference to twistor theory.

In the Carrollian case, one constructs \emph{corner charges} that are functions of $u$ and integrals over the whole celestial sphere. Via the extrapolate dictionary, they are expected to be dual to generators of the $S$-algebra in a Carrollian CFT. Eg., in section \ref{ssec:corner}, the positive helicity charge associated to a soft gluon represented by $\{\xi_s\}$ is found to be
\be
\boxed{\hspace{1cm}Q_\xi(u) = \int_{\P^1} \d^2z\;\Tr\sum_{s=0}^\infty\xi_s R_s\,.\hspace{0.95cm}}
\ee
Using the recursion relation obeyed by $R_s$ and the dual recursion obeyed by $\xi_s$, we obtain its derivative along Bondi time
\be
\p_uQ_\xi = \int_{\P^1}\d^2z\;\Tr\,R_{-1}(\p_\bz\xi_0+[\bar\cA,\xi_0])\,.
\ee
This shows that the charge is conserved in time if either the ``radiation'' $R_{-1}$ vanishes, or if $\xi_0$ is covariantly holomorphic in $z$.

In the celestial case, the corresponding notion is that of \emph{chiral currents} of a 2d CFT. In section \ref{ssec:celestial}, we find that they can be built out of the same charge aspects and symplectic structure if one tilts the Cauchy surface. The Carrollian Cauchy surfaces, associated with twistor uplifts of null infinity, wrap the celestial spheres. Our new Cauchy surfaces are twistorial analogs of the spacetime surfaces of constant Rindler time employed by Pasterski in \cite{Pasterski:2022jzc}. Their key property is that, rather than wrapping the whole sphere, they intersect each celestial sphere in a circle of fixed $|z|$. The resulting charges contain contour integrals in $z$ around $z=0$ or $z=\infty$. They also contain contour integrals over complexified $u$ along a contour surrounding $u=\infty$. In the spirit of \cite{Freidel:2022skz}, we expect that the contour in $u$ can be deformed to run over the null generators of real $\scri$ when various fields obey Schwartz boundary conditions near the points at infinity.

In a 2d CFT, the role of contour integrals in $z$ is to extract modes of local operators. Stripping them off leaves us with the celestial currents
\be
\boxed{\hspace{1cm}J_\xi(z) = \oint_{u=\infty}\d u\;\Tr\sum_{s=0}^\infty\xi_s R_{s-1}\,.\hspace{0.95cm}}
\ee
These are built from the same charge aspects as those for the previous corner charges, but integrated over a different contour. The FPR recursion and dual recursion can again be used to derive a conservation law for these currents in the sense of a 2d CFT,
\be
\p_\bz J_\xi = \oint_{u=\infty}\d u\;\Tr\,R_{-1}(\p_\bz\xi_0+[\bar\cA,\xi_0])\,.
\ee
The currents $J_\xi$ are chiral precisely under the same conditions of vanishing radiation or holomorphicity of $\xi_0$. Being chiral, they are conserved in radial evolution on the celestial sphere. Hence, they give rise to the $S$-algebra as a symmetry of celestial CFT. Very similar considerations follow for the negative helicity aspects as well.

The upshot of these techniques is that both the corner charges and celestial currents are expressed in terms of quantities that live on null infinity instead of twistor space so that their spacetime interpretation becomes immediate. The price one pays for this transparency is a loss of manifest invariance under small gauge transformations; one is restricted to working in radiative gauge. However, one can instead find a gauge-invariant construction of celestial currents at the expense of staying in twistor space instead of pushing data down to null infinity.

In section \ref{sec:extrapolate}, we develop expressions for the celestial currents that are invariant under small gauge transformations. This involves finding gauge-covariant formulae for the soft gluons in the presence of a nonlinear background gauge field; even though we are no longer attached to radiative gauge, many of these recursive techniques and Noether realizations generalize to this context. The resulting charges allow us to formulate an extrapolate dictionary for celestial holography in the sense of twisted holography on twistor space. A preliminary version of such a dictionary in an explicit top-down example was constructed in \cite{Costello:2023hmi}, though the expressions for charges provided there were only the soft parts of the charges, i.e., were linear in the gauge field. Here we find nonlinear (i.e., non-abelian) formulae for these charges that are conserved and gauge invariant in tree-level sdYM. 

We also outline how correlators of such Noether charges can be related to the correlators of $S$-algebra currents in a universal defect CFT living on twistor lines in the bulk of twistor space. This allows us to make contact with the Koszul duality based approach of Costello and Paquette \cite{Costello:2022wso} for computing form factors of gauge-invariant local operators in sdYM. 

Finally, we conclude by discussing the extension of our results to the gravitational case and potential implications of our analysis for going beyond the self-dual sector.


\section{Twistors}
\label{sec:twistor}

The $S$-algebra arises as the algebra of asymptotic symmetries of the twistor uplift of self-dual Yang-Mills (sdYM) theory \cite{Costello:2022wso}. In this section, we start by reviewing the twistor formulation of sdYM.\footnote{For more details, the interested reader may consult \cite{Adamo:2017qyl}.} Then we describe how to adapt the Ward construction of self-dual gauge fields to null infinity. This aids us in constructing Penrose integral formulae for the charge aspects of \cite{Freidel:2023gue}, referred to as Bramson-Tod integrals. These charge aspects will feed into the construction of conserved charges in both the Carrollian and celestial approaches.


\subsection{Self-dual Yang-Mills}
\label{ssec:twac}

Self-dual gauge theory is an integrable subsector of Yang-Mills theory in four dimensions. Let $\bM$ denote flat space of any signature. On $\bM$, sdYM is described by a (complex) gauge field $A$ valued in a Lie algebra $\g$, and an adjoint-valued anti-self-dual 2-form $B\in\Omega^2_-(\bM,\g)$, obeying the equations
\be\label{sdeq}
F^- = 0\,,\qquad D B = 0\,.
\ee
Here $D = \d+A$ is the gauge-covariant derivative, and $F^-$ is the anti-self-dual (asd) part of the field strength $F=D^2=\d A+\frac12\,[A,A]$. These equations follow from the action principle
\be\label{sdym}
S[A,B] = \int_\bM\Tr\,B\wedge F\,.
\ee
The solutions of \eqref{sdeq}, as well as this action functional, can only be real in Euclidean or $(2,2)$ signatures. They are necessarily complex in $(1,3)$ signature.

The first equation $F^- = 0$ imposes self-duality of $A$.  The second equation $DB = 0$ is a Bianchi identity that tells us that $B$ is an asd perturbation propagating on the self-dual (sd) background described by $A$. In particular, in flat space, linear perturbations of $A$ and $B$ would precisely coincide with gluons of positive and negative helicity respectively.

\paragraph{Twistor uplift.} One way to state the integrability of the self-duality equations \eqref{sdeq} is that they can be uplifted to twistor space. Let $\CO(n)\to\P^k$ denote the line bundle of first Chern class $n$ over complex projective $k$-space $\P^k$. The twistor space of flat space is $\PT = \CO(1)\oplus\CO(1)\to\P^1$. It is useful to think of this as a subset of $\P^3$. If we place homogeneous coordinates $Z^A = (\lambda_\al,\mu^\dal)$, $\al=0,1$, $\dal=\dot0,\dot1$, on $\P^3$, we find that
\be
\PT = \P^3 - \{\lambda_\al=0\}\,.
\ee
Then $\mu^\dal$ become coordinates along the fibers of $\CO(1)\oplus\CO(1)$, and $\lambda_\al$ become homogeneous coordinates along the base $\P^1$. 

The indices $\al,\dal$ are spinor indices of the complexified Lorentz group $\SL_2(\C)_\text{left}\times\SL_2(\C)_\text{right}$. They may be raised or lowered using Levi-Civita symbols: $\lambda^\al = \veps^{\al\beta}\lambda_\beta$ and $\mu_\dal = \mu^{\dot\beta}\veps_{\dot\beta\dal}$. We will use the conventions $\veps^{\al\beta}\veps_{\gamma\beta} = \delta^\al_\gamma$ and $\veps^{\dal\dot\beta}\veps_{\dot\gamma\dot\beta} = \delta^\dal_{\dot\gamma}$. We will also make use of the spinor inner products $\la\lambda\omega\ra = \lambda^\al\omega_\al$ and $[\mu\rho] = \mu^\dal\rho_\dal$.

Every point $x^{\al\dal}$ in complexified flat space corresponds to a projective line $L_x\simeq\P^1$ in twistor space cut out by the incidence relations
\be
L_x\;:\quad \mu^\dal = x^{\al\dal}\lambda_\al\,.
\ee
The excised locus $\{\lambda_\al=0\}\subset\P^3$ may also be thought of as the line corresponding to the spatial infinity $i^0$ of Minkowski space. We will denote this line by $I$.

By a theorem of Ward, self-dual gauge fields on suitable regions in complexified flat space can be uplifted to holomorphic vector bundles on $\PT$.
\paragraph{Theorem.}(Ward \cite{Ward:1977ta}) \textit{ There is a one-to-one correspondence between:
\begin{enumerate}
    \item self-dual gauge fields on $\mathbb{M}$ up to gauge, and
    \item holomorphic vector bundles $E\rightarrow \mathbb{PT} $ with $E\vert_{L_x}$ topologically trivial for every twistor line $L_x$ corresponding to $x\in \bM$. 
\end{enumerate}}
\noindent Holomorphic vector bundles may be described as complex vector bundles carrying integrable Dolbeault operators. Working in a smooth frame, a Dolbeault operator on $E$ takes the form $\Dbar = \dbar+a$, where $\dbar = \d\bar Z^{\bar A}\p_{\bar Z^{\bar A}}$, and $a\in\Omega^{0,1}(\PT,\g)$ is a partial connection.\footnote{In this work, $\Omega^{p,q}(X,\cE)$ will denote the space of smooth -- not necessarily holomorphic -- sections of the sheaf of $(p,q)$-forms on the indicated complex manifold $X$, valued in some vector bundle $\cE$.} The self-duality equation $F^-=0$ uplifts to the condition of integrability of $\Dbar$,
\be\label{aeq}
\Dbar^2 \equiv \dbar a + \frac12\,[a,a] = 0\,.
\ee
This ensures that the complex vector bundle $E$ indeed carries a holomorphic structure.

This integrability condition can be derived as the equation of motion of a holomorphic BF theory on twistor space,
\be\label{twac}
S[a,b] = \int_{\PT}\D^3Z\wedge\Tr\,b\wedge\left(\dbar a + \frac12\,[a,a]\right)\,.
\ee
Here $\D^3Z = \frac16\,\veps_{ABCD}\, Z^A\,\d Z^B\wedge\d Z^C\wedge\d Z^D$ is a weighted holomorphic volume form on $\PT$, and $b\in\Omega^{0,1}(\PT,\CO(-4)\otimes\g)$ is a Lagrange multiplier field that imposes \eqref{aeq}. Being a section of $\CO(-4)$, $b$ has weight $-4$ in $Z^A$, ensuring that the Lagrangian is weightless. 

The equation of motion of $a$ is
\be\label{beq}
\Dbar b \equiv \dbar b + [a,b] = 0\,.
\ee
We can construct the spacetime field $B$ from $b$ through a Penrose integral formula,
\be\label{BPenrose}
B(x) = \int_{L_x}\D^3Z\wedge f^{-1}bf\,,
\ee
where $f$ is any global holomorphic frame of $E|_{L_x}$ (which exists by the assumptions of Ward's theorem). On every line $L_x$, holomorphicity of $f$ is encoded by the equation
\be\label{Fhol}
\dbar|_{L_x}f + a|_{L_x}f = 0\,.
\ee
On the support of \eqref{beq}, the integral \eqref{BPenrose} obeys the expected equation of motion $DB = 0$.

\paragraph{Gauge symmetries.} Infinitesimal gauge symmetries of the action \eqref{twac} are given by
\begin{equation} \label{gauge}
        \begin{split}
            &\delta_\xi a = \bar D \xi\,,\qquad\delta_\xi b = [b,\xi]\,,\\
            &\delta_\phi a = 0\,,\qquad\hspace{0.32cm}\delta_\phi b =\bar D \phi\,,
        \end{split}
\end{equation}
where $\xi \in \Omega^{0}(\PT , \mathfrak{g})$ and $\phi \in \Omega^{0}(\PT, \mathcal{O}(-4)\otimes \mathfrak{g})$ are the gauge parameters. The first of these are the standard gauge transformations corresponding to rotations of frames of $E$, and the second ensures that on-shell $b\in H^{0,1}(\PT,\CO(-4)\otimes\mathfrak{g})$. They form the gauge algebra
\be\label{dbracket}
[\delta_\xi,\delta_{\xi'}] = \delta_{[\xi,\xi']}\,,\qquad[\delta_\xi,\delta_\phi] = \delta_{[\xi,\phi]}\,,\qquad[\delta_\phi,\delta_{\phi'}] = 0\,.
\ee
The $S$-algebra arises as the corresponding algebra of large gauge transformations. 

In section \ref{sec:holo}, we will introduce boundary conditions with respect to which the large gauge transformations in linear theory are spanned by generators of the form
\be
\begin{split}
    \xi_{k,l,r}^a &= \frac{(\mu^{\dot0})^k(\mu^{\dot1})^l}{(\lambda_1)^r(\lambda_0)^{k+l-r}}\,\mfk{t}^a\,,\\
    \phi_{k,l,r}^a &= \frac{(\mu^{\dot0})^k(\mu^{\dot1})^l}{(\lambda_1)^r(\lambda_0)^{4+k+l-r}}\,\mfk{t}^a\,, \label{twistor-S}
\end{split}
\ee
where $k,l\in\Z_{\geq0}$ label elements in the `wedge', $r\in\Z$ is a loop parameter, and $\mfk{t}^a\in\g$ is a generator of the Lie algebra $\g$. These transformations occur in one-to-one correspondence with soft gluons of positive and negative helicity respectively. The algebra of these gauge parameters is
\be
\boxed{\hspace{1cm}\begin{split}
    [\xi^a_{k,l,r},\xi^b_{k',l',r'}] &= f^{ab}{}_c\,\xi^c_{k+k',l+l',r+r'}\,,\\
    [\xi^a_{k,l,r},\phi^b_{k',l',r'}] &= f^{ab}{}_c\,\phi^c_{k+k',l+l',r+r'}\,,
\end{split}\hspace{0.95cm}}
\ee
and the associated algebra of large gauge transformations obtained from \eqref{dbracket} is known as the $S$-algebra. We will also see how the same algebra continues to hold in nonlinear theory when the Lie bracket is modified to handle field-dependent gauge transformations.

Returning to \eqref{gauge}, the finite gauge transformations generated by $\xi$ are
\be
a \mapsto g^{-1}ag + g^{-1}\dbar g\,,\qquad b\mapsto g^{-1}bg\,,
\ee
with $g = \e^\xi$. Under such a transformation, \eqref{Fhol} requires that the frame $f$ rotate by
\be
f\mapsto g^{-1}f\,.
\ee
This ensures for instance that $f^{-1}bf$ is gauge invariant, so that \eqref{BPenrose} is well-defined. We will use this gauge symmetry to go to a $\cK$-matrix gauge when studying conserved quantities generated from the integrability of sdYM.

\paragraph{Symplectic structure and conserved charges.} The $S$-algebra constitutes the asymptotic symmetries of sdYM theory. Before closing this section, let us also review the symplectic form of the twistorial theory, which we will use in section \ref{sec:holo} to construct Noether charges of such asymptotic symmetries. 

The presymplectic potential is obtained as the boundary term in the on-shell variation of the action \eqref{twac}. Along an on-shell variation $\delta$ around a point $(a,b)$ in phase space, it is explicitly given by
\begin{equation}
    \boldsymbol\theta = \int_{\Sigma} \D^3Z \wedge \text{Tr}\,b\wedge \delta a\,, 
\end{equation} where $\Sigma$ is a real-codimension-1 ``Cauchy'' hypersurface in twistor space that will be specified in section \ref{sec:holo}. The presymplectic form is obtained by taking one more phase space variation,
\begin{equation}
   \boldsymbol{\omega} = \int_\Sigma \D^3 Z \wedge \Tr\,\delta b\wedge \delta a \,.
\end{equation}
The symplectic form would then be obtained by performing symplectic reduction by the gauge symmetries. But $\boldsymbol\omega$ is all that we will ever need.

Standard Noether arguments give boundary charge integrals associated with gauge transformations \eqref{gauge} generated by $\alpha = (\xi,\phi)$. The gauge variations act as vector fields on phase space,
\be
\delta_\al = \delta_\al a\,\frac{\delta}{\delta a} +  \delta_\al b\,\frac{\delta}{\delta b}\,.
\ee
Contracting the presymplectic form with these vector fields yields real-codimension-1 integrands that are total derivatives when the equations of motion are imposed. This gives rise to Noether charges that are codimension-2 boundary integrals \cite{Dirac:1936fq, Iyer:1994ys,Barnich:2001jy},
\be
\begin{split}
    \ndelta H_\al &= \delta_\al\ip\boldsymbol\omega = \int_\Sigma \d k_\al = \int_{\p\Sigma}k_\al\,,\\
    k_\al &= \D^3 Z \wedge \Tr\,(\xi\, \delta b + \phi\, \delta a )\,.
\end{split}
\ee
Here $k_\alpha$ is a 4-form on twistor space which can be integrated over $\partial \Sigma$.

The notation $\ndelta$ allows for the possibility that $\alpha$ depends on the fields $a,b$ (perhaps due to gauge fixing),\footnote{For instance, our asymptotic symmetry generators will get related to soft gluon wavefunctions around generic backgrounds $(a,b)$, so they will have to depend nonlinearly on $a$.} so we must subtract off their gauge variation. The resulting charges can thus be written as 
\begin{equation}
    \ndelta H_{\xi,\phi} \vcentcolon=  \delta H_{\xi, \phi} - H_{\delta \xi, \delta\phi} \,, \qquad    H_{\xi,\phi} = \int_{\partial \Sigma} \D^3 Z \wedge \Tr\,(\xi\, b + \phi\, a )\, .
\label{non-integrable charges}
\end{equation} With the subtraction of $\delta \alpha$ implied by $\ndelta$, the variation is generally \emph{not} $\delta$-exact.  The charges become integrable, i.e., $\delta$-exact, if the gauge parameters are field independent,  $\delta \xi = 0 = \delta \phi$. But generally, this is the case only around a trivial background $a=0$.

For generic non-integrable expressions, the Barnich-Troessaert bracket \cite{Barnich:2011mi}
\begin{equation}\label{BT}
    \{H_{\alpha} , H_{\alpha'}\}_{\star} = \delta_{\alpha} H_{\alpha'} - H_{\delta_{\alpha'}\alpha}
\end{equation} 
 can be used to write the charge algebra. We find
\begin{equation}
    \{H_{\alpha} , H_{\alpha'}\}_{\star}  = H_{[\alpha,\alpha']_\star} + K_{\alpha,\alpha'} \, .\label{charge algebra generic}
\end{equation}
The modified bracket \cite{Barnich:2010eb} on the right-hand side takes into account the possible field dependence of the symmetry parameters $\xi$ and $\phi$ and reads explicitly as 
\begin{equation}\label{alal'}
    [\alpha, \alpha ']_\star = \left([\xi,\xi'] + \delta_{\alpha} \xi' - \delta_{\alpha'}\xi \,, [\phi,\xi']+[\xi,\phi'] + \delta_{\alpha}\phi' - \delta_{\alpha'}\phi  \right)\, .
\end{equation}
The (possibly field-dependent) cocycle in the right-hand side is given by
\begin{equation} \label{cocycle}
    K_{\alpha,\alpha'} = \int_{\partial\Sigma}\D^3 Z \wedge\text{Tr}\left(\xi' \bar\partial \phi + \phi' \bar\partial \xi \right)\,.
\end{equation} It satisfies the generalized cocycle condition \cite{Barnich:2011mi}
\begin{equation}
    K_{[\alpha_1,\alpha_2]_* , \alpha_3} - \delta_{\alpha_3}K_{\alpha_1,\alpha_2} + \text{cyclic(1,2,3)} = 0
\end{equation} which extends the standard cocycle condition to the field-dependent case. The BRST interpretation of field-dependent cocycles has been discussed in \cite{Barnich:2017ubf,Baulieu:2023wqb}.

Another subtlety associated with these charges is that even though the integrand of $H_\xi$ is adjoint-valued, the integrand of $H_\phi$ is $\phi\,a$, which isn't adjoint-valued because it contains a connection. Because of this, it does not appear to obey any gauge-covariant conservation or flux-balance law in nonlinear theory. A simple way around this is to replace $a$ by any Lie derivative $\cL_\ell a$ of $a$ along a holomorphic vector field $\ell$ on $\PT$.\footnote{Holomorphic here means that $\ell$ is a $(1,0)$-vector that commutes with the $(0,1)$-vectors $\p/\p\bar Z^{\bar A}$.} In later sections, we will often choose to project twistor space onto null infinity and take $\ell$ to be $\p_u$, the generator of Bondi time translations.

This defines a new type of charge that we will label $\til H_\phi$. In summary, we will take the definition of our boundary charges to be
\be\label{HHt}
\begin{split}
    H_\xi &= \int_{\p\Sigma}\D^3Z\wedge\Tr\,\xi\,b\,,\\
    \til H_\phi &= \int_{\p\Sigma}\D^3Z\wedge\Tr\,\phi\,\cL_\ell a\,.
\end{split}
\ee
As usual, variations of a connection transform in the adjoint. Explicitly, under $a\mapsto a+\Dbar\xi$, we get $\cL_\ell a\mapsto\cL_\ell a + [\cL_\ell a,\xi] + \Dbar(\cL_\ell\xi)$, so $\cL_\ell a$ transforms just like $b$. The resulting charge $\til H_\phi$ will be seen to be invariant under small gauge transformations. We can also obtain $\til H_\phi$ by applying a field-dependent parameter redefinition $\phi\mapsto-\cL_\ell\phi$ (sometimes referred to as a ``change of slicing'', see e.g. \cite{Adami:2020ugu,Barnich:2019vzx,Ruzziconi:2020wrb,Adami:2021nnf,Geiller:2021vpg,Grumiller:2023ahv,Fiorucci:2024ndw}).

These charges are conserved in the following sense. Let $\bar m$ denote a $(0,1)$-vector field normal to $\p\Sigma$. Acting with $\bar m$ on the charges produces (we will see many explicit examples of this general computation in the text)
\be
\begin{split}
    \bar m(H_\xi) &= \int_{\p\Sigma}\D^3Z\wedge\bar m\ip\Tr(\Dbar\xi\wedge b + \xi\,\Dbar b)\,,\\
    \bar m(\til H_\phi) &= \int_{\p\Sigma}\D^3Z\wedge\bar m\ip\Tr(\Dbar\phi\wedge \cL_\ell a + \phi\,\Dbar\cL_\ell a)\,.
\end{split}
\ee
On-shell, $b$ satisfies $\Dbar b=0$. Similarly, acting on $\Dbar^2=0$ with $\cL_\ell$ shows that $\cL_\ell a$ satisfies $\Dbar(\cL_\ell a) = 0$, which reconfirms that it transforms in the adjoint (up to $\Dbar$-exact terms as seen above). So the fluxes along $\bar m$ reduce to
\be
\begin{split}
    \bar m(H_\xi) &= \int_{\p\Sigma}\D^3Z\wedge\bar m\ip\Tr(\Dbar\xi\wedge b)\,,\\
    \bar m(\til H_\phi) &= \int_{\p\Sigma}\D^3Z\wedge\bar m\ip\Tr(\Dbar\phi\wedge \cL_\ell a)\,.
\end{split}
\ee
These will vanish if $\Dbar\xi=\Dbar\phi = 0$ holds in the region of interest, which will be a strong version of our definition of large gauge parameters. 

We will also encounter weaker versions of this definition where instead of vanishing identically, $\Dbar\xi$ and $\Dbar\phi$ reduce to gauge transformations of radiative data on null infinity. From the twistor perspective, these will be residual gauge symmetries of radiative gauge. In this case, we will find that the charges are conserved if radiation vanishes. If the radiation doesn't vanish, it will give us flux-balance laws. We return to these considerations in sections \ref{sec:holo} and \ref{sec:extrapolate}.

The algebra of $H_\xi$ and $\til H_\phi$ can be computed by noting that $\til H_\phi$ generates the gauge transformations
\be
\til\delta_\phi a = 0\,,\qquad\til\delta_\phi b = -\Dbar(\cL_\ell\phi)\,.
\ee
Using the definition \eqref{BT}, a short calculation shows that
\be
\begin{split}
    \{H_\xi,\til H_\phi\}_\star &= \til H_{[\xi,\phi]_\star} + \til K_{\xi,\phi}\,,\\
    \{\til H_\phi,\til H_{\phi'}\}_\star &= 0
\end{split}
\ee
where $[\xi,\phi]_\star=[\xi,\phi]+\delta_\xi\phi$ as before, and the new field-dependent cocycle can be written as 
\be
    \til K_{\al,\al'} = \int_{\p\Sigma}\D^3Z\wedge\Tr\left(\cL_\ell\xi'\Dbar\phi-\cL_\ell\xi \Dbar\phi' \right)
\ee after some integration by parts. It satisfies the 
generalized cocycle condition
\begin{equation}
    \til K_{[\alpha_1,\alpha_2]_\star , \alpha_3} - \til \delta_{\alpha_3} \til K_{\alpha_1,\alpha_2} + \text{cyclic}(1,2,3) = 0 \,.
\end{equation} In short, the algebra of the adjoint-valued charges is also governed by the same Barnich-Troessaert bracket \eqref{BT} together with the modified bracket $[\al,\al']_\star$ of the gauge parameters \eqref{alal'}. 

\subsection{Asymptotic twistor geometry}
\label{ssec:q}

In section \ref{sec:holo}, we will see that Noether charges of the $S$-algebra are naturally expressible as combinations of certain building blocks called \emph{charge aspects}. These were first formulated in a series of remarkable papers by Freidel-Pranzetti-Raclariu (FPR) in gauge theory and gravity \cite{Freidel:2021dfs, Freidel:2021ytz, Freidel:2022skz,Freidel:2023gue}, and then further explored in \cite{Geiller:2024bgf,Cresto:2024fhd,Cresto:2024mne,Cresto:2025bfo,Cresto:2025ubl}.

In this section, we provide the background geometry needed to obtain these charge aspects from twistor space. In the sections that follow, we will put this to use in attaining radiative gauge and constructing integral formulae for the aspects.

\paragraph{Asymptotic twistor lines.} We begin by studying the geometry of twistor lines associated to points at conformal infinity. To do this, we first need to compactify spacetime by adding in the conformal boundaries.

Complexified compactified Minkowski space is the Grassmannian $\Gr_2(\C^4)$ of 2-planes in 4 dimensions. An element of $\Gr_2(\C^4)$ is described by a $2\times4$ matrix $X^{\al A}$, taken up to $\GL_2(\C)$ rotations on the left. The spinor indices $\al=0,1$, $\dal=\dot0,\dot1$, and the twistor index $A=(\al,\dot\alpha)$ are the same as before. So we have the decomposition $X^{\al B} = (X^\al{}_\beta,X^{\al\dot\beta})$. Each such point corresponds to a parametrized twistor line in $\P^3$,
\be
Z^A = X^{\al A}\sigma_\al\,,
\ee
where $\sigma_\al\in\P^1$ is the parameter. Penrose defines a conformal structure on $\Gr_2(\C^4)$ by declaring two points to be null separated if and only if their twistor lines intersect.

By a $\GL_2(\C)$ rotation, matrices $X^{\al B}$ satisfying $\det(X^\al{}_\beta)\neq0$ may be brought to the form $X^{\al B} = (\delta^\al_\beta,x^{\al\dot\beta})$. Such points are just points $x^{\al\dal}$ in the interior of complexified flat space $\C^4$. The associated twistor line becomes $\lambda_\al = \sigma_\al$, $\mu^\dal = x^{\al\dal}\sigma_\al$, which is equivalent to $\mu^\dal = x^{\al\dal}\lambda_\al$.

The remaining points in $\Gr_2(\C^4)$ constitute ``boundary'' divisors. Complexified null infinity is defined to be the set of points for which $\det(X^\al{}_\beta) = 0$ and $X^\al{}_\beta$ has rank 1. In this case, $X^\al{}_\beta$ can be split as a product of two undotted spinors. We will find it useful to parametrize the resulting $X^{\al B}$ as
\be
X^{\al B} = (T^{\al\dal}\tilde\zeta_\dal\zeta_\beta, uT^{\al\dot\beta})\,,\qquad T^{\al\dal} = \begin{pmatrix}
    1&&0\\0&&1
\end{pmatrix}\,.
\ee
$\frac{1}{\sqrt2}\,T^{\al\dal}$ plays the role of the unit timelike vector in spacetime. $(u,\zeta_\al,\tilde\zeta_\dal)$ are coordinates on complexified null infinity. $u$ plays the role of (homogeneous) Bondi time, and $\zeta_\al\tilde\zeta_\dal$ is a null vector, taken up to scale, that describes a point on the complexified celestial sphere $\P^1\times\P^1$. Due to the residual $\GL_1(\C)$ symmetry acting on $X^{\al B}$, the points $(u,\zeta_\al,\tilde\zeta_\dal)$ and $(t\tilde t u,t\zeta_\al,\tilde t\tilde\zeta_\dal)$ get identified for any $t,\tilde t\in\C^\times$.

In this work, we will only encounter partially complexified null infinity, denoted $\scri_\C$. This is the subset on which $\tilde\zeta_\dal$ is the complex conjugate of $\zeta_\al$,
\be
\scri_\C\;:\quad \tilde\zeta_\dal = \bar\zeta_\dal\,.
\ee
$\scri_\C$ has the topology of the line bundle $\CO(1)\otimes\overline{\CO(1)}\to\P^1$, with $\zeta_\al$ being coordinates on the base $\P^1$ and $u$ being the fiber coordinate. The twistor lines of points on $\scri_\C$ are found to be
\be\label{Luzeta}
L_{u,\zeta}\;:\quad\lambda_\al = \zeta_\al\bar\zeta_\dal q^\dal\,,\quad\mu^\dal = uq^\dal\,,
\ee
where we have introduced a new parameter $q^\dal = T^{\al\dal}\sigma_\al\in\P^1$. We will refer to these as asymptotic twistor lines.

Similarly, the point at infinity $i^0$ is defined as the point at which $X^\al{}_\beta$ has rank 0. This means that $X^\al{}_\beta$ vanishes identically, and $\GL_2(\C)$ rotations bring the remaining entries of $X^{\al B}$ to the form
\be
i^0\;:\quad X^{\al B} = (0,T^{\al\dot\beta})\,.
\ee
The associated twistor line is seen to be precisely $I = \{\lambda_\al=0\}$.

All asymptotic twistor lines intersect the line at infinty $I$, with the intersection point being
\be
L_{u,\zeta}\cap I = \{[q\bar\zeta] = 0\}\,.
\ee
So, the point $[q\bar\zeta]=0$ on each asymptotic twistor line gets removed when $I$ is excised from $\P^3$ to obtain $\PT$. Because of this, the projective asymptotic twistor lines become non-compact affine lines in $\PT$. Moreover, the twistor lines of two points $(u_1,\zeta_1)$ and $(u_2,\zeta_2)$ intersect iff $\zeta_1^\al\propto\zeta_2^\al$, and this intersection point is once again of the form $[q\bar\zeta_1]=[q\bar\zeta_2]=0$. Importantly, the asymptotic twistor lines never intersect within $\PT$. In spacetime language, this is just saying that two points on $\scri_\C$ are null separated iff they lie on the same null generator of $\scri_\C$, and they are also simultaneously null separated from $i^0$.

\paragraph{Affine coordinates.} In the rest of this work, we will mostly employ a system of affine coordinates on $\scri_\C$ and the associated twistor lines. A useful choice is to introduce pairs of basis spinors $o_\al=o_\dal=(1,0)$ and $\iota_\al=\iota_\dal=(0,1)$ and decompose
\be\label{affine}
\zeta_\al = o_\al + z\iota_\al\,,\qquad q^\dal = \iota^\dal - \frac{q}{u}\,\bar\zeta^\dal\,.
\ee
We will interchangeably use the notation $L_{u,\zeta}$ or $L_{u,z}$ for the same asymptotic twistor lines, whose equations now read
\be\label{Luz}
L_{u,z}\;:\quad\lambda_\al = \begin{pmatrix}
    1\\z
\end{pmatrix}\,,\quad\mu^\dal = \begin{pmatrix}
    u-q\bz\\q
\end{pmatrix}\,.
\ee
The affine coordinate $q$ is a close cousin of the coordinate of the same name used in \cite{Kmec:2024nmu}. Our only change in convention in this work is to use flat Bondi coordinates on $\scri_\C$ instead of the standard Bondi coordinates, as these aid in making contact with 2d CFT notation. The coordinate $q$ here is also adapted to flat Bondi coordinates.

In this language, the intersections $[q\bar\zeta]=0$ of asymptotic twistor lines with $I$ move to the points $q=\infty$ on each line. Removing these points, we get affine twistor lines that foliate $\PT=\P^3-I$. So the coordinates $(u,q,z)$ can actually be used as an alternative set of coordinates on $\PT$ itself. The change of coordinates from $Z^A$ to $(u,q,z)$ is given by
\be\label{pmap}
z = \frac{\lambda_1}{\lambda_0}\,,\qquad u = \frac{[\mu\bar\lambda]}{|\lambda_0|^2}\,,\qquad q = \frac{\mu^{\dot1}}{\lambda_0}\,,
\ee
in the patch $\lambda_0\neq0$. One can define a similar set of coordinates in the $\lambda_1\neq0$ patch using the transition functions
\be
z\mapsto\frac{1}{z}\implies u\mapsto \frac{u}{|z|^2}\,,\quad q\mapsto\frac{q}{z}\,.
\ee
This tells us that $u$ is a section of $\CO(1)\otimes\br{\CO(1)}\to\P^1$, and $q$ a section of $\CO(1)$.

Fixing the scale by setting $\lambda_0=1$, we may invert this change of coordinates to find $\lambda_\al = (1,z)$ and $\mu^\dal=(u-q\bz,q)$. In these coordinates, the asymptotic twistor lines \eqref{Luz} are manifestly the lines of fixed $u,z$. The vectors on $\PT$ that annihilate these holomorphic coordinates are spanned by
\be\label{Li}
e_\bq = \p_\bq\,,\qquad e_\bz = \p_\bz+q\,\p_u\,,\qquad e_\bu = \p_\bu\,.
\ee
Collectively, the $e_\bi$, $\bi = \bq,\bu,\bz$, span the antiholomorphic tangent bundle $T^{0,1}_{\PT}$. They commute with each other, which ensures that twistor space is a complex manifold. The dual basis of $(0,1)$-forms is
\be\label{theta}
\theta^\bq = \d\bq\,,\qquad\theta^\bz = \d\bz\,,\qquad\theta^\bu = \d\bu - \bq\,\d z\,.
\ee
We collectively denote them as $\theta^\bi$. They span the same 1-forms as $\d\bar\lambda_\dal$ and $\d\bar\mu^\al$ after projectivization.


\subsection{Radiative gauge}
\label{ssec:ward}

In section \ref{ssec:twac}, we recollected the usual Ward construction. In this section, we expand upon it in the context of extracting the asymptotic data $A|_{\scri_\C}$ of the spacetime gauge field $A$. 

The general idea is fairly universal in twistor constructions. We will construct quantities on $\PT$ that are holomorphic along asymptotic twistor lines. Then Liouville's theorem will show that they are constant along the twistor lines $L_{u,z}$ and depend only $(u,z)$. In the $(u,q,z)$ coordinates, twistor space admits a projection onto $\scri_\C$,
\be
p:\PT\to\scri_\C\,,\qquad(u,q,z)\mapsto(u,z)\,.
\ee
Via this projection, these quantities will be interpreted as asymptotic data pulled back from null infinity.

\medskip

We start with the partial connection $a$ expanded upon the basis \eqref{theta},
\be
a = a_\bi\theta^\bi = a_\bq\theta^\bq + a_\bu\theta^\bu + a_\bz\theta^\bz\,.
\ee
The twistor lines $L_{u,z}$ are simply lines of constant $u,z$. A holomorphic frame $f(u,q,z)$ on $L_{u,z}$ is a gauge transformation that trivializes the partial connection $a_\bq$ pointing along $L_{u,z}$,
\be
a_{\bq} = -\p_\bq f\,f^{-1}\,.
\ee
Extraction of spacetime data proceeds by using this frame to construct the gauge transforms of the remaining components of $a$,
\be\label{atil}
\begin{split}
    \til a_\bi &= e_\bi\ip f^{-1}(\dbar+a) f \\
    &= f^{-1}(e_\bi + a_\bi)f\,,\qquad\bi = \bu,\bz\,,
\end{split}
\ee
where $e_\bi$ are the $(0,1)$-vectors given in \eqref{Li}, and we applied $a_\bi = e_\bi\ip a$.

In this gauge, the partial connection reduces to
\be
\til a = \til a_\bu\theta^\bu + \til a_\bz\theta^\bz
\ee
and contains no $\theta^\bq$ component. The corresponding field strength $\Dbar^2=(\dbar+\til a)^2$ reads
\be
\dbar\til a + \frac12\,[\til a,\til a] = \p_\bq\til a_\bu\,\theta^\bq\wedge\theta^\bu + \p_\bq\til a_\bz\,\theta^\bq\wedge\theta^\bz + \left(e_\bz\til a_\bu-e_\bu\til a_\bz+[a_\bz,a_\bu]\right)\theta^\bz\wedge\theta^\bu\,.
\ee
As this must vanish on-shell, we obtain\footnote{Strictly speaking, we are only using a mixed tangential-normal component $\p_\bq\ip e_\bi\ip\Dbar^2=0$ of the equation of motion instead of the full equation $\Dbar^2=0$. So we can work partially off-shell. This won't make a difference for us as we will work in the Heisenberg picture where all fields are treated as operators obeying the equations of motion. Moreover, we will confine ourselves to tree-level considerations.}
\be
\p_\bq\til a_\bi = 0\,,\qquad\bi = \bu,\bz\,.
\ee
Hence, the $\til a_\bi$ are globally holomorphic on the portion of asymptotic twistor lines contained in $\PT$. In the previous section, we found that each asymptotic twistor line intersects $\PT$ in a copy of $\C$ given by $q\neq\infty$, as the point $q=\infty$ belonged to the line at infinity. Applying Liouville's theorem for $q\in\C$, we conclude that the $\til a_\bi$ must take the form
\be\label{badeq}
\til a_{\bz} \equiv \bar\cA + \sum_{n\geq1}q^n\bar\cA_n\,,\quad \til a_\bu \equiv \sum_{n\geq0}q^n\cA_{\bu,n}\,,
\ee
where $\bar\cA$, $\bar\cA_n$ and $\cA_{\bu,n}$ are functions of $u,\bu,z,\bz$ only.

We can constrain the expansion of $\til a$ in $q$ by imposing appropriate boundary conditions on $a$ as $q\to\infty$. In section \ref{ssec:bdry}, we will argue that the minimal boundary condition we need is that $a$ should vanish to first order as one approaches the line at infinity. So in particular, we will assume that $a\to 0$ as $q\to\infty$.

In fact, if allow ourselves to work partially on the support of the self-duality equations, we can further reduce the data in $\til a_\bi$ by a judicious choice of gauge. In the literature, this is referred to variously as \emph{radiative gauge}, \emph{Newman gauge} or \emph{scri gauge}. It will be an adaptation of Plebanski-type gauges to asymptotic data.

Gauge rotations $a\mapsto g^{-1}(\dbar+a) g$ act on the frame as $f\mapsto g^{-1}f$. Because $\p_\bq f\equiv -a_\bq f\to0$ as $q\to\infty$, we can use this gauge freedom to fix
\be
\lim_{q\to\infty}f = 1\,.
\ee
In this gauge, we observe that
\begin{align}
    &\lim_{q\to\infty}e_{\bz}f = \lim_{q\to\infty}(\p_\bz f+q\,\p_u f) = \lim_{q\to\infty}q\,\p_uf\,,\label{L1F}\\
    &\lim_{q\to\infty} e_{\bu}f = 0\,,\label{L2F}
\end{align}
where the limit trivially commutes through the derivatives as they do not involve $\p_q$ or $\p_\bq$. We will make the assumption that the limit $\lim_{q\to\infty}q\,\p_uf$ is well-defined. It is finite because $\p_uf\to0$ as $q\to\infty$. It will also get justified when we describe how to obtain radiative gauge directly on spacetime in section \ref{ssec:Bondi}. 

Applying \eqref{L1F} -- \eqref{L2F} to \eqref{atil} generates the limits of $\til a_\bi$ as $q\to\infty$,
\be
\lim_{q\to\infty}\til a_\bz = \p_u\big(\lim_{q\to\infty}qf\big)\,,\qquad\lim_{q\to\infty}\til a_\bu = 0\,,
\ee
where we used the boundary condition that $a_\bi\to0$ as $q\to\infty$. Comparing this with the output of Liouville's theorem \eqref{badeq}, we see that all the coefficients $\bar\cA_n,\cA_{\bu,n}$ of positive powers of $q$ must vanish. Upon identifying $\lim_{q\to\infty}q\,\p_u f\equiv\bar\cA$, we obtain the drastically simpler result
\be\label{radeq}
\til a_\bz = \bar\cA\,,\qquad\til a_\bu = 0\,.
\ee
In terms of $\bar\cA$, the asymptotic data of the spacetime gauge field takes the form
\be
A|_{\scri_\C} = \bar\cA\,\d\bz\,,
\ee
where we identified $u,z$ with coordinates on $\scri_\C$.

Comparing \eqref{radeq} with \eqref{atil}, we obtain a pair of crucial identities that will play a recurring role in many calculations,
\begin{align}
    &f^{-1}(e_\bz+a_\bz)f = \bar\cA\,,\label{F1}\\
    &f^{-1}(\p_\bu+a_\bu)f = 0\,.\label{F2}
\end{align}
They will allow us to translate between generic gauges obeying our boundary conditions and radiative gauge which is convenient for calculations.

A priori, $\bar\cA$ is a function of $u,\bu,z,\bz$, i.e., it is not necessarily holomorphic in either $u$ or $z$. The gauge transformation to radiative gauge sends the partial connection to $a\mapsto\til a = f^{-1}(\dbar+a)f=\bar\cA\,\d\bz$. The leftover curvature of $\Dbar$ in this gauge is
\be
\dbar\til a+\frac12\,[\til a,\til a] = \p_\bu\bar\cA\;\theta^\bu\wedge\theta^\bz\,.
\ee
Demanding that this vanishes imposes holomorphicity in $u$ over the domain of applicability of radiative gauge,
\be
\p_\bu\bar\cA=0\,.
\ee
So on-shell, $\bar\cA$ turns out to be a function of $u,z,\bz$ only.

Radiative gauge will be our gauge of choice in the rest of this work. In general, it can only be attained globally on null infinity if there are no Coulombic sources present in spacetime. So we will think of it as being a local gauge fixing, valid in suitably small regions of $\scri_\C$. Furthermore, this gauge fixing requires a partial imposition of the self-duality equations. This will become clearer when we formulate self-dual gauge theory in Bondi coordinates in section \ref{sec:space}.


\subsection{Charge aspects and recursion relations}
\label{ssec:fpr}

Having set up radiative gauge, we can begin to construct the charge aspects studied by FPR. The simplest way to understand the positive helicity charge aspects is as $\scri$ analogs of the Penrose integral \eqref{BPenrose}. They will be moments of $b$ evaluated along the asymptotic twistor lines. 

By virtue of being Penrose-type integrals, the charge aspects will satisfy certain equations of motion. These equations have been identified as a certain truncation of the asymptotic Bianchi identities of Yang-Mills theory \cite{Freidel:2023gue}. Our analysis confirms that the self-dual subsector is properly contained within this truncation.

The negative helicity charge aspects will then be constructed by analogy as moments of the time derivative $\p_ua$ of the partial connection. Here and in what follows, we will sometimes suppress Lie derivative notation. So for instance $\p_ua$ should be read as the Lie derivative $\cL_{\p_u}a$, etc. In section \ref{sec:space}, we will confirm that these charge aspects emerge naturally from the classical integrability of sdYM.

Before delving deeper, let us briefly summarize the results of this section for the reader's convenience. In affine coordinates $(u,q,z)$ on twistor space, the charge aspects for `spins' $s=\{-1,0,1,2,\cdots\}$ will be given by the following \emph{Bramson-Tod integrals}
\begin{equation} \label{Tod formulae}
\boxed{\begin{split}
    \hspace{1cm}R_{s}(u,z,\bar z) &= \int_{L_{u,z}} q^{s+1}\,\d q \wedge f^{-1} b f\,,\hspace{0.95cm}\\
    \hspace{1cm}\til R_{s}(u,z,\bar z) &= \int_{L_{u,z}} q^{s+1}\,\d q\wedge f^{-1}\,\partial_{u}a\,f\,.\hspace{0.95cm}
\end{split}}
\end{equation}
They will be holomorphic in $u$, so will naturally live on real null infinity. Additionally, for $s\geq0$, they will obey the \emph{Freidel-Pranzetti-Raclariu recursion relations} in flat Bondi coordinates,
\be\label{fprrec}
\boxed{\begin{split}
    \hspace{1cm}&\p_u R_s + \cD_\bz R_{s-1} = 0\,,\hspace{0.95cm}\\
    \hspace{1cm}&\p_u\til R_s + \cD_\bz\til R_{s-1} = 0\,,\hspace{0.95cm}
\end{split}}
\ee
written in terms of the gauge covariant derivative on the celestial sphere,
\be
\cD_\bz = \p_\bz + [\bar\cA,-]\,.
\ee
These represent truncations of the asymptotic Bianchi identities of Yang-Mills theory \cite{Freidel:2023gue}.

\paragraph{Positive helicity aspects.} Let us begin by constructing the positive helicity aspects $R_s$. Working in homogeneous coordinates \eqref{Luzeta} on asymptotic twistor lines, for $s\geq-1$ a first guess might be to write a spin $\frac12(s+1)$ moment of the form
\be
R^{\dal_0\dal_1\cdots\dal_s}(u,\zeta) \overset{?}{=} \int_{L_{u,\zeta}}\D q\;q^{\dal_0}q^{\dal_1}\cdots q^{\dal_s}\,f^{-1}bf\,,
\ee
where $\D q = q_\dal\d q^\dal$ and $f$ is the frame on the line $L_{u,\zeta}$. A priori, this is a function of $u,\zeta_\al$ as well as their complex conjugates. Unfortunately, in this guess, the integrand of the $q$ integral is generically \emph{not} weightless, so it is ill-defined as a projective integral.

A natural way out is suggested by working on $\PT$ instead of $\P^3$. On $\PT$, the asymptotic twistor lines are affine lines and do not include the points $[q\bar\zeta]=0$. So we can use powers of $[q\bar\zeta]$ to saturate the weight in $q$. This motivates us to work with the moments
\be\label{Rdal}
\boxed{\hspace{1cm}R^{\dal_0\dal_1\cdots\dal_s}(u,\zeta) = \int_{L_{u,\zeta}}\D q\;[q\bar\zeta]^{1-s}\,q^{\dal_0}q^{\dal_1}\cdots q^{\dal_s}\,f^{-1}bf\,.\hspace{0.95cm}}
\ee
Because $b$ is a section of $\CO(-4)$, its restriction to $L_{u,\zeta}$ carries weight $-4$ in $q^\dal$ and $0$ in $\bar q^\al$; similarly, the frame $f$ is weightless. So the integrand becomes weightless and the $q$-integral is well-defined. This is an asymptotic analog of the Penrose integral transform, cf.\ that for the gravitational asymptotic Bianchi identities in \cite{Tod:2001}.

In what follows, for ease of notation we will refer to the spin $\frac12(s+1)$ charge aspect as simply the spin $s$ charge aspect.

FPR choose to write their charge aspects in components with respect to a basis of dotted spinors. The basis used by them and also our previous work \cite{Kmec:2024nmu} is the dyad $\bar\zeta^\dal$ and $\hat{\bar\zeta}^\dal\equiv T^{\al\dal}\zeta_\al$. This corresponds to working with stereographic Bondi coordinates on flat space. In this paper, we adopt flat Bondi coordinates, which translates to working with the spinor basis formed from $\bar\zeta^\dal$ and a constant spinor $\iota^\dal = (1,0)$. In this spinor basis, one may define the spin $s$ charge aspects as\footnote{To obtain FPR's original charge aspects, replace each factor of $\iota_{\dal_k}$ by a factor of $\hat{\bar\zeta}_{\dal_k}$.}
\begingroup
\allowdisplaybreaks
\begin{align}
    R_s(u,\zeta) &= \frac{u^{s+2}}{[\iota\bar\zeta]^{s+1}}\,\iota_{\dal_0}\iota_{\dal_1}\cdots\iota_{\dal_s}R^{\dal_0\dal_1\cdots\dal_s}(u,\zeta)\nonumber\\
    &= u^{s+2}\int_{L_{u,\zeta}}\frac{\D q}{[q\bar\zeta]^{s-1}}\,\frac{[q\iota]^{s+1}}{[\iota\bar\zeta]^{s+1}}\,f^{-1}bf\,.\label{Rstospinor}
\end{align}
\endgroup
They are sections of the line bundle $\CO(s-2)\otimes\br{\CO(-s-2)}\to\scri_\C$. That is, they carry weights $(s-2,-s-2)$ in $(\zeta,\bar\zeta)$. 

In the affine coordinates $(u,q,z)$ introduced in \eqref{affine}, these reduce to
\be\label{Rs}
\boxed{\hspace{1cm}R_s(u,z) = \int_{L_{u,z}}\d q\,q^{s+1}f^{-1}bf\,.\hspace{0.95cm}}
\ee
This is the gauge theory cousin of the Bramson-Tod integrals for gravitational charge aspects described in \cite{Kmec:2024nmu}. Notice that the pole at $[q\bar\zeta]=0$ has moved to $q=\infty$.

Due to the obvious identity
\be
\bar\zeta_{\dal_0}R^{\dal_0\dal_1\cdots\dal_s} = R^{\dal_1\cdots\dal_s}\,,
\ee
components of $R^{\dal_0\dal_1\cdots\dal_s}$ obtained from contractions involving $\bar\zeta_\dal$ do not contain any new data on top of the $R_s$ components. So it suffices to study the $R_s$ aspects defined by FPR. 

\medskip

The integral over $q$ runs over $\C=\P^1-\{q=\infty\}$. Of course, as indicated in \eqref{Rs}, we would still like to integrate over all of $L_{u,z}$, including the point $q=\infty$. For instance, we will shortly see that compactness of the integration domain is necessary to obtain the asymptotic Bianchi identities. 

In order for the $q$ integral to extend across $q=\infty$, $b$ must exhibit ``good'' boundary behavior near the line at infinity. That is, it should extend from $\PT$ to $\P^3$ in some reasonably smooth manner. This is not a problem for the leading, subleading and subsubleading spins $s=-1,0,1$, as our integrals have a pole at $q=\infty$ only for $s>1$. For the higher spin aspects to be well-defined, at spin $s$ we need $b$ to contain a zero of order $s-1$ at $q=\infty$ to cancel the pole in \eqref{Rs}. The more charge aspects that are well-defined, the more conserved quantities we will be able to construct in Carrollian and celestial CFTs.

For \emph{all} the charge aspects to be well-defined, one needs $b$ to die off faster than any polynomial in $q=\infty$. That is, it lives in a space of Schwartz functions in $q$. We will assume that $b$ exhibits such a Schwartz behavior. The reader may pick their favorite alternative fall-off conditions, which will depend on the situation under study. Depending on the fall-off conditions, a certain subset of the charge aspects will remain well-defined.

\medskip

With these assumptions, let us derive the equations of motion of these charge aspects. For $s\geq0$, applying $q\,\p_u+\p_\bz\equiv e_\bz$ along with the gauge condition $e_{\bz}f = f\bar\cA-a_\bz f$ derived in \eqref{F1}, we see that
\begin{align}
        \p_u R_s + \cD_\bz R_{s-1} &= \int_{L_{u,z}}\d q\,q^s\left(\cL_{e_\bz}(f^{-1}bf) + [\bar\cA,f^{-1}bf]\right)\nonumber\\
        &= \int_{L_{u,z}}\d q\,q^s\,f^{-1}(\cL_{e_{\bz}}b + [a_\bz,b])f\nonumber\\
        &=  \int_{L_{u,z}}\d q\,q^s\,f^{-1}\left(e_{\bz}\ip\Dbar b + \Dbar(e_\bz\ip b)\right)f\label{dzetaR}\,.
\end{align}
On-shell, one imposes $\Dbar b = 0$. As the frame obeys $f^{-1}\Dbar|_{L_{u,z}} f = \dbar|_{L_{u,z}}$, this leaves us with almost a total derivative,
\be\label{dzetaR1}
 \p_u R_s + \cD_\bz R_{s-1} = \int_{L_{u,z}}\d q\,q^s\,\dbar\!\left(e_{\bz}\ip f^{-1}bf\right)\,.
\ee
Except that for $s\geq2$, there is a pole at $q=\infty$ staring back at us! At this stage, we invoke Schwartz regularity conditions at $q=\infty$, which soaks up the pole. This allows us to deduce the equation of motion
\be\label{fprb}
\boxed{\hspace{1cm}\p_uR_s + \cD_\bz R_{s-1} = 0\,,\quad s\geq0\,.\hspace{0.95cm}}
\ee
These are the \emph{Freidel-Pranzetti-Raclariu recursion relations} that relate the spin $s$ charge aspects to the spin $s-1$ aspects.


Similarly, we can obtain the action of $e_{\bu}=\p_\bu$ on the aspects. Using the radiative gauge identity $e_\bu f = -a_\bu f$, it follows that
\begin{align}
\p_\bu R_s &= \int_{L_{u,z}}\d q\,q^{s+1}\,f^{-1}(\cL_{e_{\bu}}b + [a_\bu,b])f\,.
\end{align}
By the same steps as taken in \eqref{dzetaR} and \eqref{dzetaR1}, the right hand side vanishes by virtue of being a total derivative when $b$ is sufficiently regular at infinity. In the end, we land on
\be
\p_\bu R_s = 0\,,\quad s\geq-1\,.
\ee
So the charge aspects are only functions of $u,z,\bar z$. Note that they can still have poles or other singularities at $u=0$, as our description of asymptotic twistor lines in \eqref{Luzeta} degenerates at $u=0$.

With our conventions, real null infinity sits inside partially complexified null infinity as the subset $\scri\subset\scri_\C$ given by $\text{Re}(u) = 0$. Analyticity in $u$ tells us that the charge aspects on $\scri_\C$ may be defined as analytic continuations of their values along $\scri$. So their only nontrivial information content resides in their values along $\scri$.

As pointed out in \cite{Freidel:2023gue,Cresto:2025bfo}, for spins $s=0,1$, the recursion \eqref{fprb} can be derived in a large $r$ expansion of the Bianchi identity of full Yang-Mills theory. For higher spins, they continue to hold in sdYM, though it remains unclear whether they hold in the non-self-dual theory, see also \cite{Nagy:2024jua,Nagy:2024dme}. As we will see in section \ref{sec:space}, this is reflecting the integrability of sdYM.

\paragraph{Negative helicity aspects.} sdYM theory has $S$-algebra charges of both helicities. So it is also desirable to construct negative helicity charge aspects $\til R_s$ out of $a$ that mirror FPR's original positive helicity aspects.

One naive idea would be to replace $b$ by $a$ in the formula \eqref{Rdal}. But one of the key properties of \eqref{Rdal} is that it is invariant under twistor space gauge transformations $a\mapsto g^{-1}(\dbar+a)g$ and $b\mapsto g^{-1}bg$ because the frame transforms as $f\mapsto g^{-1}f$. Replacing $b$ by $a$ would violate this gauge invariance.

Instead, since $a$ is a connection, we can use the idea that variations of a connection transform in the adjoint. So consider for $s\geq-1$ the aspects defined by
\be\label{Rtdal}
\boxed{\hspace{1cm}\til R^{\dal_0\dal_1\cdots\dal_s}(u,\zeta) = \int_{L_{u,\zeta}}\D q\;[q\bar\zeta]^{-3-s}\,q^{\dal_0}q^{\dal_1}\cdots q^{\dal_s}\,f^{-1}\,\p_ua\,f\,,\hspace{0.95cm}}
\ee
where $\p_ua$ is to be read as $\cL_{\p_u}a$. Under a gauge transformation $a\mapsto g^{-1}(\dbar+a)g$ and $f\mapsto g^{-1}f$, one can show that
\be
f^{-1}\,\p_ua\,f\bigr|_{L_{u,\zeta}}\mapsto f^{-1}\,\p_ua\, f\bigr|_{L_{u,\zeta}} + \dbar|_{L_{u,\zeta}}\!\left(f^{-1}\,\p_ug \,g^{-1}f\right)\,.
\ee
Assuming Schwartz boundary conditions on $\p_ua$ at $[q\bar\zeta]=0$ (i.e., at $q=\infty$), the $\dbar$-exact term drops from the variation of \eqref{Rtdal}. Thus, the aspects \eqref{Rtdal} remain gauge invariant.

They satisfy the same property as the $R^{\dal_0\cdots\dal_s}$,
\be
\bar\zeta_{\dal_0}\til R^{\dal_0\dal_1\cdots\dal_s} = \til R^{\dal_1\cdots\dal_s}\,.
\ee
So their invariant data is encoded in the contracted aspects
\begingroup
\allowdisplaybreaks
\begin{align}
    \til R_s(u,\zeta) &= \frac{u^{s+2}}{[\iota\bar\zeta]^{s+1}}\,\iota_{\dal_0}\iota_{\dal_1}\cdots\iota_{\dal_s}\til R^{\dal_0\dal_1\cdots\dal_s}(u,\zeta)\nonumber\\
    &= u^{s+2}\int_{L_{u,\zeta}}\frac{\D q}{[q\bar\zeta]^{s+3}}\,\frac{[q\iota]^{s+1}}{[\iota\bar\zeta]^{s+1}}\,f^{-1}\,\p_ua\,f\,.\label{Rtstospinor}
\end{align}
\endgroup
Their expression in affine coordinates reads
\be\label{Rts}
\boxed{\hspace{1cm}\til R_s(u,z) = \int_{L_{u,z}}\d q\,q^{s+1}f^{-1}\,\p_ua\,f\,.\hspace{0.95cm}}
\ee
These constitute negative helicity Bramson-Tod integrals. They are sections of the line bundle $\CO(s+1)\otimes\br{\CO(-s-3)}\to\scri_\C$. That is, they carry weights $(s+1,-s-3)$ in $(\zeta,\bar\zeta)$.

For $s\geq0$, an analog of the FPR recursion can be derived for these negative helicity aspects. We compute
\begingroup
\allowdisplaybreaks
\begin{align}
        \p_u\til R_s + \cD_\bz\til R_{s-1} &= \int_{L_{u,z}}\d q\,q^s\,f^{-1}(\cL_{e_{\bz}}\cL_{\p_u}a + [a_\bz,\cL_{\p_u}a])f\nonumber\\
        &=  \int_{L_{u,z}}\d q\,q^s\,f^{-1}\left(\cL_{\p_u}(e_{\bz}\ip\Dbar^2) + \Dbar(e_{\bz}\ip\cL_{\p_u} a)\right)f\nonumber\\
        &= \int_{L_{u,z}}\d q\,q^s\,\dbar\!\left(e_{\bz}\ip f^{-1}\p_ua\,f\right)\,,\label{dzetaRt}
\end{align}
\endgroup
where we used $[\p_u,e_{\bz}]=0$ to commute various Lie derivatives, as well as invoked the self-duality equation of motion $\Dbar^2=0$. This vanishes if $\p_ua$ vanishes to order $s+3$ at $q=\infty$, giving us the negative helicity recursion relation
\be\label{fpra}
\boxed{\hspace{1cm}\p_u\til R_s + \cD_\bz\til R_{s-1} = 0\,,\quad s\geq0\,.\hspace{0.95cm}}
\ee
Similarly, in radiative gauge we obtain the holomorphicity condition
\be
\p_\bu\til R_s = 0\,,\quad s\geq-1\,,
\ee
so that the negative helicity aspects are again found to only depend on $(u,z,\bz)$ but not $\bu$.

In the next section, we will find direct spacetime evidence that $R_s$ and $\til R_s$ are deeply rooted in the integrability of sdYM. This will give further motivation for studying the $\til R_s$ aspects defined here.


\section{Spacetime}
\label{sec:space}

In this section, we uncover the origins of these charge aspects in the integrability of self-dual Yang-Mills theory. To do this, we work on spacetime and gauge-fix the sdYM equations to radiative gauge. It turns out that radiative gauge is a $\cJ$ or $\cK$-matrix-type gauge adapted to Bondi coordinates. 

In their original works, FPR considered aspects of only one helicity, which we have labeled positive helicity. They provide a spacetime interpretation of their charge aspects as certain coefficients in the large $r$ expansions of Newman-Penrose scalars. In the previous section, we were able to build aspects of both helicities. On spacetime, we will find that the negative helicity aspects have a somewhat different interpretation, but one that relates directly to integrability. They are built out of frames obeying a Lax equation, which is reminiscent of the construction of Lax monodromy in 2d integrable systems. 

Following this, we show that the charge aspects act as generators of an infinite hierarchy of commuting flows on the moduli space of solutions of sdYM theory. A helpful review of integrability, hierarchies and related ideas may be found in \cite{Mason:1991rf}.


\subsection{Gauge choices and Lax integrability}
\label{ssec:Bondi}

In this section, we will work in Lorentzian signature and with complex fields $A,B$. In deriving asymptotic charges, it will be helpful to adapt our formalism to null infinity. To this end, we choose to work in Bondi coordinates. We will work in flat Bondi coordinates $u,r,z,\bz$, in which the Minkowski metric takes the form
\be
\d s^2 = -\,2\,\d u\,\d r + 2\,r^2\,\d z\,\d\bz\,.
\ee
In contrast to the standard Bondi coordinates, both $u$ and $r$ here range over the full real line $\R$. The null infinities $\scri^\pm$ are reached as $r\to\pm\infty$. 

Self-dual gauge theory is conformally invariant, so we can introduce a new coordinate
\be
w = r^{-1}
\ee
and work with the conformally rescaled metric
\be\label{met}
\d\tilde s^2 = w^2\d s^2 = 2\,\d u\,\d w + 2\,\d z\,\d\bz\,.
\ee
In this coordinate system, $\scri^\pm$ is approached in the limits $w\to0^\pm$.


Let us study the component expansion of the equations of motion \eqref{sdeq} of sdYM theory. The components of the self-duality equation $F^-=0$ read
\begin{equation}\label{sdymeq}
    F_{uz} = F_{w\bz} = F_{uw}+F_{z\bz} = 0\,.
\end{equation}
Similarly, the 2-form $B$ is asd, so its components are constrained by
\be
B_{u\bz} = B_{wz} = B_{uw}-B_{z\bz} = 0\,.
\ee
The Bianchi identity $DB = 0$ then admits the component expansion
\be\label{bianchi}
\begin{alignedat}{2}
    &D_uB_{z\bz} + D_\bz B_{uz} = 0\,,\qquad &&D_uB_{w\bz} + D_\bz B_{z\bz} = 0\,,\\
    &D_zB_{z\bz} - D_wB_{uz} = 0\,,\qquad &&D_zB_{w\bz} - D_wB_{z\bz} = 0\,,
\end{alignedat}
\ee
in terms of the remaining three independent components $B_{uz},B_{w\bz},B_{z\bz}$.

The component action of sdYM theory reads
\be\label{compac}
S[A,B] = \int_{\bM}\vol\;\Tr\,\big\{B_{z\bz}(F_{uw}+F_{z\bz})-B_{uz}F_{w\bz}-B_{w\bz}F_{uz}\big\}\,,
\ee
where we have set $\vol=\d u\wedge\d w\wedge\d z\wedge\d\bz$. We will reduce this to radiative gauge by partially imposing some of the equations of motion. To accomplish this, we integrate out the Lagrange multiplier $B_{w\bz}$. This imposes $F_{uz}=0$. On the support of $F_{uz}=0$, the part $A_u\d u+A_z\d z$ of the spacetime connection is flat. So in a sufficiently small, simply connected region of $\bM$, we can go to a gauge in which
\be
\text{Gauge choice}\;:\quad A_u = A_z = 0\,.
\ee 
This is one major technique that is available in sdYM theory but not in full Yang-Mills. We now describe a pair of formulations for sdYM that are adapted to this gauge choice.

\paragraph{$\cJ$-matrix formulation.} We can also integrate out one of the remaining Lagrange multipliers $B_{uz}$ and $B_{z\bz}$. Integrating out the former imposes 
\be
F_{w\bz} \equiv \p_w A_\bz-\p_\bz A_w+[A_w,A_\bz] = 0\,.
\ee
This can be solved in terms of a group-valued function $\cJ$ by setting
\be
A_w = \cJ^{-1}\p_w\cJ\,,\qquad A_\bz = \cJ^{-1}\p_z\cJ\,.
\ee
Substituting these back into the Lorentz invariant action \eqref{compac} yields a gauge-fixed formulation of sdYM that breaks part of the Lorentz symmetry,
\be
S[\cJ,\til\cJ] = \int_\M\vol\;\Tr\,\til\cJ\Big(\p_u(\cJ^{-1}\p_w\cJ)+\p_z(\cJ^{-1}\p_\bz\cJ)\Big)\,,
\ee
having set $B_{z\bz}\equiv\til\cJ$.

The $F_{uw}+F_{z\bz}=0$ equation becomes Yang's equation for the $\cJ$-matrix \cite{Yang:1977zf},
\be\label{Jeq}
\p_u(\cJ^{-1}\p_w\cJ)+\p_z(\cJ^{-1}\p_\bz\cJ)=0\,.
\ee
The equation of motion of $\cJ$ yields the equation governing $\til\cJ$,
\be
\Box\til\cJ + [\cJ^{-1}\p_w\cJ,\p_u\til\cJ] + [\cJ^{-1}\p_\bz\cJ,\p_z\til\cJ] = 0\,.
\ee
Here, $\Box=\p_u\p_w+\p_z\p_\bz$ denotes the scalar wave operator. This is a gauge-fixing of the gauge-covariant wave equation $(D_uD_w+D_z D_\bz)B_{z\bz}=0$ obtained by eliminating $B_{uz}$ from the first and third Bianchi identities listed in \eqref{bianchi} on the support of $F_{uz}=0$. It is also a linearization of \eqref{Jeq}.

\paragraph{$\cK$-matrix formulation.} Alternatively, we can integrate out $B_{z\bz}$ to impose $F_{uw}+F_{z\bz}=0$. In the gauge $A_u=A_z=0$, this reduces to
\be
\p_uA_w + \p_zA_\bz = 0\,.
\ee
This can be solved by introducing a potential $\cK$ for the remaining components of $A$,
\be
A_w = \p_z\cK\,,\qquad A_\bz = -\p_u\cK\,.
\ee
Plugging these values of $A$ back into the action \eqref{compac} yields an action for sdYM in terms of the scalar $\cK$ and its partner $\til \cK \equiv B_{uz}$,
\be\label{leznov}
S[\cK,\til \cK] = \int_\bM\vol\;\Tr\,\til \cK\left(\Box \cK + [\p_z\cK,\p_u\cK]\right)\,.
\ee
This provides another formulation of sdYM that partially breaks the Lorentz symmetry.

The Lie algebra valued scalar $\cK$ is often referred to as the $\cK$-\emph{matrix} in integrability literature. The $F_{w\bz}=0$ equation gives an effective equation obeyed by $\cK$,
\be\label{Keq}
\Box \cK + [\p_z\cK,\p_u\cK] = 0\,.
\ee
The associated action \eqref{leznov} is known as the \emph{Leznov action}, used for various applications in \cite{leznov1987deformation,leznov1987equivalence,parkes1992cubic,Chalmers:1996rq}. Varying it with respect to $\cK$ gives the equation obeyed by $\til \cK$,
\be\label{Kteq}
\Box\til \cK + [\p_z\cK,\p_u\til \cK]-[\p_u\cK,\p_z\til \cK] = 0\,.
\ee
This is a gauge-fixing of $(D_uD_w+D_z D_\bz)B_{uz}=0$ obtained by eliminating $B_{z\bz}$ from the first and third Bianchi identities in \eqref{bianchi}. It can also be obtained by linearizing \eqref{Keq}.

\paragraph{Classical Integrability.} Next introduce a formal parameter $q\in\P^1$, known as a Lax parameter. The self-duality equations $F^-=0$ are classically integrable. To manifest this, we introduce a Lax pair
\be\label{LaxPair}
L = D_\bz+qD_u\,,\qquad M = D_w-qD_z
\ee
whose commutativity is equivalent to $F^-=0$. Indeed, we observe that
\be
[L,M] = - q^2F_{uz} + q(F_{uw}+F_{z\bz}) - F_{w\bz}\,,
\ee
which vanishes for all $q\in\P^1$ iff \eqref{sdymeq} holds.

The condition $[L,M]=0$ is the integrability constraint for solving for a $\g$-valued matrix $\cU$ obeying the pair of Lax equations
\be\label{laxU}
L\,\cU = M\,\cU = 0\,.
\ee
It is helpful to make these equations more explicit,
\be\label{laxexp}
\begin{split}
    &(\p_\bz+q\,\p_u)\,\cU + (A_\bz+q\,A_u)\,\cU = 0\,,\\
    &(\p_w-q\,\p_z)\,\cU + (A_w-q\,A_z)\,\cU = 0\,.
\end{split}
\ee
We can solve for $\cU$ patch by patch as a function of $q$ on the Riemann sphere. On any patch, $\cU$ will depend holomorphically in $q$. 

The matrix $\cU$ acts as a frame for the Yang-Mills bundle on spacetime. We will be interested in solving the first-order PDEs \eqref{laxexp} in a neighborhood of the north pole $q=\infty$. So assume that $\cU$ admits a Taylor expansion
\be
\cU = \sum_{s=0}^\infty\frac{\cU_s}{q^s}\,,
\ee
where $\cU_s$ are functions of spacetime coordinates but not $q$. 

Under a gauge transformation of the connection, $A\mapsto h^{-1}Ah+h^{-1}\d h$, the frame $\cU$ rotates as $\cU\mapsto h^{-1}\cU$. Previously, we attained the $\cK$-matrix gauge by solving $F_{uz}=F_{uw}+F_{z\bz}=0$. An equivalent way to attain it is to go to a gauge in which the order $q^0$ coefficient $\cU_0$ is set to $1$ by an appropriate choice of gauge rotation $h$. Therefore, we expand $\cU$ as
\be\label{Uexp}
\cU = 1 + \frac{\cK}{q} + \sum_{s=2}^\infty\frac{\cU_s}{q^s}\,,
\ee
where we have set $\cU_0=1$ and renamed $\cU_1$ as $\cK$. Then imposing the Lax equations \eqref{laxexp} up to terms of order $q^{-1}$ gives us
\be
A_u=A_z=0\,,\qquad A_w = \p_z\cK\,,\qquad A_\bz=-\p_u\cK\,,
\ee
which is precisely the $\cK$-matrix gauge. This is the formalism we will use to derive the FPR charge aspects.

We can also attain the $\cJ$-matrix gauge in this way. In \eqref{Uexp}, we presented a solution for $\cU$ in the patch $q\neq0$. Similarly, we can solve the Lax equations in the patch $q\neq\infty$ to obtain $\cU$ as a holomorphic function around $q=0$,
\be
\cU = \cJ^{-1} + \sum_{r=1}^\infty\cV_r q^r\,.
\ee
We have identified the first term in this expansion as the inverse of the $\cJ$-matrix. This is because evaluating the Lax equations at $q=0$ immediately outputs
\be
A_w = \cJ^{-1}\p_w\cJ\,,\qquad A_\bz = \cJ^{-1}\p_\bz\cJ\,.
\ee
Keeping the same gauge fixing condition $\cU|_{q=\infty}=1$ as before, we again land on $A_u=A_z=0$. This reproduces the $\cJ$-matrix gauge.


\subsection{Charge aspects on spacetime}

Let us briefly review the spacetime origin of the positive helicity charge aspects described in \cite{Freidel:2023gue}, following which we will provide the corresponding spacetime interpretation for the negative helicity charge aspects.

Let $\bar\cA$ denote the leading term in the large $r$ (small $w$) expansion of $A_\bz$,
\be
\bar\cA = \lim_{r\to\infty}A_\bz = \lim_{w\to0}A_\bz\,.
\ee
As before, let $\cD_\bz = \p_\bz + [\bar\cA,-]$ denote the associated covariant derivative. The original motivation for the charge aspects came from looking at the asymptotic $w\to0$ limits of the Bianchi identities \eqref{bianchi}. Sending $w\to0$ and working in the gauge $A_u=A_z=0$, the two Bianchi identities in the first line of \eqref{bianchi} reduce to
\be\label{fprbase}
\p_u R_s + \cD_\bz R_{s-1}\,,\qquad s=1,0\,,
\ee
where we have defined the leading, subleading and subsubleading charge aspects
\be\label{R10-1}
R_{-1} = \lim_{w\to0}B_{uz}\,,\qquad R_{0} = \lim_{w\to0}B_{z\bz}\,,\qquad R_1 = \lim_{w\to0}B_{w\bz}\,.
\ee
The relations \eqref{fprbase} are known as asymptotic Bianchi identities.

FPR conjectured that they can find certain natural quantities in Yang-Mills theory for which the pattern \eqref{fprbase} continues \cite{Freidel:2023gue}. Briefly, what we have called $B_{w\bz}$ coincides with the Newman-Penrose scalar $\Phi_0$ in our coordinate system. FPR expand this Newman-Penrose scalar in a large $r$ expansion, which for us takes the form
\be
B_{w\bz} = \sum_{s=0}^\infty\frac{w^s}{s!}\left(\p_z^sR_{s+1} + P_{s+1}\right)\,.
\ee
At each order, they split the coefficient of $w^s$ into a term $\p_z^sR_{s+1}$ that is in the image of $\p_z^s$, and a term $P_{s+1}$ that is in the cokernel of $\p_z^s$. The quantities $P_s$, $s\geq1$, are gauge theory cousins of the conserved quantities studied by Newman and Penrose in \cite{Newman:1968uj}. FPR conjecture that the aspects $R_s$ so defined continue to obey the recursion \eqref{fprbase} for all higher spins $s$.

However, they do not construct charge aspects of both helicities. In this work, we have filled in this gap in the self-dual sector by means of its twistor uplift. But we would also like to find a spacetime interpretation of the $\til R_s$ aspects, just as we reviewed here for the $R_s$ aspects. This can be accomplished using the Lax formulation of sdYM, as we now show.

\medskip

Suppose the expansion coefficients $\cU_s$ in the expansion \eqref{Uexp} of the frame admit well-defined small $w$ limits,
\be
K = \lim_{w\to0}\cK\,,\qquad U_s = \lim_{w\to0}\cU_s\,.
\ee
We are being agnostic as to whether we are approaching $\scri^+$ or $\scri^-$, as our formalism is quite general. In terms of these, we can define the $w\to0$ limit of $\cU$:
\be\label{Uinf}
U = \lim_{w\to0}\cU = 1 + \frac{K}{q} + \sum_{s=2}^\infty\frac{U_s}{q^s}\,.
\ee
In particular, the $w\to0$ limit of the $\cK$-matrix gauge condition $A_\bz=-\p_u\cK$ tells us that
\be\label{Abar}
\bar\cA = -\p_uK\,.
\ee
This will be useful shortly.

To make progress, we will extract an equation that $U$ needs to obey from the asymptotics of the Lax equations \eqref{laxexp}. In $\cK$-matrix gauge, we set $A_u=0$. So the first Lax equation simplifies to
\be
(\p_\bz+q\,\p_u+A_\bz)\,\cU = 0\,.
\ee
Taking the $w\to0$ limit of this equation yields
\be\label{Ueq}
(\p_\bz+q\,\p_u+\bar\cA)\,U = 0\,.
\ee
This is very close to the kind of PDE we need. Except that it contains $\bar\cA\,U$ instead of $[\bar\cA,U]$. To construct a quantity from which we can extract adjoint-valued charge aspects, we simply take derivatives of $U$.

In particular, we will use the derivative $\p_uU$. We define the aspects $\til R_s(u,z,\bz)$ as the coefficients in the large $q$ expansion of the quantity $-\p_uU\,U^{-1}$,
\be\label{dUU}
-\p_uU\,U^{-1} = \sum_{s=-1}^\infty\frac{\til R_s}{q^{s+2}}\,. 
\ee
We can invert this expansion to obtain a contour integral formula for the aspects,
\be
\til R_s = -\frac{1}{2\pi i}\oint_{q=\infty}\d q\,q^{s+1}\,\p_uU\,U^{-1}\,,\qquad s\geq-1\,,
\ee
where the contour surrounds $q=\infty$. To show that they obey the recursion \eqref{fprrec}, start by deriving a PDE for $\p_u U\,U^{-1}$ by differentiating \eqref{Ueq} with respect to $u$,
\be
(\p_\bz+q\,\p_u)(\p_u U\,U^{-1}) + [\bar\cA,\p_uU\,U^{-1}] = -\p_u\bar\cA\,.
\ee
As a result, for $s\geq0$ we obtain
\begin{align}
    \p_u\til R_s + \p_\bz\til R_{s-1} + [\bar\cA,\til R_{s-1}] &= \frac{\p_u\bar\cA}{2\pi i}\oint_{q=\infty}\d q\,q^s = 0\,.
\end{align}
The range $s\geq0$ manages to avoid precisely the point $s=-1$ where this contour integral would have failed to vanish. Thus, the $\til R_s$ aspects obey the recursion \eqref{fprrec}.

In principle, once we have a seed for the recursion, we would no longer need the explicit expression for $U$, as we can solve the recursion iteratively. The seed value is the value of the lowest aspect $\til R_{-1}$. This can be explicitly determined by plugging in the expansion \eqref{Uinf} of $U$ into $\p_uU\,U^{-1}$,
\be
-\p_uU\,U^{-1} = -\frac{\p_uK}{q} + O(q^{-2})\,.
\ee
Comparing this with \eqref{dUU} and recalling that $\bar\cA=-\p_uK$, we find
\be
\til R_{-1} = \bar\cA\,.
\ee
So we see that the leading negative helicity charge aspect is just the radiative data $\bar\cA$.


\subsection{Hierarchies and the recursion operator}
\label{ssec:integrable}

The integrability of self-dual gauge theory manifests itself in the form of an infinite hierarchy of commuting flows on its moduli space of classical solutions. We now show that the charge aspects generate this hierarchy at the level of the asymptotic data of classical solutions. This imparts a geometric meaning to the action of the $S$-algebra.

In section \ref{ssec:Bondi}, we formulated sdYM in flat Bondi coordinates in terms of $\cJ$ and $\cK$ matrices. In the $\cJ$-matrix gauge, one sets $A=\cJ^{-1}\p_w\cJ\,\d w + \cJ^{-1}\p_\bz\cJ\,\d\bz$, where $w=r^{-1}$. In the $\cK$-matrix gauge, one instead uses $A=\p_z\cK\,\d w-\p_u\cK\,\d\bz$. Now, let $\Phi(x)$ be an adjoint-valued scalar field. Perturbations of $\cJ,\cK$ of the form
\be
\delta\cJ = \cJ\Phi\,,\qquad\delta\cK = \Phi
\ee
generate the following perturbations of $A$:
\begin{align}
    \cJ\text{-matrix case}\;&:\quad \delta A = D_w\Phi\,\d w + D_\bz\Phi\,\d\bz\,,\label{Jpert}\\
    \cK\text{-matrix case}\;&:\quad \delta A = \p_z\Phi\,\d w - \p_u\Phi\,\d\bz\,.\label{Kpert}
\end{align}
These furnish symmetries of the equations of motion \eqref{Jeq} and \eqref{Keq} of $\cJ,\cK$ if $\Phi$ satisfies
\be
\Box\Phi + [A_w,\p_u\Phi] + [A_\bz,\p_z\Phi] = 0\,.
\ee
On the support of $F_{uw}+F_{z\bz}=0$, this can be identified with the gauge-covariant linearized wave equation $D^2\Phi = 0$ written in the gauge $A_u=A_z=0$.

The recursion operator $\mathscr{R}$ of sdYM theory arises as follows \cite{Mason:1991rf}. We start with a perturbation $\delta A = D_w\Phi\,\d w + D_\bz\Phi\,\d\bz$ of the $\cJ$-matrix type, generated by a scalar $\Phi$ satisfying $D^2\Phi=0$. Then we define a scalar $\sR\Phi$ associated to $\Phi$ that expresses this as a perturbation $\delta A = \p_z\sR\Phi\,\d w - \p_u\sR\Phi\,\d\bz$ of the $\cK$-matrix type. That is, we solve for $\sR\Phi$ by imposing
\be\label{recop}
\p_u\sR\Phi = -D_\bz\Phi\,,\qquad \p_z\sR\Phi = D_w\Phi\,.
\ee
The integrability condition for these two PDEs to admit a simultaneous solution is precisely $D^2\Phi=0$ on the support of $F_{uw}+F_{z\bz}=0$. And one can show that $\sR\Phi$ again satisfies $D^2\sR\Phi = 0$ on the support of $F_{w\bz}=0$. Therefore, $\sR$ provides a map $\Phi\mapsto\sR\Phi$ on the space of solutions of $D^2\Phi=0$. Since we could obtain $D^2\Phi=0$ as a linearization of the $\cJ$ or $\cK$-matrix equations, this map generates a recursion operator on tangent spaces of the moduli space of self-dual gauge fields.

It is this recursion operator that appears to be the true origin of FPR recursion relations, at least in the self-dual sector. Suppose one takes a seed solution $\Phi_{-1}$ of the gauge-covariant wave equation. Then one can generate higher solutions $\Phi_s$ by setting
\be
\Phi_s = \sR^{s+1}\Phi_{-1}\,,\qquad s\geq0\,.
\ee
It follows from \eqref{recop} that these obey the recursion relations
\be
\begin{split}
    &\p_u\Phi_s + D_\bz\Phi_{s-1} = 0\,,\\
    &\p_z\Phi_s - D_w\Phi_{s-1} = 0\,.
\end{split}
\ee
Recursions like \eqref{fprrec} are obtained by evaluating the first of these at $w=0$, i.e., $r=\infty$. Defining
\be
\vphi_s = \lim_{w\to0}\Phi_s\,,\qquad s\geq-1\,,
\ee
one readily obtains
\be
\p_u\vphi_s + \cD_\bz\vphi_{s-1} = 0\,,
\ee
where we recalled that $\cD_\bz = \p_\bz+[\bar\cA,-]$ and $\bar\cA = A_\bz|_{w=0}$.

There are two extremely natural candidates in the $\cK$-matrix formulation that we can use to seed these recursions: $\til\cK$ and $\p_u\cK$. These get paired with each other through the symplectic structure
\be
\int_{u=\text{const.}}\d w\,\d^2z\;\Tr\,\delta\til\cK\;\delta\p_u\cK
\ee
associated to the $\cK$-matrix action \eqref{leznov}. They give rise to the two types of charge aspects. 

To obtain positive helicity charge aspects, we take our cue from the leading order guesses \eqref{R10-1} and seed the recursion with $\Phi_{-1}=\til\cK \equiv B_{uz}$. In equation \eqref{Kteq}, we used the Bianchi identity $DB=0$ along with self-duality $F^-=0$ to show that this obeys $D^2\til\cK = 0$. So we obtain the positive helicity aspects
\be
R_s = \lim_{w\to0}\sR^{s+1}\til\cK\,,\qquad s\geq-1\,.
\ee
The other candidate is the seed $\Phi_{-1}=-\p_u\cK\equiv A_\bz$, which is motivated from requiring $\til R_{-1}=\bar\cA$. That this obeys $D^2\p_u\cK=0$ follows from acting with $\p_u$ on the $\cK$-matrix equation \eqref{Keq}. Hence, we obtain the negative helicity aspects
\be
\til R_s = -\lim_{w\to0}\sR^{s+1}\p_u\cK\,,\qquad s\geq-1\,.
\ee
By our general considerations above, these two families of charge aspects will obey the recursion relations \eqref{fprrec}.

Each $\Phi_s$ solves $D^2\Phi_s=0$, whereby it can be thought of as a tangent vector generating on-shell perturbations $\delta A$ mentioned in \eqref{Jpert} or \eqref{Kpert}. So it can be used to generate a flow on the moduli space of classical solutions. Explicitly, let $\cK(x,t_s)$ denote a family of matrix-valued potentials labeled by a parameter $t_s$. Then the flow generated by $\Phi_s$ is obtained by solving
\be
\p_{t_s}\cK = \Phi_s\,.
\ee
A collection $\{\Phi_s\}$ generates a hierarchy of flows on the moduli space of solutions. Twistor theory allows us to show that these flows commute with each other. We refer the interested reader to \cite{Mason:1991rf} for further details.

Asymptotically, we find that the charge aspects generate flows on the space of radiative data. In terms of $K=\cK|_{w=0}$, these flows are described by
\be
\p_{t_s}K = \vphi_s\,.
\ee
Eg., $R_{-1}$ generates constant gauge rotations that are non-vanishing at $\scri$. The higher aspects generate similar flows in the space of radiative data. These will be related to residual gauge symmetries of radiative gauge in section \ref{ssec:softwave}.

\section{Holography}
\label{sec:holo}

In this section, we finally present the conservation laws arising from the $S$-algebra symmetries of sdYM. We will do this directly on twistor space. Placing boundary conditions on the fields on twistor space will allow us to classify large gauge symmetries. We will find that these are in one-to-one correspondence with soft gluon states. The latter arise precisely as the overleading modes that violate our boundary conditions. This confirms that $S$-algebra transformations are indeed asymptotic symmetries.

Plugging the soft gluon wavefunctions into the symplectic form of sdYM on twistor space will result in conserved $S$-algebra charges. Depending on which ``Cauchy surface'' one picks to quantize the twistorial theory, this will result in either charges of a Carrollian CFT or those of a celestial CFT. The first will be conserved along the null generators of $\scri$, i.e., annihilated by $\p_u$. The second will be chiral currents of a 2d CFT, i.e., annihilated by $\p_\bz$ on the celestial sphere. Both of these conservation laws will be subject to the vanishing of radiation at $\scri$.

Perhaps the deepest unifying feature of our analysis is that both types of charges will be built out of the same charge aspects $R_s$ and $\til R_s$. This provides another check of the equivalence of Carrollian and celestial holography -- at least in the self-dual sector -- building on previous approaches of \cite{Donnay:2022aba, Donnay:2022wvx}.


\subsection{Boundary conditions}
\label{ssec:bdry}

Asymptotic symmetries of sdYM may be realized as large gauge symmetries of the twistorial theory \eqref{twac}. To do this, we will need to impose boundary conditions on $a$ and $b$ near the ``boundary'' $I$ of $\PT\subset\P^3$. Our analysis is adapted from a similar procedure that works in twisted holography \cite{Costello:2018zrm, Costello:2022wso, Costello:2023hmi}.

Our choices of boundary conditions will ensure that soft gluons constitute a basis of all the states that \emph{violate} them, i.e., are ``overleading'' on twistor space. The wavefunctions of soft gluons will be pure gauge away from the boundary of $\PT$, and the associated gauge transformation parameters will be in one-to-one correspondence with generators of the $S$-algebra.

\paragraph{Boundary of twistor space.} Recall that $\PT=\P^3-I$. This removal of the line at infinity from $\P^3$ creates a natural boundary in twistor space. But since $I$ had complex-codimension 2, this is not a standard boundary. In real geometry, the usual notion of a boundary is that of a real-codimension-1 subspace at which the geometry ``ends''. In complex geometry, the natural analog of a boundary would be a \emph{boundary divisor}. 

A divisor of a complex variety is an analytic subvariety of \emph{complex}-codimension 1. Simple examples would be hyperplanes in projective spaces. A boundary divisor is a divisor that one attaches to a non-compact variety to make it compact. Equivalently, it is a divisor at which we declare that a given variety ``ends''. Such a boundary divisor is what we need in order to study flat space holography from a twistorial standpoint. But where do we find such a divisor in $\PT$?

A useful resolution of this puzzle was presented in a closely related context in \cite{Costello:2023hmi}. Any construction of the boundary must somehow incorporate $I$. The resolution is that we will create our boundary divisor by blowing up $\P^3$ along $I$.

In homogeneous coordinates $Z^A=(\lambda_\al,\mu^\dal)$, the line $I$ was cut out by the pair of equations $\lambda_\al=0$. The blowup of $\P^3$ along $I$ is the following space:
\be\label{ptbar}
\br\PT = \big\{(Z^A,\pi_\al)\in\P^3\times\P^1\;|\;\la\lambda\pi\ra = 0\big\}\,.
\ee
I.e., it is the subvariety of $\P^3\times\P^1$ cut out by the equation $\la\lambda\pi\ra=0$, where the $\P^3$ has homogeneous coordinates $Z^A$ and the auxiliary $\P^1$ has homogeneous coordinates $\pi_\al$. 

When $\lambda_\al\neq0$, the equation $\la\lambda\pi\ra=0$ tells us that the new coordinate $\pi_\al$ is proportional to $\lambda_\al$. In this case, the point $\pi_\al\in\P^1$ is completely fixed in terms of $Z^A\in\P^3$. So we obtain a biholomorphism
\be
\br{\PT}-\{\lambda_\al=0\} \simeq \PT-\{\lambda_\al=0\}\,.
\ee
On the other hand, when $\lambda_\al$ is zero, $\la\lambda\pi\ra=0$ is trivially satisfied for any $\pi_\al$. Therefore, the line at infinity $\{\lambda_\al=0\}$ gets replaced by an exceptional divisor
\be
\cE = \P^1_\mu\times\P^1_\pi\,,
\ee
where $\P^1_\mu$ and $\P^1_\pi$ are Riemann spheres carrying independent homogeneous coordinates $\mu^\dal$ and $\pi_\al$ respectively.

$\cE$ will be our holographic boundary divisor. For instance, the holographic plate for celestial holography will be the $\P^1_\pi$ factor generated by the blowup. This is where celestial CFT lives.

\paragraph{Boundary conditions on $\cE$.} We have presented $\br\PT$ as a compact space by embedding it as a projective subvariety of $\P^3\times\P^1$. Let $\CO(r,s)\to\P^3\times\P^1$ denote the line bundle whose sections carry weight $r$ in $Z^A$ and $s$ in $\pi_\al$, and let $\CO_X(r,s)$ denote its restriction to any subvariety $X\subset\P^3\times\P^1$. Eg., the canonical bundle of $\br\PT$ is $K_{\br\PT}=\CO_{\br\PT}(-3,-1)$.

The volume form on $\PT$ was $\Omega_{\PT} = \D^3Z$ and was valued in $\CO_{\PT}(4,0)$. The volume form on $\br\PT$ is
\be
\Omega_{\br\PT} = \Res{\la\lambda\pi\ra=0}\frac{\D^3Z\wedge\D\pi}{\la\lambda\pi\ra}
\ee
and is indeed valued in $\CO_{\br\PT}(3,1)$, the dual of $K_{\br\PT}$. On the variety $\la\lambda\pi\ra=0$, one can define a coordinate $n$ normal to $\cE=\{\lambda_\al=0\}$ by setting 
\be
\lambda_\al=n\pi_\al\,.
\ee
$\cE$ becomes the surface $n=0$. A computation of residues shows that the volume form on $\PT$ is proportional to that on $\br\PT$ with proportionality factor $n$,
\be\label{ono}
\Omega_{\PT} = n\,\Omega_{\br\PT}\,.
\ee
Since $n$ is a section of $\CO_{\br\PT}(1,-1)$, and $\Omega_{\br\PT}$ a section of $\CO_{\br\PT}(3,1)$, their product comes out a section of $\CO_{\br\PT}(4,0)$ exactly as desired.

To study the behavior of the sdYM twistor action \eqref{twac} near the boundary $\cE$, we want to extend it to an action over $\br\PT$. To do this, we follow \cite{Costello:2018zrm} and try to construct a holomorphic BF action of the form
\be
\int_{\br\PT}\Omega_{\br\PT}\,\Tr\!\left(\cB\,\dbar\cA + \frac12\,\cB\,[\cA,\cA]\right)\,,
\ee
where $\cA$ and $\cB$ are $(0,1)$-forms on $\br\PT$. It is convenient to focus attention on the cubic interaction term to read off the weights of various fields. Since $\Omega_{\br\PT}$ is valued in $\CO_{\br\PT}(3,1)$, the interaction $\Tr\,\cB\,[\cA,\cA]$ must be valued in $\CO_{\br\PT}(-3,-1)$. Naively, any choice of weights of $\cA$ and $\cB$ that accomplishes this is allowed. But we will make the choice that makes all soft gluons overleading, which is the physically relevant case. We assign $\cA$ and $\cB$ the following weights:
\be
\begin{split}
    &\cA \in \Omega^{0,1}(\br\PT,\CO_{\br\PT}(-1,1)\otimes\mathfrak{g})\,,\\
    &\cB \in \Omega^{0,1}(\br\PT,\CO_{\br\PT}(-1,-3)\otimes\mathfrak{g})\,.
\end{split}
\ee
This ensures that $\Tr\,\cB\,[\cA,\cA]$ is valued in $\CO_{\br\PT}(-3,-1)$.\footnote{To make the kinetic term well-defined, one would need to modify the $\dbar$ operator on $\br\PT$ so that it maps sections of $\CO(-1,1)$ to sections of $\CO(-2,2)$. This may be done by replacing $\dbar$ by $n^{-1}\dbar$, and asking that $\dbar\cA$ have a first-order zero at $n=0$ to cancel the pole in $n^{-1}\dbar\cA$. We will not pursue this further here, as we are only using the holomorphic BF theory on $\br\PT$ to motivate our boundary conditions. But the interested reader may look at appendix H of \cite{Costello:2018zrm} for the subtleties involved.} 

Just as we were able to do for the volume form in equation \eqref{ono}, we can restrict $\cA$ and $\cB$ to $\PT$ and obtain fields $a$ and $b$ of sdYM by setting
\be\label{Atoa}
    a = n\cA\,,\qquad b = n^{-3}\cB\,.
\ee
Because $n$ was valued in $\CO_{\br\PT}(1,-1)$, these come out valued in $\CO_{\PT}(0,0)$ and $\CO_{\PT}(-4,0)$ respectively. Now it is easy to see that when its integration domain is restricted to $\PT$, the cubic interaction on $\br\PT$ reduces to the cubic interaction of sdYM on $\PT$:
\be
\int_{\PT}\Omega_{\br\PT}\,\Tr\,\cB\,[\cA,\cA] = \int_{\PT}\frac{\Omega_{\PT}}{n}\,\Tr\bigg(n^3b\cdot\frac{[a,a]}{n^2}\bigg) = \int_{\PT}\Omega_{\PT}\,\Tr\,b\,[a,a]\,.
\ee
This confirms that we have correctly extended the interactions of our sdYM twistor action from $\PT$ to its compactification $\br\PT$.

Having extended the fields $a$ and $b$ to $\br\PT$, we can study their behavior as one approaches the boundary $\cE$. Equation \eqref{Atoa} associates boundary conditions to $a$ and $b$ as $n\to0$. Indeed, demanding that $\cA$ and $\cB$ are smooth on $\cE$, we obtain the boundary conditions
\be\label{nbdry}
\boxed{\hspace{1cm}a\sim O(n)\,,\quad b \sim O(n^{-3})\quad\text{as}\;n\to0\,.\hspace{0.95cm}}
\ee
That is, $a$ needs to have a first-order zero along $\cE$, while $b$ can at most have a third-order pole along $\cE$.

The $n\to0$ behavior can also be correlated with the $u\to\infty$ behavior, giving a Carrollian version of our boundary conditions. To do this, we can introduce a map sending the $\scri$-adapted coordinates $(u,q,z)$ on $\PT$ to coordinates on $\br\PT$,
\be\label{pb}
    \lambda_\al = \frac{1}{u}\begin{pmatrix}
        1\\z
    \end{pmatrix}\,,\quad\mu^\dal = \frac{1}{u}\begin{pmatrix}
        u-q\bz\\q
    \end{pmatrix}\,,\quad\pi_\al = \begin{pmatrix}
        1\\z
    \end{pmatrix}\,,\quad n = \frac{1}{u}\,.
\ee
This clearly embeds into $\lambda_\al=n\pi_\al$. The boundary divisor is approached as $u,q\to\infty$ with $q/u$ held fixed. From this, we deduce the boundary conditions
\be\label{ubdry}
\boxed{\hspace{1cm}a\sim O(u^{-1})\,,\quad b \sim O(u^3)\quad\text{as}\;u\to\infty\hspace{1cm}}
\ee
that hold along null infinity.


\subsection{Asymptotic symmetries}
\label{ssec:large}

In the previous section, we saw that $a$ must have a first-order zero at the boundary of twistor space, and $b$ can have a pole of order up to $3$. Equipped with these boundary conditions, we can divide the twistorial gauge transformations into those that preserve boundary conditions vs.\ those that violate them. This provides us with a notion of \emph{small} vs.\ \emph{large} gauge transformations.

In fact, the celestial approach provides the quickest way to classify asymptotic symmetries in twistor space. The blown up space $\br\PT$ fibers over the sphere $\P^1_\pi$ carrying coordinate $\pi_\al$ via the projection map $(Z^A,\pi_\al)\mapsto\pi_\al$. This fibration admits a sphere's worth of sections contained in the boundary divisor $\cE=\P^1_\mu\times\P^1_\pi$ at $n=0$. Celestial CFT lives on the $\P^1_\pi$ factor of $\cE$, and the $\P^1_\mu$ factor is analogous to the internal $S^5$ factor in the standard example of AdS$_5$/CFT$_4$. This is fairly natural: indeed, the map \eqref{pb} identifies the celestial sphere coordinate $z$ precisely with the coordinate $\pi_1/\pi_0$.

Away from $n=0$, because $\pi_1/\pi_0 = \lambda_1/\lambda_0$, we can identify the celestial sphere coordinate $z$ as the affine coordinate $\lambda_1/\lambda_0$ on twistor space. Two-dimensional CFTs can be studied in radial quantization, with $|\text{in}\ra$ and $|\text{out}\ra$ vacua placed at $z=0,\infty$, and quantization surfaces taken to be the contours of constant $|z|$. On the bulk side, this motivates working with ``Cauchy'' surfaces given by the 5-real-dimensional slices of constant $|z|$ in twistor space. 

By construction, these 5d bulk Cauchy surfaces intersect the celestial sphere $\P^1_\pi$ in 1d boundary Cauchy surfaces. This ensures that radial evolution on the boundary sphere coincides with bulk evolution, much like in AdS/CFT. This idea was first described in twisted holography in \cite{Costello:2018zrm}, and in celestial holography in Pasterski's prescient work \cite{Pasterski:2022jzc}. Pasterski proposed a change of Cauchy surface in spacetime to surfaces that cut the celestial sphere in circles. We simply employ the same idea in twistor space, where it becomes even more natural.

The $|\text{in}\ra$ and $|\text{out}\ra$ vacua of the bulk theory are described by field configurations of $a,b$ at the divisors $z=0,\infty$ of twistor space. Having understood this, we can define small and large gauge transformations (see also section 4 of \cite{Costello:2022wso}):
\begin{itemize}
    \item \emph{Large gauge transformations} (LGT's) are those gauge symmetries that violate the boundary conditions in $n$ precisely at $z=0,\infty$. They remain legal symmetries of the theory because we are quantizing it on $\PT-\{z=0,\infty\}$. But they can act nontrivially on the $|\text{in}\ra$ and $|\text{out}\ra$ vacuum states as they can alter the values of $a,b$ at $z=0,\infty$.
    \item \emph{Small gauge transformations} (SGT's) are those gauge symmetries that preserve the boundary conditions in $n$ everywhere including $z=0,\infty$. By definition, they do not change the $|\text{in}\ra$ and $|\text{out}\ra$ vacua.
    \item The group of \emph{asymptotic symmetries} is the quotient group $\text{ASG}=\text{LGT's}/\text{SGT's}$.
\end{itemize}
Roughly speaking, asymptotic symmetries are generated by gauge transformations that are allowed to be singular at the boundary $n=0$. 

An equivalent and perhaps more pertinent definition of asymptotic symmetries is that they are gauge transformations that generate nonzero Noether charges. Since the Noether charges \eqref{non-integrable charges} are supported purely on the boundary of the Cauchy surface, the gauge transformations have to be non-vanishing near the boundary to accomplish this. Let us now see this in some explicit examples.

\paragraph{Soft gluons in linear theory.} With these definitions, we can show that soft gluons generate a basis of such large gauge symmetries. The algebra of these symmetries is known as the $S$-algebra. In this section, we work it out in the linearized theory whose equations of motion are $\dbar a = \dbar b = 0$. The extension to nonlinear theory is studied in the next section.

Start with the linearized gauge symmetry associated to $a$:
\be
\delta a = \dbar\xi\,,\qquad\delta b = 0\,.
\ee
Small gauge transformations are generated by parameters $\xi$ that have a first-order zero at $n=0$ for all values of $z\in\P^1$. If $\xi$ is a large gauge transformation, then it will have a first-order zero at $n=0$ everywhere except along $z=0,\infty$. So we can always shift it by small gauge transformations to ensure that $\delta a=0$ away from small neighborhoods of $z=0,\infty$. This means that $\xi$ must be holomorphic in $z$ away from $z=0,\infty$, and holomorphic in $\mu^\dal$ away from the boundary $n=0$ (which is projectively equivalent to $\mu^\dal\to\infty$). 

Such functions $\xi$ are allowed to be polynomials in $\mu^\dal$ and Laurent in $z=\lambda_1/\lambda_0$.\footnote{Keep in mind that since $z=\lambda_1/\lambda_0$, the locus $z=0$ is the surface $\lambda_1=0$ whereas $z=\infty$ is the surface $\lambda_0=0$ in homogeneous coordinates.} A useful basis for them is given by the monomials
\be\label{xikl}
\xi = \frac{(\mu^{\dot0}/\lambda_0)^k(\mu^{\dot1}/\lambda_0)^l}{(\lambda_1/\lambda_0)^{r}}\,\mfk{t}^a\,,\qquad k,l\in\Z_{\geq0}\,,\;\;r\in\Z\,,\;\;\mfk{t}^a\in\g\,.
\ee
In the affine coordinates $(u,q,z)$, these read
\be
\xi = \frac{(u-q\bz)^kq^l}{z^{r}}\,\mfk{t}^a\,,
\ee
obtained by substituting $\mu^{\dot0}/\lambda_0=u-q\bz$ and $\mu^{\dot1}/\lambda_0=q$.

The associated gauge transformations contain derivatives of delta functions $\bar\delta(\lambda_0)$, $\bar\delta(\lambda_1)$.\footnote{For any complex variable $w$, the $(0,1)$-form delta function used here is defined as $\bar\delta(w) \equiv \delta^2(w)\,\d\bar w$.} For instance, when $r\geq0$, we obtain
\be
\delta a = \dbar\xi = \frac{2\pi i(-1)^{r-1}}{(r-1)!} \frac{(\mu^{\dot0})^k(\mu^{\dot1})^l}{\lambda_0^{k+l-r}}\,\bar\delta^{(r-1)}(\lambda_1)\,\mfk{t}^a\,.
\ee
Setting $\lambda_\al=n\pi_\al$, we see that
\be
\delta a \sim O(n^{-k-l})\,.
\ee
So this is a gauge transformation that violates the boundary condition in $n$ to order $k+l$ in a small neighborhood of $z=0$. This is a genuine violation of our boundary conditions \eqref{nbdry} iff $k,l\geq0$. This range of $k,l$ is known as the \emph{wedge} in the subject.

We can also move the singularities from $z=0,\infty$ to generic $z$. Doing so, we can ask for asymptotic symmetries generated by large gauge transformations that violate the boundary conditions in $n$ in a small neighborhood of a generic point $\lambda_\al\propto\kappa_\al$. The associated gauge transformations admit the representatives (dropping normalization factors)
\be\label{softa}
\delta a = \int_{\C^\times}\d s\,s^{k+l-1}\,\bar\delta^2(\kappa-s\lambda)\,(\mu^{\dot0})^k(\mu^{\dot1})^l\,\mfk{t}^a\,.
\ee
The delta function here is a projective delta function that sets $\lambda_\al\propto\kappa_\al$. As usual, allowing for generic $z$ in this way means that we no longer need to work with derivatives of delta functions at $z=0,\infty$.

The large gauge transformations \eqref{softa} are not actually pure gauge, because they are $\dbar$ exact only away from $\lambda\propto\kappa$. Instead, they are linear perturbations of $a$ that correspond to positive helicity \emph{soft gluons} when Penrose transformed to spacetime. The $k+l=0$ wavefunctions correspond to leading soft gluons, the $k+l=1$ wavefunctions to subleading soft gluons, and so on. In this way, our asymptotic symmetry analysis lands us on precisely the $S$-algebra of soft gluon states.

We can do the same analysis to obtain negative helicity soft modes. The linearized gauge transformation associated to $b$ is
\be
\delta a = 0\,,\qquad\delta b = \dbar\phi\,.
\ee
A basis of asymptotic symmetries is provided by the gauge parameters
\be\label{phikl}
\phi = \frac{1}{\lambda_0^4}\frac{(\mu^{\dot0}/\lambda_0)^k(\mu^{\dot1}/\lambda_0)^l}{(\lambda_1/\lambda_0)^r}\,\mfk{t}^a\,,\qquad k,l\in\Z_{\geq0}\,,\;\;r\in\Z\,,\;\;\mfk{t}^a\in\g\,.
\ee
These carry weight $-4$ in $Z^A$, as necessitated by the weight of $b$. In affine coordinates, they can be expressed
\be
\phi = \frac{(u-q\bz)^kq^l}{z^r}\,\mfk{t}^a\,,
\ee
where we have chosen to fix the scale by setting $\lambda_0=1$.

For $r\geq0$, the associated large gauge modes of $b$ are
\be
    \delta b = \dbar\phi = \frac{2\pi i(-1)^{r-1}}{(r-1)!} \frac{(\mu^{\dot0})^k(\mu^{\dot1})^l}{\lambda_0^{k+l-r+4}}\,\bar\delta^{(r-1)}(\lambda_1)\,\mfk{t}^a\,.
\ee
Near the boundary, these behave like
\be
\delta b\sim O(n^{-k-l-4})\,.
\ee
Our boundary conditions \eqref{nbdry} would have allowed perturbations of $b$ to have at most a third-order pole at $n=0$. So once again, this set of gauge transformations violate our boundary conditions in the vicinity of $z=0,\infty$ precisely in the wedge $k,l\geq0$.

Generic negative helicity soft gluon wavefunctions take a similar form,
\be\label{softb}
\delta b = \int_{\C^\times}\d s\,s^{k+l+3}\,\bar\delta^2(\kappa-s\lambda)\,(\mu^{\dot0})^k(\mu^{\dot1})^l\,\mfk{t}^a\,.
\ee
These generate asymptotic symmetries that extend the $S$-algebra by currents dual to negative helicity gluons.


\subsection{Soft gluons and dual recursions} 
\label{ssec:softwave}

In equations \eqref{softa} and \eqref{softb}, we wrote down a basis of asymptotic symmetry generators $\xi,\phi$ in a trivial background $(a,b)=(0,0)$. In the same way, we would like to find useful bases of large gauge transformations in the nonlinear theory, i.e., around a general background $(a,b)$. These will enter the associated Noether charges \eqref{non-integrable charges}. Since $b$ is always a linear field propagating on the background described by $a$, it will be sufficient for our purposes to compute soft gluons in a nontrivial background with only $a$ turned on. 

Infinitesimal gauge transformations of the nonlinear theory are displayed in equation \eqref{gauge}. Again, by using small gauge transformations, we can hope to set any large gauge transformation to zero away from small neighborhoods of the boundaries of interest. In the nonlinear theory, this is tantamount to imposing
\be\label{Dbarxi}
\Dbar\xi = \Dbar\phi = 0
\ee
away from $z=0,\infty$ and $n=0$. 

Gauge transformations satisfying these conditions form a closed algebra under the Barnich-Troessaert bracket. To see this, recall that the bracket of two field-dependent gauge transformations was
\be
\begin{split}
    [\xi,\xi']_\star &= [\xi,\xi'] + \delta_\xi\xi'-\delta_{\xi'}\xi\,,\\
    [\xi,\phi]_\star &= [\xi,\phi] + \delta_\xi\phi\,,\\
    [\phi,\phi']_\star &= 0\,.
\end{split}
\ee
Acting on this with $\Dbar$ and noting that $\delta_\xi(\Dbar\xi') = \Dbar(\delta_\xi\xi') + [\Dbar\xi,\xi']$, we get
\be\label{DbarBT}
\begin{split}
    \Dbar\big([\xi,\xi']_\star\big) &= \Dbar(\delta_\xi\xi') + [\Dbar\xi,\xi'] - \Dbar(\delta_{\xi'}\xi) - [\Dbar\xi',\xi]\\
    &= \delta_\xi\Dbar\xi' - \delta_{\xi'}\Dbar\xi\\
    &= [\delta_\xi,\delta_{\xi'}]a\,.
\end{split}
\ee
This corroborates the expectation that $[\delta_\xi,\delta_{\xi'}]=\delta_{[\xi,\xi']_\star}$. Similarly, we get $[\delta_\xi,\delta_\phi]=\delta_{[\xi,\phi]_\star}$:
\be
\Dbar\big([\xi,\phi]_\star\big) = [\xi,\Dbar\phi] + \delta_\xi\Dbar\phi = [\delta_\xi,\delta_\phi]b\,.
\ee
As a corollary, we obtain the implications
\be\label{Sclose}
\begin{split}
    \Dbar\xi=\Dbar\xi'=0&\implies\Dbar[\xi,\xi']_\star = 0\,,\\
    \Dbar\xi=\Dbar\phi=0&\implies\Dbar[\xi,\phi]_\star = 0\,.
\end{split}
\ee
This confirms that the algebra of asymptotic symmetries of sdYM indeed closes.

\paragraph{Soft gluons in radiative gauge.} We will solve the soft gluon equations of motion \eqref{Dbarxi} in radiative gauge. Denote the solutions in radiative gauge by $\tilde\xi$ and $\tilde\phi$ respectively. The gauge transformation back to a general gauge is
\be
\xi = f\tilde\xi f^{-1}\,,\qquad\phi = f\tilde\phi f^{-1}\,,
\ee
where $f$ was the frame on asymptotic twistor lines $L_{u,z}$. In terms of $\tilde\xi,\tilde\phi$, \eqref{Dbarxi} reduces to
\be
\begin{split}
    &\p_\bq\tilde\xi = \p_\bu\tilde\xi = e_{\bz}\tilde\xi + [\bar\cA,\tilde\xi] = 0\,,\\
    &\p_\bq\tilde\phi = \p_\bu\tilde\phi = e_{\bz}\tilde\phi + [\bar\cA,\tilde\phi] = 0\,.
\end{split}
\ee
The $\p_{\bar q},\p_\bu$ conditions, along with the ability to quotient by small gauge transformations, tell us that $\tilde\xi,\tilde\phi$ can be taken to be analytic in $u,q$ around $u=q=0$. As $e_\bz(u-q\bz)=0$, this is equivalent to analyticity in $\mu^\dal=(u-q\bz,q)$, so it agrees with the result in a trivial background.

Being holomorphic in $q\in\C$, the parameters $\tilde\xi, \tilde\phi$ admit Taylor expansions:
\be
\begin{split}
    \tilde\xi &= \sum_{s=0}^\infty\xi_s(u,z,\bz) \,q^{s}\,,\\
    \tilde\phi &= \sum_{s=0}^\infty\phi_s(u,z,\bz) \,q^{s}\,.
\end{split}
\ee
The $e_{\bz}$ conditions can be used to constrain the coefficients $\xi_s,\phi_s$. The lowest coefficients obey
\be\label{extra}
\cD_\bz\xi_0 = \cD_\bz\phi_0 = 0\,,
\ee
while for $s\geq1$ the higher coefficients obey the \emph{dual FPR recursion relations},
\be\label{dualrec}
\boxed{\hspace{1cm}\begin{split}
    &\cD_\bz\xi_s + \p_u\xi_{s-1} = 0\,,\\
    &\cD_\bz\phi_s + \p_u\phi_{s-1} = 0\,.
\end{split}\hspace{0.95cm}}
\ee
These run in the reverse direction of the usual recursions \eqref{fprrec}.

\medskip

In the next few sections, we will construct the Noether charges associated with these large gauge symmetries. We will weaken our definition of large gauge transformations by dropping the condition \eqref{extra}, leading to flux-balance laws at null infinity. The strict conservation laws will be recovered after assuming the absence of radiation. Instead of \eqref{Dbarxi}, the corresponding gauge parameters $\xi,\phi$ will satisfy the weaker conditions
\be\label{CarrollianGaugeParameters}
\Dbar\xi = f\,\cD_\bz\xi_0\,f^{-1}\,,\qquad \Dbar\phi = f\,\cD_\bz\phi_0\,f^{-1}\,.
\ee
These are precisely the residual gauge symmetries of radiative gauge. Indeed, if we gauge transform both sides of
\be
\bar\cA = f^{-1}(e_\bz+a_\bz)f
\ee
using $\delta a=\Dbar\xi$ and holding the frame (i.e., the gauge choice) fixed, we find that $\bar\cA$ must transform as 
\be
\delta\bar\cA = e_\bz\tilde\xi+[\bar\cA,\tilde\xi] = \cD_\bz\xi_0\,.
\ee
The first equality here is general, while the second equality holds when $\xi_s$ satisfy the dual recursion. This physical interpretation of $\xi_0$ as a gauge transformation on $\scri$ motivates us to try solving for a basis of soft gluons without imposing \eqref{extra}.

\paragraph{Positive helicity soft gluons.} When $\bar\cA=0$, we saw that soft gluons are polynomial in $\mu^\dal=(u-q\bz,q)$. Let $\rho_m(\lambda,\bar\lambda)$, $m\geq0$, be a smooth section of $\CO(-m)\otimes\br{\CO(m)}\to\P^1_\lambda$ obeying the \emph{wedge condition},
\begin{equation}
    \partial_{\bar z}^{m+1}\rho_m = 0\,,
\end{equation}
at least away from $z=0,\infty$. In a trivial background, to each such $\rho_m$ we could associate a soft gluon symmetry generator $\xi = \xi(\rho_m)$ given by a monomial in $\mu^\dal$ of the form
\begin{equation}
    \begin{split}
        \xi(\rho_m) &= \frac{1}{m!}\,\frac{\partial^m \rho_m}{\partial\bar\lambda^{\dot \alpha_1} \cdots \partial\bar\lambda^{\dot \alpha_m} }\,\mu^{\dot \alpha_1} \cdots \mu^{\dot \alpha_s} \\
        &= \sum_{s=0}^{m} \frac{(-1)^s}{(m-s)!}\,u^{m-s}\,q^s\, \partial_{\bar z}^{m-s}\rho_m\,.
    \end{split}
\end{equation}
Eg., $\rho_m = (-1)^lz^{-r}\bz^k\mfk{t}^a$ for $0\leq k\leq m$ corresponds to a soft gluon of order $m$. This trivial background result can be used as a seed to solve the dual recursion relations. We can correct it order-by-order in $q$ to find soft gluons in the presence of a nontrivial background.

Setting $\xi(\rho_m) = f\,\tilde\xi(\rho_m)\,f^{-1}$, we expand $\tilde\xi(\rho_m)=\sum_{s=0}^m\xi_s\,q^s$ as before. But now we are assuming that this expansion truncates at $s=m$. This allows us to solve the dual recursions. We see that $\p_u\xi_m=0$, so we set $\xi_m = \rho_m$ as in flat space. Then we solve for $\xi_{m-1}$ in terms of $\xi_m$ by setting $\xi_{m-1}=-\p_u^{-1}\cD_\bz\xi_m$, and so on.  We content ourselves with only stating the results for the leading, subleading and subsubleading soft gluons, i.e., $m=0,1,2$. The first few examples of $\tilde \xi(\rho_m)$ are found to be
\begin{equation}\label{xilist}
\begin{split}
    \tilde \xi(\rho_0) &= \rho_0\,,\\
    \tilde \xi(\rho_1) &= q\,\rho_1 - u\,\partial_{\bar z}\rho_1 - [\partial_u^{-1}\bar{\mathcal{A}} , \rho_1]\,, \\
    \tilde \xi(\rho_2) &=  q^2\rho_2 - q\,\big(u\,\partial_{\bar z}\rho_2 +[\partial_u^{-1}\bar{\mathcal{A}}, \rho_2]\big) + \frac{u^2}{2}\,\partial_{\bar z}^2\rho_2 \\
    & \qquad\qquad \qquad+ \partial_{\bar z}[\partial_{u}^{-2}\bar{\mathcal{A}}, \rho_2] + [\partial_u^{-1}(u\bar{\mathcal{A}}), \partial_{\bar z}\rho_2] + \partial_u^{-1}[\bar{\mathcal{A}},[\partial_u^{-1}\bar{\mathcal{A}},\rho_2]]\,.
\end{split}
\end{equation}
Higher-order results may be systematically obtained using the techniques developed in appendix A of \cite{Kmec:2024nmu}.

Using the modified bracket \eqref{alal'}, one can check with brute force that these generators satisfy the $S$-algebra
\begin{equation}\label{xialg}
    [\xi(\rho_m) , \xi(\rho'_{m'})]_\star = \xi([\rho_m, \rho'_{m'}])\,.
\end{equation}
We can also give a slightly more abstract proof of this algebra by directly working in radiative gauge. In \eqref{DbarBT}, we showed that
\be
    \Dbar\big([\xi,\xi']_\star\big) = [\delta_\xi,\delta_{\xi'}]a\,.
\ee
Transitioning to radiative gauge, we find that
\be
\cD_\bz([\tilde\xi,\tilde\xi']_\star) = [\delta_{\tilde\xi},\delta_{\tilde\xi'}]\bar\cA\,.
\ee
Transformations of the form $\delta_{\tilde\xi}\bar\cA = \cD_\bz\tilde\xi$ map scri data to scri data, so they never generate any dependence on $q$. Hence, the commutator $[\delta_{\tilde\xi},\delta_{\tilde\xi'}]\bar\cA$ cannot contain any dependence in $q$ either. It follows that the coefficients in a $q$-expansion of $[\tilde\xi,\tilde\xi']_\star$ must again satisfy the dual recursions. Moreover, it is easily checked that
\be
[\xi(\rho_m),\xi(\rho'_{m'})]_\star = f\left(q^{m+m'}[\rho_m,\rho'_{m'}] + O(q^{m+m'-1})\right)f^{-1}\,.
\ee
Thus, at least by a formal uniqueness of solutions of the dual recursion relations, we conclude that $[\xi(\rho_m),\xi'(\rho_{m'})]_\star$ must be of the form $\xi(\rho''_{m+m'})$ with $\rho''_{m+m'} = [\rho_m,\rho'_{m'}]$.

To obtain the more familiar form of the $S$-algebra, choose
\be
\rho_m = \rho_{k,l,r}^a \equiv (-1)^lz^{-r}\bz^k\mfk{t}^a
\ee
with $m = k+l$, for some non-negative integers $k,l$. This defines for us a more standard basis for the generators:
\begin{equation}\label{xirhokl}
    \xi^a_{k,l,r} = \xi(\rho^a_{k,l,r})\,,
\end{equation}
which results in the commutator
\begin{equation}
\label{S algebra}
    [\xi_{k,l,r}^a , \xi_{m,n,s}^b]_{\star} = f^{ab}{}_c\,\xi_{k+m,l+n,r+s}^c \,.
\end{equation} 
This coincides with the commutators of the $S$-algebra in its original form \cite{Strominger:2021mtt,Costello:2022wso}. 

\paragraph{Negative helicity soft gluons.} A similar discussion holds for the gauge parameter $\phi$. In a trivial background, we introduce the following basis for the large gauge transformations:
\begin{equation}
    \begin{split}
        \phi(\chi_m) &= \frac{1}{m!}\,\frac{\partial^m \chi_m(\lambda,\bar\lambda)}{\partial\bar\lambda^{\dot \alpha_1} \cdots \partial\bar\lambda^{\dot \alpha_m} }\,\mu^{\dot \alpha_1} \cdots \mu^{\dot \alpha_m}  \\
        &= \sum_{s=0}^{m} \frac{(-1)^s}{(m-s)!}\,u^{m-s}\,q^s\, \partial_{\bar z}^{m-s}\chi_{m}(z,\bar z)\,,
    \end{split}
\end{equation}
where $\chi_m(\lambda, \bar\lambda)$ is a section of $\CO(-m-4)\otimes\br{\CO(m)}$ and satisfies the wedge condition $\partial_{\bar z}^{m+1}\chi_m = 0$. In the second line, we are working in the coordinate patch $\lambda_0=1$. In a nontrivial background, we write $\phi = f\tilde\phi f^{-1}$ and solve the dual recursion \eqref{dualrec} for the coefficients in the expansion $\tilde\phi=\sum_{s\geq0}\tilde\phi_sq^s$ by setting $\phi_m = \chi_m$ and $\phi_s=0$ for $s>m$. Denote the resulting solution by $\tilde\phi(\chi_m)$. Its first few examples are
\begin{equation}\label{philist}
\begin{split}
    \tilde \phi(\chi_0) &= \chi_0\,,\\
    \tilde \phi(\chi_1) &= q\,\chi_1 - u\,\partial_{\bar z}\chi_1 - [\partial_u^{-1}\bar{\mathcal{A}} , \chi_1]\,, \\
    \tilde \phi(\chi_2) &=  q^2\chi_2 - q\,\big(u\,\partial_{\bar z}\chi_2 +[\partial_u^{-1}\bar{\mathcal{A}}, \chi_2]\big) + \frac{u^2}{2}\,\partial_{\bar z}^2\chi_2 \\
    & \qquad\qquad \qquad+ \partial_{\bar z}[\partial_{u}^{-2}\bar{\mathcal{A}}, \chi_2] + [\partial_u^{-1}(u\bar{\mathcal{A}}), \partial_{\bar z}\chi_2] + \partial_u^{-1}[\bar{\mathcal{A}},[\partial_u^{-1}\bar{\mathcal{A}},\chi_2]]\,.
\end{split}
\end{equation}
In our choice of affine coordinates, these have taken the same form as the $\tilde\xi(\rho_m)$ in \eqref{xilist}.

The algebra of positive and negative helicity soft gluons can again be computed using the Barnich-Troessaert bracket. We expect that the commutator $[\xi(\rho_m),\phi(\chi_{m'})]_\star$ will result in a negative helicity soft gluon $\phi([\rho_m,\chi_{m'}])$. Generalizing our abstract argument to this case should be possible by introducing the radiative data for $b$. However, we will find it easier to circumvent this issue by working with soft gluons satisfying the stronger condition \eqref{Dbarxi}, as will be the case in section \ref{ssec:gicc}.


\subsection{Corner charges}
\label{ssec:corner}

In the previous section, we defined large gauge symmetries as gauge transformations that violated the boundary conditions of the twistorial theory. They are nontrivial symmetries of the theory, so they give rise to non-vanishing Noether charges. Such Noether charges may be constructed using the symplectic structure on the phase space of the twistorial theory, as outlined in section \ref{ssec:twac}.

As shown in \eqref{HHt}, the Hamiltonians $H_\xi$ and $\til H_\phi$ generating large gauge transformations are integrals over the boundary of a chosen Cauchy surface. In this section, we will evaluate them on a choice of Cauchy surfaces adapted to null infinity. To contrast them with the celestial charges, we will denote the resulting corner charges by $Q_\xi$ and $\til Q_\phi$,
\be
\begin{split}
    Q_{\xi} &= \int_{\partial \Sigma} \D^3 Z \wedge \Tr\,\xi\, b\,,\\
    \til Q_{\phi} &= \int_{\partial \Sigma} \D^3 Z \wedge \Tr\,\phi\, \p_ua \,,
\end{split}
\ee
In these expressions, $\Sigma$ denotes the quantization `Cauchy' surface in twistor space, and $\p\Sigma$ is its boundary. The vector $\ell$ used to construct the negative helicity charges in \eqref{HHt} has been taken to be $\ell=\p_u$. This will be the choice that helps us in expressing these charges in terms of the negative helicity charge aspects.

In this section, we follow \cite{Freidel:2023gue} in using the charge aspects to build corner charges obeying conservation laws or flux-balance laws in Bondi time $u$. These may be thought of as generators of the $S$-algebra in a Carrollian dual to sdYM. In the next section, we will use the same charge aspects to build conserved currents of a celestial dual. The charge aspects are the perfect middlemen for relating the two approaches.

\medskip

To construct corner charges, one needs to choose a Cauchy surface adapted to the Carrollian picture. Such Cauchy surfaces can be found by remembering the projection $p:\PT\to\scri_\C$ given by $(u,q,z)\mapsto(u,z)$. Real null infinity embeds into $\scri_\C$ as the subset $\scri=\{u+\bu=0\}$ on which $u$ is pure imaginary.\footnote{This is a convention that arose from working with the incidence relations $\mu^\dal=x^{\al\dal}\lambda_\al$, which are the convenient choice for most purposes. Had we worked with the alternative incidence relations $\mu^\dal=ix^{\al\dal}\lambda_\al$, real $\scri$ would have embedded as the set of real $u$.} So $u$ is related to real Bondi time $u_B$ as $u=iu_B$. To discuss the translation of the phase space from twistor space to spacetime at $\mathscr{I}$, a convenient choice of codimension-$1$ hypersurface is $\Sigma = \{u = -\bar u \}$. This projects to a subset of real $\scri$ under $p$. The surface charges will then be integrated over the codimension-$2$ surfaces $\partial \Sigma = \{ u = -\bar u = \text{constant} \}$.

The holomorphic measure on twistor space is
\be\label{d3z}
\D^3Z = (\d u-q\,\d\bz)\wedge\d z\wedge\d q\,.
\ee
On a cut of constant of $u$, only the $q\,\d z\wedge\d\bz\wedge\d q$ factor need be kept. So, dropping normalization factors, our surface charges take the form
\be
\begin{split}
    Q_{\xi}(u) &= \int_{\P^1} \d^2z \int_{L_{u,z}}q\,\d q\wedge \Tr\,\xi\, b\,,\\
    \til Q_{\phi}(u) &= \int_{\P^1} \d^2z \int_{L_{u,z}}q\,\d q\wedge \Tr\,\phi\, \p_ua \,.
\end{split}
\ee
These are examples of the more general concept of \emph{corner charges}. 

Plugging in $\xi = \sum_{s\geq0}f\xi_sf^{-1}q^s$ and $\phi=\sum_{s\geq0}f\phi_s f^{-1}q^s$, the charges $Q_\xi$ and $\til Q_\phi$ expand out as
\be
\boxed{\hspace{1cm}\begin{split}
    Q_\xi(u) &= \int_{\P^1} \d^2z\;\Tr\sum_{s=0}^\infty\xi_s R_s\,,\\
    \til Q_\phi(u) &= \int_{\P^1} \d^2z\;\Tr\sum_{s=0}^\infty\phi_s \til R_s\,,
\end{split}\hspace{0.95cm}}
\ee
where we invoked the definitions \eqref{Tod formulae} of the charge aspects. The merit of expressing the corner charges in this way is that they only include quantities like $\xi_s$, $R_s$, etc.\ that live on null infinity instead of twistor space. In practice, we determine the charge aspects and gauge parameters purely by solving the FPR recursion and dual recursion relations, which are purely spacetime concepts and do not require twistor theory! This makes manifest the connection with spacetime physics, albeit at the expense of needing to work in radiative gauge and losing manifest invariance under small gauge transformations.

We now show that when $\xi$ and $\phi$ obey the dual recursion relations \eqref{dualrec}, these charges are conserved in $u$,
\be
\p_uQ_\xi = \p_u\til Q_\phi = 0\,.
\ee
This will also require us to assume the absence of radiation $R_{-1}$ and $\til R_{-1}$.

We can compute the time derivative of $Q_\xi$ with the help of the recursion \eqref{fprb} and the first dual recursion in \eqref{dualrec}:
\begingroup
\allowdisplaybreaks
\begin{align}
    \p_uQ_\xi &= \int_{\P^1} \d^2z\;\Tr\,\bigg(\sum_{s=0}^\infty\p_u\xi_s\,R_s + \sum_{s=0}^\infty\xi_s\,\p_uR_s\bigg)\nonumber\\
    &= -\int_{\P^1}\d^2z\;\Tr\,\bigg(\sum_{s=0}^\infty\cD_\bz\xi_{s+1}\,R_s + \sum_{s=0}^\infty\xi_s\,\cD_\bz R_{s-1}\bigg)\nonumber\\
    &= -\int_{\P^1}\d^2z\;\p_\bz\bigg(\Tr\,\sum_{s=0}^\infty\xi_{s+1}R_s\bigg) - \int_{\P^1}\d^2z\;\Tr\;\xi_0\cD_\bz R_{-1}\,.
\end{align}
\endgroup
Dropping the total derivative term gives an expression for the local flux,
\be\label{fluxxi}
\cF_\xi \equiv \p_uQ_\xi = \int_{\P^1}\d^2z\;\Tr\,R_{-1}\cD_\bz\xi_0\,.
\ee
As promised, this vanishes if we assume that $R_{-1}=0$, which is the condition for the absence of radiation. An identical calculation also gives the negative helicity flux,
\be\label{fluxphi}
\til\cF_\phi \equiv \p_u\til Q_\phi = \int_{\P^1}\d^2z\;\Tr\,\til R_{-1}\cD_\bz\phi_0\,.
\ee
This vanishes when $\til R_{-1}\equiv\bar\cA = 0$. 

In conclusion, in self-dual gauge theory, we obtain positive helicity conserved quantities $Q_\xi$ if $R_{-1}=0$. But generically the negative helicity charges $\til Q_\phi$ fail to be conserved unless one is in a trivial background $\bar\cA=0$. One way to obtain conserved quantities of both helicities is to develop gauge parameters satisfying the additional constraint $\cD_\bz\xi_0=\cD_\bz\phi_0=0$. We will do this in the celestial formalism in section \ref{ssec:gicc}.

Writing $\xi(\rho_s) = f\,\tilde\xi(\rho_s)\,f^{-1}$ and $\phi(\chi_s)=f\,\tilde\phi(\chi_s)\,f^{-1}$, we provided a basis of soft gluon generators in \eqref{xilist} and \eqref{philist} in terms of functions $\rho_s(z,\bz),\chi_s(z,\bz)$ on the sphere satisfying the wedge condition $\p_\bz^{s+1}\rho_s=\p_\bz^{s+1}\chi_s=0$. Using these, we can define a basis of Noether charges. Soft gluons at order $s\geq0$ in the soft expansion give rise to the corner charge
\be\label{Qbasis}
Q_s \equiv Q_{\xi(\rho_s)}\,,\qquad\til Q_s \equiv Q_{\phi(\chi_s)}\,.
\ee
We provide below some explicit expressions of positive helicity charges for the lowest spins:
\begingroup
\allowdisplaybreaks
\begin{align}\label{Qegs}
        Q_0(u) &= \int_{\P^1}\d^2z\; \text{Tr}\,(\rho_0 R_0)\nonumber\\
         Q_1(u) &= \int_{\P^1}\d^2z\;  \text{Tr}\,\big(\rho_1 R_1 - u\,\partial_{\bar z} \rho_1 R_0 - [\partial_u^{-1}\bar{\mathcal{A}}, \rho_1]R_0\big)\\
          Q_2(u) &= \int_{\P^1}\d^2z\;  \text{Tr}\,\bigg(\rho_2 R_2 - u\,\partial_{\bar z} \rho_2 R_1+\frac{u^2}{2}\,\partial_{\bar z}^2 \rho_2 R_0 - [\partial_u^{-1}\bar{\mathcal{A}}, \rho_2]R_1 \nonumber\\
          &\qquad \qquad\qquad+ \partial_{\bar z} [\partial_u^{-2}\bar{\mathcal{A}}, \rho_2]R_0 + [\partial_u^{-1}(u\bar{\mathcal{A}}), \partial_{\bar z}\rho_2]R_0 + \partial_u^{-1}[\bar{\mathcal{A}},[\partial_u^{-1}\bar{\mathcal{A}}, \rho_2]]R_0\bigg)\,.\nonumber
\end{align}
\endgroup
These can be tested by verifying \eqref{fluxxi} by brute force. The corresponding negative helicity charges take a similar form,
\begingroup
\allowdisplaybreaks
\begin{align}\label{Qtegs}
        \til Q_0(u) &= \int_{\P^1}\d^2z\; \text{Tr}\,(\chi_0 \til R_0)\nonumber\\
         \til Q_1(u) &= \int_{\P^1}\d^2z\;  \text{Tr}\,\big(\chi_1 \til R_1 - u\,\partial_{\bar z} \chi_1 \til R_0 - [\partial_u^{-1}\bar{\mathcal{A}}, \chi_1]\til R_0\big)\\
          \til Q_2(u) &= \int_{\P^1}\d^2z\;  \text{Tr}\,\bigg(\chi_2 \til R_2 - u\,\partial_{\bar z} \chi_2 \til R_1+\frac{u^2}{2}\,\partial_{\bar z}^2 \chi_2 \til R_0 - [\partial_u^{-1}\bar{\mathcal{A}}, \chi_2]\til R_1 \nonumber\\
          &\qquad \qquad\qquad+ \partial_{\bar z} [\partial_u^{-2}\bar{\mathcal{A}}, \chi_2]\til R_0 + [\partial_u^{-1}(u\bar{\mathcal{A}}), \partial_{\bar z}\chi_2]\til R_0 + \partial_u^{-1}[\bar{\mathcal{A}},[\partial_u^{-1}\bar{\mathcal{A}}, \chi_2]]\til R_0\bigg)\,.\nonumber
\end{align}
\endgroup
Higher-order charges can be obtained in a similar fashion.


\subsection{Celestial currents}
\label{ssec:celestial}

The construction of $S$-algebra currents in celestial CFT follows the same pattern as the corner charge construction. The only change in perspective that the reader needs to make is a change of Cauchy surface.

As discussed at the beginning of section \eqref{ssec:large}, the natural Cauchy surface pertinent to celestial holography is the surface $\Sigma = \{|z|=\text{constant}\}$, where $z=\lambda_1/\lambda_0$ is an affine coordinate on the base of the fibration $\PT\to\P^1$. Essentially, celestial CFT lives on the $\P^1$ base, and radial evolution in $z$ along $\P^1$ coincides with evolution along these Cauchy surfaces in $\PT$. In this way, bulk and boundary `times' get identified.

The boundary of $\Sigma$ is the surface $\p\Sigma = \{|u|=1/\eps\}\cap\{|z|=\text{constant}\}$ in the limit as $\eps\to0$. This is because the boundary of twistor space is located at $u\to\infty$ (see equation \eqref{pb}). An integral over $\p\Sigma$ contains a contour integral capturing the poles around $u=\infty$, as well as a contour integral around $z=0$ (or $z=\infty$) capturing modes of a loop algebra. The contour integral around $u=\infty$ may also be understood as an integral along the null generators of real $\scri$ when the integrands obey appropriate boundary conditions; see for instance \cite{Freidel:2022skz}.

Using the same holomorphic measure \eqref{d3z} as before, we obtain the expressions
\be
\begin{split}
    H_{\xi} &= \oint_{z=0} \d z\oint_{u=\infty}\d u \int_{L_{u,z}}\d q\wedge \Tr\,\xi\, b\,,\\
    \til H_{\phi} &=  \oint_{z=0}\d z\oint_{u=\infty}\d u\int_{L_{u,z}}\d q\wedge \Tr\,\phi\,\p_ua \,,
\end{split}
\ee
for the associated Noether charges. The contour integrals over $z$ are clearly picking out modes of the ``2d operators'' defined by
\be\label{Jdef}
\begin{split}
    J_{\xi}(z) &= \oint_{u=\infty}\d u\int_{L_{u,z}}\d q\wedge \Tr\,\xi\, b\,,\\
    \til J_{\phi}(z) &= \oint_{u=\infty}\d u\int_{L_{u,z}}\d q\wedge \Tr\,\phi\,\p_ua \,.
\end{split}
\ee
These are abstract versions of the $S$-algebra currents of celestial CFT. They are the celestial counterparts of the corner charges $Q_\xi,\til Q_\phi$ that were conserved under evolution in $u$. 

Plugging either $\xi = \sum_{s\geq0}f\xi_sf^{-1}q^s$ or $\phi = \sum_{s\geq0}f\phi_sf^{-1}q^s$ into the definitions \eqref{Jdef} yields
\be
\boxed{\hspace{1cm}\begin{split}
    J_\xi(z) &= \oint_{u=\infty}\d u\;\Tr\sum_{s=0}^\infty\xi_s R_{s-1}\,,\\
    \til J_\phi(z) &= \oint_{u=\infty}\d u\;\Tr\sum_{s=0}^\infty\phi_s \til R_{s-1}\,.
\end{split}\hspace{0.95cm}}
\ee
The contour integral over $u$ is well-defined because all the quantities $\xi_s$ and $R_s$ are holomorphic in $u$. Quite beautifully, we find that the charge aspects $R_s$ that were used to construct corner charges -- along with their negative helicity counterparts $\til R_s$ -- also end up being the building blocks of celestial currents.


Next let us study the conditions under which these currents behave as chiral currents of a 2d CFT. As always, we start with the positive helicity currents $J_\xi$. Now we can use the recursions \eqref{fprrec} and dual recursions \eqref{dualrec} in reverse to compute
\begingroup
\allowdisplaybreaks
\begin{align}
    \p_\bz J_\xi &= \oint_{u=\infty}\d u\;\Tr\,\bigg(\sum_{s=0}^\infty\cD_\bz\xi_s\,R_{s-1} + \sum_{s=0}^\infty\xi_s\,\cD_\bz R_{s-1}\bigg)\nonumber\\
    &= \oint_{u=\infty}\d u\;\Tr\,\bigg(\cD_\bz\xi_0\,R_{-1} - \sum_{s=1}^\infty\p_u\xi_{s-1}\,R_{s-1} - \sum_{s=0}^\infty\xi_s\,\p_uR_{s}\bigg)\nonumber\\
    &= \oint_{u=\infty}\d u\;\Tr\,R_{-1}\cD_\bz\xi_0 - \oint_{u=\infty}\d u\;\p_u\bigg(\Tr\sum_{s=0}^\infty\xi_sR_s\bigg)\,.
\end{align}
\endgroup
This holds away from $z=0,\infty$ where the large gauge parameters were allowed to be singular. Dropping the total $u$-derivative term, we find that
\be
\p_\bz J_\xi = \oint_{u=\infty}\d u\;\Tr\,R_{-1}\cD_\bz\xi_0\,.
\ee
An analogous calculation also gives
\be
\p_\bz\til J_\phi = \oint_{u=\infty}\d u\;\Tr\,\til R_{-1}\cD_\bz\phi_0
\ee
for the negative helicity charges.

Once again, we obtain the conservation laws
\be
\p_\bz J_\xi = \p_\bz\til J_\phi = 0
\ee
if either the radiative data $R_{-1}$ and $\til R_{-1}$ vanish, or if $\xi_0$ and $\phi_0$ obey the extra constraints $\cD_\bz\xi_0=\cD_\bz\phi_0=0$. This mirrors the structure of the corner charge fluxes \eqref{fluxxi}, \eqref{fluxphi}, whose vanishing was also subject to the vanishing of radiation.

As for the corner charges \eqref{Qbasis}, we can define a basis of soft gluon celestial currents graded by the order of softness $s$,
\be
J_s \equiv J_{\xi(\rho_s)}\,,\qquad\til J_s\equiv \til J_{\chi(\rho_s)}\,.
\ee
A more standard basis of modes may be found by using the basis of generators given in \eqref{xirhokl} and \eqref{phirhokl},
\be
\begin{split}
    J^a_r[k,l] &\equiv \oint_{z=0}\d z\;J_{\xi^a_{k,l,r}}(z)\,,\\
    \til J^a_r[k,l] &\equiv \oint_{z=0}\d z\;J_{\phi^a_{k,l,r}}(z)\,.
\end{split}
\ee
The explicit expressions of $J_s$ at leading, subleading and subsubleading orders are
\begingroup
\allowdisplaybreaks
\begin{align}\label{Jegs}
        J_0(z) &= \oint_{u=\infty}\d u\; \text{Tr}\,(\rho_0 R_{-1})\nonumber\\
         J_1(z) &= \oint_{u=\infty}\d u\;  \text{Tr}\,\big(\rho_1 R_0 - u\,\partial_{\bar z} \rho_1 R_{-1} - [\partial_u^{-1}\bar{\mathcal{A}}, \rho_1]R_{-1}\big)\\
          J_2(z) &= \oint_{u=\infty}\d u\;  \text{Tr}\,\bigg(\rho_2 R_1 - u\,\partial_{\bar z} \rho_2 R_0+\frac{u^2}{2}\,\partial_{\bar z}^2 \rho_2 R_{-1} - [\partial_u^{-1}\bar{\mathcal{A}}, \rho_2]R_0 \nonumber\\
          &\qquad \qquad\qquad+ \partial_{\bar z} [\partial_u^{-2}\bar{\mathcal{A}}, \rho_2]R_{-1} + [\partial_u^{-1}(u\bar{\mathcal{A}}), \partial_{\bar z}\rho_2]R_{-1} + \partial_u^{-1}[\bar{\mathcal{A}},[\partial_u^{-1}\bar{\mathcal{A}}, \rho_2]]R_{-1}\bigg)\,.\nonumber
\end{align}
\endgroup
And the corresponding negative helicity currents are found to be
\begingroup
\allowdisplaybreaks
\begin{align}\label{Jtegs}
        \til J_0(z) &= \oint_{u=\infty}\d u\; \text{Tr}\,(\chi_0 \til R_{-1})\nonumber\\
         \til J_1(z) &= \oint_{u=\infty}\d u\;  \text{Tr}\,\big(\chi_1 \til R_0 - u\,\partial_{\bar z} \chi_1 \til R_{-1} - [\partial_u^{-1}\bar{\mathcal{A}}, \chi_1]\til R_{-1}\big)\\
          \til J_2(z) &= \oint_{u=\infty}\d u\;  \text{Tr}\,\bigg(\chi_2 \til R_1 - u\,\partial_{\bar z} \chi_2 \til R_0+\frac{u^2}{2}\,\partial_{\bar z}^2 \chi_2 \til R_{-1} - [\partial_u^{-1}\bar{\mathcal{A}}, \chi_2]\til R_0 \nonumber\\
          &\qquad \qquad\qquad+ \partial_{\bar z} [\partial_u^{-2}\bar{\mathcal{A}}, \chi_2]\til R_{-1} + [\partial_u^{-1}(u\bar{\mathcal{A}}), \partial_{\bar z}\chi_2]\til R_{-1} + \partial_u^{-1}[\bar{\mathcal{A}},[\partial_u^{-1}\bar{\mathcal{A}}, \chi_2]]\til R_{-1}\bigg)\,.\nonumber
\end{align}
\endgroup
Their integrands are basically the same as those of the corner charges \eqref{Qegs} and \eqref{Qtegs}, except that every $R_s$ or $\til R_s$ is replaced by $R_{s-1}$ or $\til R_{s-1}$ respectively. Notably, the lowest aspects $R_{-1},\til R_{-1}$ never make an explicit appearance in the corner charges, but do show up in the celestial currents.

\medskip

Until now, the gauge parameters used have been those that satisfy \eqref{CarrollianGaugeParameters}, which give celestial currents that are conserved only in the absence of radiation. We can instead solve \eqref{Dbarxi}, which will give currents that are conserved even in the presence of radiation. One formal way to do this is to use the frame $U$ of the Yang-Mills bundle at $\scri$ that satisfies \eqref{Ueq}.

We can formally solve \eqref{Dbarxi} by dressing the linear results
\begin{equation} \label{dressing parameters}
\begin{split}
    \xi_{k,l} &= (u-\bar z q)^k q^l f \,U \rho(z)\,U^{-1} f^{-1}\,, \\  
    \phi_{k,l} &=(u-\bar z q)^k q^l  f\,U\chi(z)\,U^{-1} f^{-1}\, .
\end{split}    
\end{equation}
 Using the Cauchy kernel on $\mathbb{C}$, we can write the generators as an infinite sum of integrals
\begin{equation}
    \xi_{k,l} = (u-\bar z q)^k q^l f\left[\rho(z) +\sum_{n=1}^{\infty}(-1)^{n}\int_{\mathbb{C}^n}\frac{\d^2 z_n \cdots \d^2 z_1}{(z_n-z_{n-1})\cdots (z_1 - z)}E_{1}\cdots E_n \rho(z)\right]f^{-1}\, ,
\end{equation}
where we have introduced the operator $E_n = q\partial_u + [\bar{\mathcal{A}}(u,z_n,\bar z_n) ,-]$.

The corresponding Noether currents take the form
\be
\begin{split}
     J_{\xi_{k,l}}(z) &= \oint_{u=\infty}\d u\int_{L_{u,z}}\d q\,(u-q\bz)^k\,q^l\,\Tr\,(U\rho\,U^{-1}f^{-1}bf)\,,\\
    \til J_{\phi_{k,l}}(z) &= \oint_{u=\infty}\d u\int_{L_{u,z}}\d q\,(u-q\bz)^k\,q^l\,\Tr\,(U\chi\,U^{-1}f^{-1}\p_uaf)\,.
\end{split}
\ee
We may again express these charges in terms of the charge aspects by expanding $U$ in a power series in $q$ around $q=0$. An example of a subleading soft gluon current constructed in this way is given by
\begin{equation}
     \til J[0,1] = \oint_{u=\infty}\d u\; \text{Tr}\bigg(\chi \til R_{0} - \sum_{n=0}^{\infty } (-1)^n [\partial_u^n \partial_{\bar z}^{-(n+1)}\bar{\mathcal{A}}, \chi]\til R_{n}\bigg)
\end{equation}
As such expansions seem somewhat unilluminating, we will prefer not to do this in general. Instead, in the next section, we will provide a more direct technique to construct soft gluons obeying $\Dbar\xi=\Dbar\phi=0$, inspired from Burns holography \cite{Costello:2023hmi}. The resulting celestial currents will not contain any frames and will be much more explicitly expressible in terms of $a,b$.


\section{The extrapolate dictionary}
\label{sec:extrapolate}

In previous sections, we presented Noether charges of the $S$-algebra in terms of charge aspects adapted to radiative gauge. But large gauge charges should be invariant under small gauge transformations. In this section, we improve upon the charge aspect formulation by describing a scheme to build manifestly gauge-invariant expressions for celestial currents of the $S$-algebra. This procedure was carried out in linear theory (in a 4d WZW model of sdYM) in \cite{Costello:2023hmi}, and we extend it to nonlinear theory in what follows. 

As opposed to the previous sections, we also choose to work with soft gluons obeying the extra constraint \eqref{extra}. This will ensure that the resulting $S$-algebra currents are conserved in radial evolution even when the radiation is non-vanishing. Unfortunately, this might mean that we are working in a slightly different basis of soft modes than FPR. Our bases of soft modes coincide in a trivial background, and possibly only differ by field-dependent terms. Furthermore, the field-dependent terms might be gauge equivalent, but only via large gauge transformations that do not leave the charges invariant. We leave a detailed comparison to the future.

In the subsequent section, we will study the effect of inserting these charges in the path integral of sdYM. We will show that when computing form factors, inserting a soft charge supported at $\mu^\dal\to\infty$ is equivalent to inserting a soft current in the universal defect CFT living on the line $\mu^\dal=0$. This provides a precise relationship between our construction of celestial charges and the Koszul duality approach of Costello and Paquette \cite{Costello:2020jbh,Costello:2022wso,Paquette:2021cij}. It is an instantiation of the extrapolate dictionary.


\subsection{Gauge invariant celestial currents}
\label{ssec:gicc}

In section \ref{ssec:bdry}, we created the celestial sphere by blowing up twistor space along the line at infinity $\lambda_\al=0$. The blowup was a subspace of $\P^3_Z\times\P^1_\pi$ cut out by $\la\lambda\pi\ra=0$, where the $\P^3$ carried twistor coordinates $Z^A$, and the auxiliary $\P^1$ carried homogeneous coordinates $\pi_\al$. We defined a coordinate $n\in\C$ by solving $\la\lambda\pi\ra=0$ as $\lambda_\al=n\pi_\al$. The divisor $\cE$ at $n=0$ that replaced the line at infinity was a copy of $\P^1_\mu\times\P^1_\pi$ with independent homogeneous coordinates $\mu^\dal$ and $\pi_\al$.

Celestial CFT lives on the $\P^1_\pi$ factor of the boundary divisor $\cE$. So our gauge invariant charges will be local operators on $\P^1_\pi$ and will be integrated along $\P^1_\mu$. To study these, it is again useful to introduce a judicious choice of affine coordinates:
\be\label{nyz}
\eta = \frac{\pi_0}{\mu^{\dot0}}\,n\,,\qquad y = \frac{\mu^{\dot1}}{\mu^{\dot0}}\,,\qquad z = \frac{\pi_1}{\pi_0}\,.
\ee
$\eta$ now plays the role of the coordinate normal to the boundary divisor $\cE$; whereas $y$ and $z$ are affine coordinates on the $\mu^\dal$ and $\pi_\al$ spheres comprising $\cE=\P^1_\mu\times\P^1_\pi$. These coordinates are written in the patch $\mu^{\dot0}\neq0$ and $\pi_0\neq0$. They can be extended to other patches via the transition functions
\be
y\mapsto\frac{1}{y}\implies\eta\mapsto \frac{\eta}{y}\,,\qquad z\mapsto\frac{1}{z}\implies\eta\mapsto z\eta
\ee
and their combinations.

In these new coordinates, the volume form on $\PT$ reads
\be
\begin{split}
    \D^3Z &= (\mu^{\dot0})^4\,\d\bigg(\frac{\lambda_0}{\mu^{\dot0}}\bigg)\wedge\d\bigg(\frac{\lambda_1}{\mu^{\dot0}}\bigg)\wedge\d\bigg(\frac{\mu^{\dot1}}{\mu^{\dot0}}\bigg) \\
    &= \eta\,\d\eta\wedge\d z\wedge\d y\,,
\end{split}
\ee
which is obtained by setting $\lambda_\al=n\pi_\al$ away from $n=0$, and fixing the scales of $Z^A,\pi_\al$ by setting $\mu^{\dot0}=\pi_0=1$. As expected, it has a first-order zero at $\eta=0$.

Away from $\lambda_\al=0$, these coordinates are related to our non-holomorphic affine coordinates $u,q,z$ by
\be
\eta = \frac{1}{u-q\bz}\,,\qquad y = \frac{q}{u-q\bz}\,,
\ee
as may be seen through the map \eqref{pb}. But as opposed to $u,q,z$, the new coordinates $\eta,y,z$ are genuine holomorphic coordinates on twistor space. This simplifies calculations on twistor space, at the expense of complicating the transition to spacetime.

Away from $\eta=0$, we expand the fields $a,b$ as
\be
\begin{split}
    a &= a_\bz\,\d\bz + a_\by\,\d\by + a_{\bar\eta}\,\d\bar\eta\,,\\
    b &= \frac{1}{\eta^4}\left(b_\bz\,\d\bz + b_\by\,\d\by + b_{\bar\eta}\,\d\bar\eta\right)\,.
\end{split}
\ee
Their equations of motion are $[\bar D_\bi,\bar D_\bj] = 0$ and $\bar D_{[\bi}b_{\bj]}=0$, where $\bi$ now runs over $\{\bz,\by,\bar\eta\}$, and $\bar D_\bi = \p_\bi + a_\bi$. We have extracted a factor of $\eta^{-4}$ from the components of $b$. In this convention, the components $a_\bi$ and $b_\bi$ appear on equal footing as they both vanish to first order at $\eta=0$. This incorporates our boundary conditions $a\sim O(\eta)$ and $b\sim O(\eta^{-3})$ as $\eta\to0$.

\paragraph{Positive helicity currents.} Once again, let's start by tackling the positive helicity case. Our charges will be local in $z$, integrals over $y\in\P^1$, and placed at the boundary $\eta=0$. The charge associated to a gauge transformation $\xi$ was
\be
H_\xi = \int_{\p\Sigma}\D^3Z\wedge\Tr\,\xi\,b\,.
\ee
Using the celestial Cauchy surfaces discussed in the last section, this reduces to
\be
H_\xi = \oint_{z=0}\d z\oint_{\eta=0}\d\eta\,\eta\int_{\P^1}\d y\wedge\Tr\,\xi\,b\,,
\ee
These Hamiltonians are modes of the positive helicity celestial currents
\be\label{Jxin}
J_\xi(z) = \oint_{\eta=0}\frac{\d\eta}{\eta^3}\int_{\P^1}\d^2y\;\Tr\,\xi\,b_\by\,,
\ee
where $\d^2y=\d y\wedge\d\bar y$. Having plugged in $b=\eta^{-4}b_\by\d\by+\cdots$, the measure now exhibits a cubic pole at $\eta=0$. Large gauge transformations $\xi$ will violate our boundary conditions and contribute further poles at $\eta=0$. Our boundary conditions demand that $b_\by$ has a first-order zero at $\eta=0$, which is insufficient to cancel all poles in $\eta$. So the contour integral in $\eta$ will be non-vanishing in general.

Using the equations of motion, we can reduce the antiholomorphic derivative of $J_\xi$ to
\be
\p_\bz J_\xi = \oint_{\eta=0}\frac{\d\eta}{\eta^3}\int_{\P^1}\d^2 y\;\Tr\left(\Dbar_\bz\xi\,b_\by - \Dbar_\by\xi\,b_\bz\right)\,.
\ee
We would like this to vanish. So, away from the $|\text{in}\ra$ and $|\text{out}\ra$ vacua placed at $z=0,\infty$, we impose holomorphicity
\be
\Dbar\xi = 0\,,
\ee
i.e., $\Dbar_\bi\xi = 0$ for all $\bi$. This also ensures that away from $\eta=0$, the integrand of the $\eta$ integral is holomorphic in $\eta$,
\be
\p_{\bar\eta}\int_{\P^1}\d y\wedge\Tr\,\xi\,b = \int_{\P^1}\d^2 y\;\Tr\left(\Dbar_{\bar\eta}\xi\,b_\by - \Dbar_{\by}\xi\,b_{\bar\eta}\right) = 0\,.
\ee
Such holomorphicity is of course crucial for the $\eta$ integral to be independent of the precise contour of integration.

Next, we build an ansatz for $\xi$ that can represent soft gluon modes. Since the integral over $y$ gives rise to a holomorphic function of $\eta$, we may as well expand $\xi$ (and also $b_\by$) purely in $\eta$ but not $\bar\eta$. Then we will solve for the coefficients in this expansion using a recursive technique inspired from the construction of charge aspects.

In a trivial background, a soft gluon symmetry at order $s\in\Z_{\geq0}$ in the soft expansion is generated by
\be
\xi = \eta^{-s}(1+\tilde zy)^sz^{-r}\mfk{t}^a\,,
\ee
where $\tilde z\in\C$ is an auxiliary parameter that is \emph{not} the complex conjugate of $z$. Expanding this as a polynomial in $\tilde z$ gives
\be
\xi = \sum_{k+l=s}\frac{s!}{k!\,l!}\,\tilde z^l\,\eta^{-k-l}y^lz^{-r}\mfk{t}^a\,.
\ee
The term $\eta^{-k-l}y^lz^{-r}\mfk{t}^a$ is a soft gluon of order $k,l$ as previously encountered in \eqref{xikl}.

In a nontrivial background, this will receive corrections that are less singular in $\eta$. We may study an ansatz of the form
\be\label{xieta}
\xi = \sum_{j=-1}^s\eta^{-j}\xi_j + O(\eta^2) + \text{terms with $\bar\eta$}\,,
\ee
where the coefficients $\xi_j$ only depend on $y,\by$ and $z,\bz$. We have truncated the expansion in $\eta$ at first order in $\eta$. This is because terms of order $\eta^2$ or higher would drop from the contour integral in \eqref{Jxin}. Plugging this expansion into \eqref{Jxin} and performing the $\eta$ integrals by picking out Taylor coefficients of $b_\by$ yields
\be\label{Jxigi}
\boxed{\hspace{1cm}J_\xi(z) = \sum_{j=-1}^s\frac{1}{(j+2)!}\int_{\P^1}\d y\wedge\Tr\;\xi_j\p_\eta^{j+2}(\eta^4b)\bigr|_{\eta=0}\,,\hspace{0.95cm}}
\ee
which we have reexpressed in terms of the differential form $b$. Since $\eta^4b\sim O(\eta)$, the $\eta$ derivatives are necessary for extracting a nonzero integrand at $\eta=0$.

This is a slightly distinct idea from the expansion of $\xi$ in the coordinate $q$ employed in section \ref{ssec:softwave}. The coordinate $q$ is a coordinate on the lines of fixed $u,z$. The analog of $q$ in our analysis here is the coordinate $y$. To construct a gauge invariant version of FPR's analysis, one would instead choose to expand $\xi$ in powers of $y$. However, an expansion in $\eta$ is more natural from the viewpoint of classifying large gauge transformations. Eg., truncating the expansion at $\eta^{-s}$ tells us that we are dealing with a soft gluon symmetry that sits at order $\omega^{s}$ in the soft expansion of a gluon of energy $\omega$ on spacetime. It would also correspond to an expansion in $u$ around $u=\infty$. A term at order $\eta^{-s}$ carries over to a soft mode of order $u^s$, which is overleading on spacetime when $s\geq0$.

We now solve $\Dbar\xi=0$ for the $\xi_j$. Due to integrability $[\Dbar_\bi,\Dbar_\bj]=0$, it suffices to solve for $\xi$ using any one of the three conditions $\Dbar_\bi\xi=0$. We choose to solve the $\by$ equation
\be\label{ybeq}
\Dbar_\by\xi\equiv\p_\by\xi + [a_\by,\xi] = 0\,.
\ee
$a_\by$ has a first-order zero at $\eta=0$, so its expansion around $\eta=0$ takes the form
\be\label{aeta}
a_\by = \sum_{p=1}^\infty\frac{\eta^p}{p!}\,\p_\eta^p a_\by\bigr|_{\eta=0} + O(\bar\eta)\,.
\ee
Again, only the terms holomorphic in $\eta$ enter the equations governing the coefficients $\xi_j$. Plugging the expansions \eqref{xieta} and \eqref{aeta} into \eqref{ybeq} gives us the recursion relation
\be\label{xirec}
\p_\by\xi_j + \sum_{p=1}^{s-j}\frac{1}{p!}\big[\p_\eta^p a_\by\bigr|_{\eta=0},\xi_{p+j}\big] = 0\,,\qquad -1\leq j\leq s\,.
\ee
This is an analog of the FPR recursion where we have expanded the gauge parameter in $\eta$ instead of $q$.

Given the coefficient of the leading singularity $\eta^{-s}$,
\be
\xi_s=(1+\tilde zy)^sz^{-r}\mfk{t}^a\,,
\ee
obeying $\p_\by\xi_s=0$, the recursion can be solved by iteratively inverting $\p_\by$ on the $y$-sphere at $\eta=0$ and fixed $z$. To take into account the spins of $\xi_j$ under $y\mapsto1/y$, it is convenient to work with the quantities
\be
\hat\xi_j = (1+\tilde zy)^{-j-1}\xi_j\,,\qquad -1\leq j\leq s\,,
\ee
that transform as sections of $\CO(-1)$ under $y\mapsto1/y$, i.e., carry spin $1/2$ under conformal transformations of the $y$-sphere. In terms of the $\hat\xi_j$, the recursion \eqref{xirec} reduces to
\be\label{xihrec}
\p_\by\hat\xi_j + \sum_{p=1}^{s-j}\frac{(1+\tilde zy)^p}{p!}\big[\p_\eta^p a_\by\bigr|_{\eta=0},\hat\xi_{p+j}\big] = 0\,.
\ee
 This is useful because $\p_\by$ can be inverted on spin $1/2$ quantities without any integration constants.

Define the operators
\be\label{bigmas}
\hat\bigma_p \vcentcolon= -\p_{\by}^{-1}\circ\frac{(1+\tilde zy)^p}{p!}\,\text{ad}\big(\p_\eta^p a_\by\bigr|_{\eta=0}\big)\,,
\ee
where ad denotes the adjoint action on $\g$, i.e., $\text{ad}(X)Y = [X,Y]$ for $X,Y\in\g$. Then the solution of \eqref{xihrec} is given by
\be\label{xisol}
\boxed{\hspace{1cm}\hat\xi_j = \sum_{p=1}^{s-j}\sum_{q_1\geq1}\cdots\sum_{q_p\geq1}\delta_{q_1+\cdots+q_p,\,s-j}\,\hat\bigma_{q_1}\cdots\hat\bigma_{q_p}\hat\xi_s\,,\qquad -1\leq j\leq s-1\,.\hspace{0.95cm}}
\ee
From this, we obtain a soft gluon symmetry generator that we will refer to as $\xi_r^a[s]$,
\begin{multline}\label{xirs}
\xi_r^a[s] = \eta^{-s}z^{-r}(1+\tilde zy)^s\mfk{t}^a\\
- \sum_{j=1}^{s+1}\frac{1}{j!}\frac{\eta^{j-s}z^{-r}(1+\tilde zy)^{s-j+1}}{2\pi i}\int_{\P^1}\frac{\d^2y'}{y-y'}\,(1+\tilde zy')^{j-1}\big[\p^j_\eta a_{\by'}\bigr|_{\eta=0},\mfk{t}^a\big] + O(a^2)\,.
\end{multline}
Here we have displayed terms of up to linear order in $a$. The associated positive helicity soft gluon will be given by $\delta a=\Dbar\xi$, which will be supported purely at $z=0,\infty$ due to the poles in $z^{-r}$.

The wedge condition says that $s\geq0$. In the wedge, we define a basis of $S$-algebra currents $J[s](z,\tilde z)$ by setting
\be
J^a[s](z,\tilde z) = J_{\xi^a_{r=0}[s]}\,.
\ee
Setting $r=0$ here is just a convenient abbreviation for stripping off the factor of $z^{-r}$ from $\xi^a_r[s]$ whose only role is to pick out a mode of the associated current under the contour integral in $z$. Putting together \eqref{Jxigi} and \eqref{xisol}, we can write out the first few terms in an expansion of these basis currents graded by the number of fields occurring in each term,
\be\label{Jkl}
\boxed{\hspace{0.2cm}\begin{split}
J[s](z,\tilde z) &= \int_{\P^1}\d^2y\;\frac{(1+\tilde zy)^s}{(s+2)!}\,\p_\eta^{s+2}b_\by\Bigr|_{\eta=0}\\
&\hspace{-1cm}+ \frac{1}{2\pi i}\int_{(\P^1)^2}\frac{\d^2y\,\d^2y'}{y-y'}\,\sum_{j=1}^{s+1}\frac{(1+\tilde zy)^{s-j+1}}{(s-j+2)!}\frac{(1+\tilde zy')^{j-1}}{j!}\big[\p^j_\eta a_{\by'},\p^{s-j+2}_\eta b_{\by}\big]\Bigr|_{\eta=0}\\ 
&\hspace{9.8cm}+ O(a^2b)\,,
\end{split}\hspace{0.2cm}}
\ee
where color indices have been suppressed, and all integrals on the right-hand side are evaluated at a fixed value of $z$ and at the boundary divisor $\eta=0$. An expansion in $\tilde z$ gives
\be
J[s](z,\tilde z) = \sum_{k+l=s}\frac{s!}{k!\,l!}\,\tilde z^lJ[k,l](z)\,,
\ee
where $J[k,l](z)$ are the basis currents
\be\label{Jklexp}
J[k,l](z) = \frac{1}{(k+l+2)!}\int_{\P^1}\d^2y\;y^l\,\p_\eta^{k+l+2}b_\by\bigr|_{\eta=0} + O(ab)\,.
\ee
By construction, one concludes conservation: $\p_\bz J[s](z,\tilde z)=0$ as well as $\p_\bz J[k,l](z)=0$.

The first term in our formula \eqref{Jkl} for $J[s]$ is linear in the fields, so it gets identified as the soft charge. Note that its integrand has no hidden poles at $y=\infty$ because under $y\mapsto1/y$, we also send $\eta\mapsto\eta/y$. So
\be
\begin{split}
    \d y\,(1+\tilde zy)^s\,\p_\eta^{s+2}&\mapsto -y^{-2}\d y\cdot y^{-s}(y+\tilde z)^s\cdot y^{s+2}\p_\eta^{s+2}\\
    &= -\d y\,(y+\tilde z)^s\,\p_\eta^{s+2}
\end{split}
\ee
which has no pole at $y=0$. The second term is quadratic in fields and acts as the hard charge. Again, it may be checked that its summands encounter no poles at $y=-\tilde z^{-1},\infty$ or $y'=-\tilde z^{-1},\infty$ precisely in the summation range $1\leq j\leq s+1$. There also appear cubic and higher order terms starting from subleading soft gluons at $s=1$. They can be systematically computed using our solution \eqref{xisol} for the large gauge generator. Due to the boundary conditions on $a,b$, the expansion of $J[s]$ will truncate at $O(a^{s+1}b)$.

Invariance of \eqref{Jkl} under small gauge transformations is guaranteed because we constructed it from a manifestly gauge invariant definition of the Noether charge and never picked a gauge. It can also be tested directly by varying $\delta b_\by = [b_\by,\xi'] + \p_\by\phi' + [a_\by,\phi']$ and $\delta a_\by=\p_\by\xi'+[a_\by,\xi']$, and assuming that $\xi',\phi'\sim O(\eta)$ as $\eta\to0$. When attempting this, one observes telescopic cancellations: the variation of the linear term is cancelled by part of a variation of the quadratic term after an integration by parts, and so on. Equivalently, demanding such a telescopic cancellation can be used to bootstrap the higher order terms in \eqref{Jkl}.

Crucially, this proof of gauge invariance does not use the equations of motion, so we expect these expressions to continue to provide gauge invariant boundary operators in the twistor uplift of the non-self-dual theory \cite{Mason:2005zm} (at least at tree level). However, we expect their conservation $\p_\bz J[k,l]=0$ to become anomalous beyond the self-dual sector. We provide a brief discussion of this possibility in section \ref{sec:discussion}.

\paragraph{Negative helicity currents.} The negative helicity case is similar. We define the celestial charges associated to negative helicity soft gluons as
\be
\til H_\phi = \int_{\p\Sigma}\D^3Z\wedge\Tr\,\phi\,\eta\,\p_\eta a\,,
\ee
which are obtained from \eqref{HHt} by making the choice $\ell=\eta\p_\eta$.\footnote{We are using $\ell=\eta\p_\eta$ instead of just $\ell=\p_\eta$ so that the currents we construct carry their expected conformal weights from past literature like \cite{Costello:2023hmi}.} Substituting $\D^3Z=\eta\,\d\eta\,\d z\,\d y$ and stripping off the contour integral in $z$, we find that these are modes of the 2d currents
\be\label{Jtxin}
\til J_\phi(z) = \oint_{\eta=0}\d\eta\,\eta^2\int_{\P^1}\d^2y\;\Tr\,\phi\,\p_\eta a_\by\,.
\ee
For $\phi$ to be a large gauge transformation, it will have a fourth or higher order pole at $\eta=0$. This will ensure that the contour integral in $\eta$ computes a nonzero residue.

Differentiating the equation of motion
\be
\p_\bz a_\by-\p_\by a_\bz + [a_\bz,a_\by] = 0
\ee
with respect to $\eta$ gives us an equation for $\p_\eta a_\by$,
\be
\bar D_{\bz}(\p_\eta a_\by) = \bar D_\by(\p_\eta a_\bz)\,.
\ee
Using this, we can compute the $\bz$ derivative of $\til J_\phi$,
\be
\p_\bz\til J_\phi = \oint_{\eta=0}\d\eta\,\eta^2\int_{\P^1}\d^2 y\;\Tr\left(\Dbar_\bz\phi\,\p_\eta a_\by - \Dbar_\by\phi\,\p_\eta a_\bz\right)\,.
\ee
Its vanishing imposes $\Dbar_\bz\phi=\Dbar_\by\phi=0$. As with the previous case, holomorphicity of the integrand of the $\eta$ contour integral imposes $\Dbar_{\bar\eta}\phi = 0$. So we find that $\til J_\phi$ is a 2d chiral current when the large gauge transformation $\phi$ satisfies
\be
\Dbar\phi = 0
\ee
away from $z=0,\infty$ and $\eta=0$.

We may solve for $\phi$ from
\be
\Dbar_\by\phi\equiv\p_\by\phi + [a_\by,\phi] = 0
\ee
by plugging in an expansion for $\phi$,
\be
\phi = \sum_{j=-1}^s\eta^{-4-j}\phi_j + O(\eta^{-2}) + \text{terms with $\bar\eta$}\,,
\ee
where we have again truncated the singularities in $\eta$ at some highest order $-4-s$, $s\geq0$. For the leading coefficient we take the trivial background solution
\be
\phi_s = (1+\tilde zy)^s z^{-r}\mfk{t}^a
\ee
which is holomorphic in $y$. For $j\leq s-1$, we again split the coefficients as
\be
\phi_j = (1+\tilde zy)^{j+1}\hat\phi_j\,,\qquad -1\leq j\leq s\,,
\ee
where $\hat\phi_j$ are $\CO(-1)$-valued on the $y$-sphere.

When the dust settles, we find the conserved charges
\be\label{Jphigi}
\boxed{\hspace{1cm}\til J_\phi(z) = \sum_{j=-1}^s\frac{1}{(j+1)!}\int_{\P^1}\d y\wedge\Tr\;\phi_j\p_\eta^{j+2}a\bigr|_{\eta=0}\,,\hspace{0.95cm}}
\ee
and the coefficients of the negative helicity soft gluon symmetry generator,
\be\label{phisol}
\boxed{\hspace{1cm}\hat\phi_j = \sum_{p=1}^{s-j}\sum_{q_1\geq1}\cdots\sum_{q_p\geq1}\delta_{q_1+\cdots+q_p,\,s-j}\,\hat\bigma_{q_1}\cdots\hat\bigma_{q_p}\hat\phi_s\,,\qquad -1\leq j\leq s-1\,.\hspace{0.95cm}}
\ee
The operators $\hat\bigma_{q_i}$ occurring here are the same ones defined in \eqref{bigmas}.

Let us write this out more explicitly up to quadratic order in the fields. The $z^{-r}$ factor can be dropped when stripping the contour integral over $z$, as its only job is to extract a mode. Doing this, we land on a basis of currents generating negative helicity soft gluon symmetries in sdYM,
\be\label{Jtkl}
\boxed{\hspace{0.2cm}\begin{split}
\til J[s](z,\tilde z) &= \int_{\P^1}\d^2y\;\frac{(1+\tilde zy)^s}{(s+1)!}\,\p_\eta^{s+2}a_\by\Bigr|_{\eta=0}\\
&\hspace{-1cm}+ \frac{1}{2\pi i}\int_{(\P^1)^2}\frac{\d^2y\,\d^2y'}{y-y'}\,\sum_{j=1}^{s+1}\frac{(1+\tilde zy)^{s-j+1}}{(s-j+1)!}\frac{(1+\tilde zy')^{j-1}}{(j-1)!}\big[\p^j_\eta a_{\by'},\p^{s-j+2}_\eta a_{\by}\big]\Bigr|_{\eta=0}\\ 
&\hspace{9.8cm}+ O(a^3)\,.
\end{split}\hspace{0.2cm}}
\ee
This expression can be tested or even bootstrapped at higher orders in $a$ by demanding invariance under $\delta a_\by = \p_\by\xi'+[a_\by,\xi']$ for $\xi'$ vanishing to first order at $\eta=0$. It can also be expanded in $\tilde z$ to obtain the basis currents $\til J[k,l](z)$,
\be
\til J[s](z,\tilde z) = \sum_{k+l=s}\frac{s!}{k!\,l!}\,\tilde z^l\til J[k,l](z)\,.
\ee
We explicitly find (cf.\ equation (6.7) of \cite{Costello:2023hmi})
\be\label{Jtklexp}
\til J[k,l](z) = \frac{1}{(k+l+1)!}\int_{\P^1}\d^2y\;y^l\,\p_\eta^{k+l+2}a_\by\bigr|_{\eta=0} + O(a^2)\,.
\ee
The obey the conservation laws $\p_\bz\til J[s](z,\tilde z) = 0$ along with $\p_\bz\til J[k,l](z)=0$.

\paragraph{Recovering the $S$-algebra.} In \eqref{Sclose}, we saw that the algebra of large gauge generators satisfying $\Dbar\xi=\Dbar\phi=0$ closes. In this section, we solved for soft gluons in a nontrivial background field $a$. The soft gluon generators associated to modes of the currents $J^a[k,l]$ and $\til J^a[k,l]$ obtained in \eqref{Jklexp} and \eqref{Jtklexp} took the form of flat space soft gluons corrected order-by-order with terms that are less singular near the boundary of twistor space,
\be
\begin{split}
    \xi^a_{k,l,r} &= \eta^{-k-l}y^lz^{-r}\mfk{t}^a + O(\eta^{1-k-l})\,,\\
    \phi^a_{k,l,r} &=\eta^{-k-l}y^lz^{-r}\mfk{t}^a + O(\eta^{1-k-l})\,.
\end{split}
\ee
Starting with the leading singularities in $\eta$ displayed here, we were able to uniquely fix the relevant subleading terms in $\eta$ by solving $\Dbar_{\by}\xi=\Dbar_{\by}\phi=0$.

Consider the positive helicity gauge transformations. The algebra of their basis generators takes the form
\be
\begin{split}
    [\xi^a_{k,l,r},\xi^b_{k',l',r'}]_\star &= \eta^{-k-k'-l-l'} y^{l+l'}z^{-r-r'}[\mfk{t}^a,\mfk{t}^b] + O(\eta^{1-k-k'-l-l'})\,,
\end{split}
\ee
where we used the fact that the leading flat space soft gluons are field-independent. Since the right-hand side has to satisfy $\Dbar_\by[\xi^a_{k,l,r},\xi^b_{k',l',r'}]_\star=0$, our recursive method of solving this PDE ensures that it has to again be a soft gluon of the type $\xi^c_{k+k',l+l',r+r'}$. Thus, we conclude that our soft gluons obey
\be
[\xi^a_{k,l,r},\xi^b_{k',l',r'}]_\star = f^{ab}{}_c\,\xi^c_{k+k',l+l',r+r'}\,.
\ee
A similar uniqueness argument also yields the remaining entries of the algebra,
\be
\begin{split}
    [\xi^a_{k,l,r},\phi^b_{k',l',r'}]_\star &= f^{ab}{}_c\,\phi^c_{k+k',l+l',r+r'}\,,\\
    [\phi^a_{k,l,r},\phi^b_{k',l',r'}]_\star &= 0\,.
\end{split}
\ee
This ensures that the associated modes of our celestial currents will satisfy the $S$-algebra.


\subsection{Source, response, and Koszul duality}

In holography, at weak coupling in the bulk, states of the bulk theory that violate boundary conditions occur in one-to-one correspondence with (gauge invariant) operators in the boundary theory. In setting up the extrapolate dictionary, one takes such a bulk state and generates a creation operator for it using the symplectic structure on the phase space of the bulk theory. Such creation operators often localize to points on the boundary, so they provide bulk proxies for the dual boundary operators. See eg.\ \cite{Harlow:2011ke} for an introduction to various operator dictionaries in conventional holography.

We would like to realize this intuition in celestial holography, and twistor space provides the perfect arena for it. As we are working bottom-up, we will attempt to make contact with the formalism of Costello and Paquette in \cite{Costello:2022wso}, which will be adapted to Euclidean signature flat space. They consider the question of computing the form factor of a gauge-invariant local operator $O(x)$ in sdYM via uplifting $O(x)$ to twistor space. Their main result states that each such spacetime operator $O(x)$ can be realized as the path integral of a 2d chiral CFT living on the twistor line $L_x$,
\be\label{Olift}
O(x) = \int\rD\psi\;\e^{-S_\text{2d}[\psi,a,b]}\,,\qquad S_\text{2d} = \int_{L_x}\cL_\text{2d}\,.
\ee
Here  $\psi$ collectively denotes the fields of the 2d CFT, and the action $S_\text{2d}$ is an integral of a Lagrangian density over the twistor line $L_x$. Importantly, the 2d theory also contains couplings of the 2d fields to the fields $a,b$ of the twistor action \eqref{twac}.

Such a representation of $O(x)$ constitutes a twistor uplift of $O(x)$, for now the 2d theory comes coupled to the 6d twistor action for sdYM. Let us denote the 6d twistor action as $S_\text{6d}$, and the 4d spacetime action of sdYM as $S_\text{4d}$. Then Costello and Paquette argue the equivalence of the path integrals
\be
\la O(x)\ra = \int\rD A\;\rD B\;\e^{-S_\text{4d}}\,O(x) = \int\rD a\;\rD b\;\rD\psi\;\e^{-S_\text{6d}-S_\text{2d}}\,.
\ee
Integrating out the 2d CFT reproduces the insertion of $O(x)$, and compactifying on the twistor lines reduces $S_\text{6d}$ to $S_\text{4d}$, establishing the equivalence. In special cases, this idea already dates back to \cite{Mason:2005zm}, where such a 2d CFT uplifting $\Tr\,B^2(x)$ to $\PT$ was constructed. Further work on finding explicit examples of such uplifts was carried out in \cite{Bu:2022dis}. The quantum extensions of this idea also required certain 6d gauge anomalies to cancel, which needed the sdYM theory to be coupled to finely tuned matter content \cite{Costello:2021bah,Costello:2022upu,Costello:2023vyy,Fernandez:2024qnu,Dixon:2024mzh,Dixon:2024tsb}.

The 2d CFT representing a given $O(x)$ may not be unique. But all such CFTs for all operators $O(x)$ share some universal features. Assuming appropriate anomaly cancellations, the matter content of all such CFTs sits in representations of the universal $S$-algebra,
\be\label{Salg}
\begin{split}
\sJ^a[k,l](z)\,\sJ^b[m,n](0) &\sim \frac{f^{ab}{}_c}{z}\,\sJ^c[k+m,l+n](0)\,,\\
\sJ^a[k,l](z)\,\til \sJ^b[m,n](0) &\sim \frac{f^{ab}{}_c}{z}\,\til \sJ^c[k+m,l+n](0)\,,\\
\til \sJ^a[k,l](z)\,\til \sJ^b[m,n](0) &\sim 0\,,
\end{split}
\ee
where we have written out the tree-level OPEs. I.e., in each 2d theory, we can build a set of chiral currents $\sJ[k,l]$ and $\til \sJ[k,l]$ obeying these OPEs. One says that the \emph{universal defect CFT} living on the line $L_x$ is the vertex algebra \eqref{Salg} generated by $\sJ[k,l]$ and $\til \sJ[k,l]$. And any specific 2d CFT uplifting a particular operator $O(x)$ is said to furnish a \emph{conformal block} for this vertex algebra.

The second universal feature is that the couplings of the 2d fields and the 6d fields always takes a universal form. Introduce another set of affine coordinates on $\PT$,
\be
z = \frac{\lambda_1}{\lambda_0}\,,\qquad v^\dal = \frac{\mu^\dal}{\lambda_0} = \left(\frac{1}{\eta}\,,\,\frac{y}{\eta}\right)\,.
\ee
Then we can always split the 2d action into a term that does not contain the bulk fields $a,b$, and a pair of universal couplings between $a,b$ and $\sJ[k,l](z),\til \sJ[k,l](z)$,
\be\label{couplings}
S_\text{2d}[\psi,a,b] = S^{(0)}_\text{2d}[\psi] + \sum_{k,l\geq0}\frac{1}{k!\,l!}\int_{L_x}\d z\,\sJ[k,l]\,\p_{\dot0}^k\p_{\dot1}^l a + \sum_{k,l\geq0}\frac{1}{k!\,l!}\int_{L_x}\d z\,\til \sJ[k,l]\,\p_{\dot0}^k\p_{\dot1}^lb\,,
\ee
where $\p_\dal \equiv \p/\p v^\dal$, and color index contractions are suppressed. The positive helicity currents $J[k,l]$ couple to derivatives of $a$ normal to $L_x$, and the negative helicity currents $\til J[k,l]$ couple to normal derivatives of $b$. The OPEs \eqref{Salg} can be bootstrapped by demanding gauge invariance of these couplings. This is a physical application of the idea of \emph{Koszul duality}; see \cite{Costello:2020jbh, Paquette:2021cij} for helpful reviews.

Due to the universal nature of these couplings, not only the 1-point function of $O(x)$ but also the form factors of $O(x)$ in the presence of scattering states can be computed uniformly. Consider the form factor of $O(x)$ in the presence of a collection of scattering states $\delta b_j$, $j=1,\dots,p$, and $\delta a_i$, $i=p+1,\dots, n$.\footnote{Not to be confused with the coordinate $n$ used earlier.} In a background $a,b$, the states satisfy the linearized field equations $\Dbar\delta a_i = \Dbar\delta b_j=0$. Define the currents
\be\label{sJi}
\begin{split}
    \sJ_i &=  \sum_{k,l\geq0}\frac{1}{k!\,l!}\int_{L_x}\d z\,\sJ[k,l]\,\p_{\dot0}^k\p_{\dot1}^l\delta a_i\,,\\
    \til\sJ_j &=  \sum_{k,l\geq0}\frac{1}{k!\,l!}\int_{L_x}\d z\,\til\sJ[k,l]\,\p_{\dot0}^k\p_{\dot1}^l\p_\eta(\eta\delta b_j)\,.\\
\end{split}
\ee
Then the form factor $\cF[O]$ of $O(x)$ in the presence of these scattering states is computed by a correlation function of these 2d currents
\be\label{formCP}
\begin{split}
    \cF[O] &= \la O(x)|\,\til\sJ_1\cdots\til\sJ_p\,\sJ_{p+1}\cdots\sJ_n\ra \\
    &\equiv \int\rD a\;\rD b\;\rD\psi\;\e^{-S_\text{6d}-S_\text{2d}}\,\til\sJ_1\cdots\til\sJ_p\,\sJ_{p+1}\cdots\sJ_n\,.
\end{split}
\ee
The notation $\la O(x)|$ indicates that we should use the 2d CFT associated to the operator $O(x)$ for this computation.

\medskip

A scattering state in our context is also an example of a boundary-condition-violating state. So an independent definition of such form factors can be framed in terms of the extrapolate dictionary. For each scattering state, we can introduce the creation operators
\be\label{creation}
\begin{split}
    J_i &= \int_\Sigma\D^3Z\wedge\Tr\,b\wedge\delta a_i\,,\\
    \til J_j &= \int_\Sigma\D^3Z\wedge\Tr\,\delta b_j\wedge\eta\p_\eta a\,,
\end{split}
\ee
where $\Sigma$ is the celestial Cauchy surface defined as the preimage of $|z|=1$ under the projection $\PT\to\P^1$. Scattering states can be decomposed in a basis comprised of soft gluon states (albeit possibly involving infinite sums). For example, we could set $\delta a_i = \Dbar\xi_i$ and $\delta b_j=\Dbar\phi_j$. In that case, $J_i$ and $\til J_j$ become the Noether charges that we associated to large gauge transformations generated by $\xi_i,\phi_j$ in the previous section. As we saw in great detail, $J_i$ and $\til J_j$ can be localized to integrals over the boundary divisor of $\br\PT$. So they may be thought of as proxies for operators of a boundary CFT living near the line at infinity. 

The standard prescription for the form factor of a spacetime operator $O(x)$ would be to compute the correlation function of $O(x)$ with the creation operators associated to the scattering states,
\be\label{form0}
\cF[O] = \la O(x)\,\til J_1\cdots\til J_p\,J_{p+1}\cdots J_n\ra\,.
\ee
We would like to show that this equals the Costello-Paquette prescription \eqref{formCP}. This would relate the extrapolate dictionary to Koszul duality.
This is useful because the Koszul duality approach gives the most efficient way of calculating such correlators, as witnessed in the remarkable two-loop calculation in \cite{Costello:2023vyy}.

The first step is to write the creation operators \eqref{creation} as integrals over all of twistor space. Since $\Sigma$ wraps the fibers of $\PT\to\P^1$, we simply need to turn the contour integral over $z$ to an integral over all of $z$. This may be arranged as follows. As we are in flat space, we can always work with states $\delta a_i$ that have a delta function support at a specific $z=z_i$, etc. The contour of integration over $z$ can be moved to a small contour surrounding $z=\infty$, or to any other point away from all the $z_i$.  Then using Stokes' theorem in reverse, we obtain
\be\label{creation1}
\begin{split}
    J_i &= \int_\PT\D^3Z\wedge\Tr\,\Dbar b\wedge \delta a_i\,,\\
    \til J_j &= \int_\PT\D^3Z\wedge\Tr\,\delta b_j\wedge\Dbar(\eta\p_\eta a)\,,
\end{split}
\ee
where the integrals over $\PT-\{z=\infty\}$ have been extended to all of $\PT$ using the fact that $\delta a_i$ or $\delta b_j$ have no support there. We have also used the linearized field equations $\Dbar\delta a_i=\Dbar\delta b_j=0$.

Next we can introduce a set of constant sources $\veps_i,\veps_j$ to write \eqref{form0} as
\be
\cF[O] = \frac{\p}{\p\veps_1}\cdots\frac{\p}{\p\veps_n}\int\D a\;\D b\;\D\psi\;\e^{-S_\text{6d}-S_\text{2d} + \sum_i\veps_i J_i + \sum_j\veps_j\til J_j}\,\Bigr|_{\veps_i=\veps_j=0}\,.
\ee
We have also replaced $O(x)$ by its twistor uplift \eqref{Olift}. Recall the structure of the 6d action,
\be
S_\text{6d} = \int_{\PT}\D^3Z\wedge\Tr\,b\wedge\Dbar^2\,,
\ee
where $\Dbar^2=\dbar a+\frac12\,[a,a]$. Performing the shifts
\be
a\mapsto a - \sum_i\veps_i\delta a_i\,,\qquad b\mapsto b - \sum_j\veps_j\p_\eta(\eta\delta b_j)
\ee
of the path integration variables, and employing the linearized field equations, we find that
\be
S_\text{6d} - \sum_i\veps_i J_i - \sum_j\veps_j\til J_j \quad\mapsto\quad  S_\text{6d}\,.
\ee
Nonlinear terms in $\delta a_i,\delta b_j$ may be dropped by assuming that they are all proportional to delta functions in $z$ and thereby contain the same differential form $\d\bz$, which wedges to zero when multiplied by itself.

At the same time, because of the linear couplings to $a$ and $b$ present in \eqref{couplings}, the 2d action shifts by
\be
S_\text{2d}\quad\mapsto\quad S_\text{2d} - \sum_i\veps_i\sJ_i - \sum_j\veps_j\til\sJ_j\,,
\ee
where we recalled the structure of the currents \eqref{sJi} in the universal defect CFT. Performing the $\veps_i$ derivatives then establishes the fundamental relation
\be
\boxed{\hspace{1cm}\la O(x)\,\til J_1\cdots\til J_p\,J_{p+1}\cdots J_n\ra = \la O(x)|\,\til\sJ_1\cdots\til\sJ_p\,\sJ_{p+1}\cdots\sJ_n\ra \,.\hspace{0.95cm}}
\ee
The correlator on the left is computed using the algebra of operators living at $\mu^\dal\to\infty$. Whereas the correlator on the right is computed using the Koszul dual to the algebra of operators living at $\mu^\dal=0$.

In summary, adding the creation operators $J_i,\til J_j$ to the 6d action sources the equations of motion near $\mu^\dal=\infty$, and the linear theory states $\delta a_i,\delta b_j$ arise as the responses. These states propagate to $\mu^\dal=0$. And by means of the universal ``open-closed couplings'' \eqref{couplings}, they in turn act as sources for the currents $J_i,\til J_j$ in the ``D1 brane CFT'' $S_\text{2d}$. 


\section{Discussion}
\label{sec:discussion}

In this paper, we have related three different approaches to the $S$-algebra: the perspectives of twistors, spacetime, and holography. Noether's theorem on twistor space forms a common thread that connects all these approaches. It produces corner charges that are relevant to spacetime physics and Carrollian holography, as well as gives rise to 2d CFT currents relevant to celestial holography. We conclude this analysis by providing some comments and directions for future work.

\paragraph{Adding matter and gravity.} Although we have focused on the Yang-Mills case, there will be a very similar story for gravity, where the celestial symmetries form the $\text{LHam}(\C^2)$ algebra. The twistor space discussion in section \ref{sec:twistor} will be very similar in the gravity case extending  \cite{Kmec:2024nmu} where only the positive helicity charges were considered. This was because working in radiative gauge from the start kills the negative helicity charge aspects. As we have seen in this work, the way around this is to relax the gauge conditions considered in \cite{Kmec:2024nmu} by introducing an analog of the frame $f$ that maps the fields from a general gauge to radiative gauge. This would allow one to find soft graviton charges of negative helicity. 

We expect the full celestial algebra for self-dual gravity to be given by the semi-direct direct sum between $\text{LHam}(\C^2)$ and an abelian algebra generated by functions with homogeneity $-6$. Similarly, the hierarchies discussed in section \ref{sec:space} will be very similar and the recursion relations will be related to Bianchi identities on the Weyl tensor. Finally, the holography discussion in section \ref{sec:holo} can be also repeated for the gravitational case, where conservation of corner charges and holomorphicity of celestial currents will be related to gravitational FPR recursion relations. And as in section \ref{sec:extrapolate}, one should be able generalize this construction to obtain soft graviton charges that are manifestly invariant under (small) gauge transformations.

Many of these ideas can also be extended to Einstein-Yang-Mills theory along the lines of the recent works \cite{Agrawal:2024sju,Cresto:2025ubl}. In that context, the $S$-algebra will provide a module of the $\text{LHam}(\C^2)$ algebra. To this, one can also add matter sectors like fermions and axions. A carefully tuned matter content is required in order to extend our results to the quantum level, see eg.\ the comprehensive works of \cite{Costello:2023vyy,Fernandez:2024qnu} for a list of possibilities in gauge theory, and also \cite{Bittleston:2022nfr, Bittleston:2023bzp} for gravitational analogs. We anticipate that classical Ward identities such as $\p_\bz J[k,l]=0$, etc.\ satisfied by our celestial currents will become anomalous at one-loop, and these anomalies will be canceled by such additional matter content flowing in the loops.

\paragraph{Beyond self-duality.} Another interesting avenue for future exploration is to go beyond the self-dual sector. Although the S-algebra was defined by its action on the full Yang-Mills S-matrix, one only expects the transformation group to act as Noether symmetries associated with conservation laws in the self-dual sector. One expects the conservation laws associated to the $S$-algebra to be modified in full Yang-Mills, at least beyond the leading and subleading soft gluon orders. With the techniques that we have developed in this paper, it becomes possible to calculate this modification.

Eg., in equation \eqref{Jtxin}, the celestial current associated to a negative helicity soft gluon $\delta b=\Dbar\phi$ was found to be
\be
\til J_\phi(z) = \oint_{\eta=0}\d\eta\,\eta^2\int_{\P^1}\d y\wedge\Tr\,\phi\,\p_\eta a
\ee
in the affine coordinates $(\eta,y,z)$ on twistor space introduced in \eqref{nyz}. The parameter $\phi$ generated a soft gluon symmetry if $\Dbar\phi=0$ away from small neighborhoods of $z=0,\infty$ and $\eta=0$. Using this condition, we can evaluate the antiholomorphic derivative of $\til J_\phi$,
\be\label{jward}
\p_\bz\til J_\phi = \oint_{\eta=0}\d\eta\,\eta\int_{\P^1}\d y\wedge\Tr\,(\phi\,\p_\bz\ip\p_\eta\Dbar^2)\,.
\ee
In the classical self-dual theory, this vanished because of $\Dbar^2=0$, leading to the Ward identity $\p_\bz\til J_\phi=0$ and any associated soft theorems. However, when non-self-dual interactions are turned on by adding a $\Tr\,B^2$ interaction along the lines of \cite{Mason:2005zm}, this zero curvature condition gets deformed to
\be
\Dbar^2 = fBf^{-1}\bigr|_{0,2}\,,
\ee
where the asd field strength $B$ is pulled back from $\R^4$ to $\PT\simeq\R^4\times\P^1$ and projected onto $(0,2)$-forms on $\PT$.


As a result, the $S$-algebra currents is unlikely to be conserved in the non-self-dual theory. Even if they continue to be conserved, their algebra should be  deformed. At the level of amplitudes, this will become visible beyond the MHV sector. Unfortunately, beyond leading and subleading orders that are governed by singular terms in BCFW recursion \cite{Casali:2014xpa}, directly working out the soft or collinear expansions of N$^k$MHV amplitudes has so far been an unwieldy task. In contrast, equation \eqref{jward} should be valid when inserted in any tree-level amplitude. 

We expect that further study of these obstructions will shed light on how the $S$-algebra symmetries act  beyond the self-dual sector. We hope to return to these questions in the future.


\acknowledgments

We would like to thank Akshay Yelleshpur Srikant for collaboration in early stages of this work. We are grateful to Wei Bu, Eduardo Casali, Kevin Costello, Nicolas Cresto, Laurent Freidel, Marc Geiller, Natalie Paquette, Sabrina Pasterski, Andrea Puhm, Ana-Maria Raclariu and Andrew Strominger for helpful conversations. A.K. is supported
by the STFC. L.M. is supported by the Simons Collaboration on Celestial Holography MP-SCMPS-00001550-08 and by the STFC consolidated grant ST/X000494/1. R.R. is supported by the Titchmarsh
Research Fellowship at the Mathematical Institute and
by the Walker Early Career Fellowship at Balliol College. A.S. is supported by the Gordon and Betty Moore Foundation and the John Templeton Foundation via the Black Hole Initiative.


\bibliographystyle{JHEP}
\bibliography{Biblio}

\end{document}